\documentclass[twocolumn,10pt,aps,pre,showpacs]{revtex4-2}
\usepackage{graphicx}
\usepackage{amsmath}
\usepackage{amsfonts}
\usepackage{amssymb}
\usepackage{color}
\usepackage{float}
\usepackage{mathrsfs}

\begin{document}
\title{Bound-state soliton gas as a limit of adiabatically growing integrable turbulence}

\author{D.\,S.~Agafontsev$^{1,2}$}
\email{dmitrij@itp.ac.ru}
\author{A.\,A.~Gelash$^{2,3}$}
\author{R.\,I.~Mullyadzhanov$^{4,5}$}
\author{V.\,E.~Zakharov$^{2,6}$}

\affiliation{\textit{$^1$Shirshov Institute of Oceanology of RAS, 117997 Moscow, Russia.\\
$^2$Skolkovo Institute of Science and Technology, 121205 Moscow, Russia.\\
$^3$Institute of Automation and Electrometry SB RAS, Novosibirsk 630090, Russia.\\
$^4$Institute of Thermophysics SB RAS, Novosibirsk 630090, Russia.\\
$^5$Novosibirsk State University, Novosibirsk 630090, Russia.\\
$^6$Department of Mathematics, University of Arizona, 857201 Tucson, AZ, USA.}}

\begin{abstract}
We study numerically the integrable turbulence in the framework of the one-dimensional nonlinear Schrodinger equation (1D-NLSE) of the focusing type using a new approach called the ``growing of turbulence''. 
In this approach, we add a small linear pumping term to the equation and start evolution from statistically homogeneous Gaussian noise. 
After reaching a certain level of average intensity, we switch off the pumping and examine the resulting integrable turbulence. 
For sufficiently small initial noise and pumping coefficient, and also for not very wide simulation box (basin length), we observe that the turbulence grows in a universal adiabatic regime, moving successively through the statistically stationary states of the integrable 1D-NLSE, which do not depend on the pumping coefficient, amplitude of the initial noise or basing length. 
Waiting longer in the growth stage, we transit from weakly nonlinear states to strongly nonlinear ones, characterized by a high frequency of rogue waves. 
Using the inverse scattering transform (IST) method to monitor the evolution, we observe that the solitonic part of the wavefield becomes dominant even when the (linear) dispersion effects are still leading in the dynamics and with increasing average intensity the wavefield approaches a dense bound-state soliton gas, whose properties are defined by the Fourier spectrum of initial noise. 
Regimes deviating from the universal adiabatic growth also lead to solitonic states, but solitons in these states have noticeably different velocities and a significantly wider distribution by amplitude, while the statistics of wavefield indicates a much more frequent appearance of very large waves.
\end{abstract}

\maketitle


\section{Introduction}
\label{Sec:Intro}

Statistical analysis of nonlinear integrable systems with random input, called in general \textit{integrable turbulence}~\cite{zakharov2009turbulence}, is a rapidly developing area of theoretical and experimental research. 
Its main difference from the ``ordinary'' weak turbulence is the absence of resonant interactions, so that the collision term in the classic kinetic equation equals zero at any order of the perturbation theory. 
The analytic approach to the integrable turbulence is possible in two opposite situations: when nonlinearity is weak and the modified kinetic equation accounting for the non-resonant interactions can be written~\cite{soh2010effect,suret2011wave}, or when the turbulence can be treated as a soliton gas~\cite{zakharov1971kinetic,el2005kinetic,el2020spectral,el2021soliton}. 
Yet a lot of many interesting types of integrable turbulence with intermediate level of nonlinearity remain out of limits of analytic theory and can be studied by implementation of massive numerical experiments only~\cite{walczak2015optical, agafontsev2015integrable, soto2016integrable, akhmediev2016breather, suret2016single, randoux2016nonlinear, agafontsev2016integrable, manvcic2018statistics, agafontsev2021extreme}. 

Particular attention in the studies of integrable turbulence is paid to the one-dimensional nonlinear Schr{\"o}dinger equation (1D-NLSE) of the focusing type, 
\begin{equation}\label{NLSE}
	i\psi_t + \psi_{xx} + |\psi|^2 \psi=0,
\end{equation}
as it is one of the basic nonlinear equations suitable for the description of rogue waves -- extremely large waves that appear unpredictably from moderate wave background~\cite{kharif2003physical,dysthe2008oceanic,onorato2013rogue,dudley2019rogue}. 
The 1D-NLSE is integrable in terms of the \textit{inverse scattering transform} (IST) method~\cite{zakharov1972exact,novikov1984theory,ablowitz1981solitons}, allowing transformation to the so-called \textit{scattering data} which is in one-to-one correspondence with the wavefield and changes trivially during the motion. 
Transformation to the scattering data includes the calculation of the eigenvalue spectrum for a specific auxiliary linear system, where the wavefield of the 1D-NLSE plays the role of the potential. 
For spatially localized potentials, the eigenvalue spectrum contains discrete (solitons) and continuous (nonlinear dispersive waves) parts, and can be used to completely characterize the wavefield; in this sense, the IST is often viewed as a nonlinear analogue to the Fourier transform~\cite{ablowitz1981solitons}. 
However, to reconstruct the wavefield from the scattering data, one needs to find a solution to the nonlinear system of integral Gelfand--Levitan--Marchenko (GLM) equations. 
For a general case, this can only be done numerically, asymptotically at large time, or in the semi-classical approximation~\cite{lewis1985semiclassical,jenkins2014semiclassical}.

One special property of an integrable system is the conservation of an infinite series of invariants. 
This means, in particular, that different types of initial conditions are characterized by different sets of integrals of motion and, during the evolution, demonstrate different statistical behavior even in the long time, see e.g.~\cite{walczak2015optical,agafontsev2015integrable}. 
Until now, the research of integrable turbulence has been concentrated on studies of evolution from different types of physically relevant initial conditions, such as perturbed condensate~\cite{agafontsev2015integrable,kraych2019statistical,perego2022complexity} and cnoidal waves~\cite{agafontsev2016integrable}, partially coherent wave~\cite{walczak2015optical,suret2016single,agafontsev2021extreme} and its superposition with the condensate~\cite{soto2016integrable,akhmediev2016breather}, and also soliton gas~\cite{anco2011interaction, pelinovsky2013two, costa2014soliton, dutykh2014numerical, el2016critical, shurgalina2016nonlinear, pelinovsky2017kdv, gelash2018strongly, gelash2019bound, redor2019experimental, didenkulova2019numerical, suret2020nonlinear}. 
In such studies, it is implied that the initial conditions are somehow prepared by an external actor, that resembles a setting of laboratory experiment. 
In nature, however, a different situation often takes place when waves (for instance, on the surface of water), currently traveling in a system, were generated from small perturbation in the same system under the influence of a time-limited external forcing. 

In the present paper, developing further the approach suggested in~\cite{agafontsev2020growing}, we mimic this situation and grow turbulence by adding a small pumping to the 1D-NLSE. 
Namely, we start from statistically homogeneous in space Gaussian noise and add a small linear pumping term to the integrable model~(\ref{NLSE}), making it nonintegrable, and wait until the average intensity $\langle|\psi|^{2}\rangle$ of the wavefield reaches a certain level. 
Then, we switch off the pumping and examine the resulting integrable turbulence.

One of the phenomena actively studied in the integrable turbulence is the \textit{statistically stationary state} -- the state, in which statistical characteristics of wavefield are independent of time. 
The existence of such state was suggested in~\cite{zakharov2009turbulence} and later corroborated in numerical simulations for some initial conditions, see e.g.~\cite{walczak2015optical,agafontsev2015integrable,soto2016integrable,akhmediev2016breather,agafontsev2016integrable,gelash2018strongly,gelash2019bound}. 
The stationary state can be considered as an analogy to thermodynamic equilibrium in non-integrable systems and, if it exists, must be defined by the infinite series of integrals of motion, so that different series of invariants determine different stationary states.

A particular design of our numerical experiments allows us to grow integrable turbulence adiabatically, moving successively through the statistically stationary states. 
We achieve this by making the pumping term small compared to the dispersion and nonlinearity terms of the 1D-NLSE, so that the dynamics during the growth stage is driven primarily by the 1D-NLSE while the pumping slowly changes the integrals of motion. 
Also, we start growth stage from small initial noise, so that the initial turbulence is very weakly nonlinear and, therefore, practically stationary. 
As a natural analogy to our setup, one can consider an ideal gas in a box, in which one molecule is added from time to time, so that the system remains in thermodynamic equilibrium state, and this state changes slowly over time.

The 1D-NLSE model with a weak linear dumping or pumping term is a common object of studies in nonlinear physics. 
For instance, this equation appears when studying slowly varying wave packets on the surface of deep water under the action of wind, in the context of Bose-Einstein condensates and in fiber optics; see e.g.~\cite{leblanc2007amplification,onorato2012approximate,chabchoub2013experiments,slunyaev2015wave,konotop2005nonlinear,agrawal2001nonlinear} and references therein. 
In contrast to these studies, which were mainly concentrated on the effects of dissipation or pumping on relatively large waves of various types over a relatively short time, we focus on the very specific problem of adiabatic growth of turbulence. 
As we show, this problem leads to a formulation, in which we start evolution from small noise, use very small pumping and keep the system in the growth stage for a very long time until the wavefield is amplified by hundreds of times in terms of characteristic amplitude $\langle|\psi|\rangle$.

In the present paper, we demonstrate that for sufficiently small initial noise and pumping coefficient, and also for not very wide simulation box (basin length), the turbulence grows in a universal adiabatic regime. 
In this regime, when we switch off the pumping, the resulting integrable turbulence turns out to be stationary and does not depend on the pumping coefficient, amplitude of the initial noise or basin length. 
However, it depends significantly on the Fourier spectrum of initial noise and the current level of average intensity. 
In particular, increasing intensity with a longer evolution in the growth stage, we transit from weakly nonlinear states to intermediate nonlinearity and then to strongly nonlinear states characterized by a high frequency of rogue waves. 
Monitoring this evolution with the IST method, we observe that the solitonic part of the wavefield becomes dominant even when the (linear) dispersion effects are still leading in the dynamics and with increasing average intensity the wavefield approaches a dense bound-state soliton gas, whose properties are defined by the Fourier spectrum of initial noise. 
Regimes deviating from the universal adiabatic growth also lead to solitonic states, but solitons in these states have noticeably different velocities and a significantly wider distribution by amplitude, while the statistics of wavefield indicates a much more frequent appearance of very large waves.


The paper is organized as follows. 
In the next Section~\ref{Sec:TheProblem} we give a general overview for the problem of adiabatically growing integrable turbulence. 
In Section~\ref{Sec:NumMethods} we discuss our numerical methods and parameters. 
In Section~\ref{Sec:Results1} we study statistics of the ``grown-up'' turbulence and in Section~\ref{Sec:Results2} we examine the corresponding wavefields with the IST method. 
The final Section~\ref{Sec:Conclusions} contains conclusions and discussions. 
The paper has also several Appendixes discussing the theoretical formalism and numerical approaches to the IST method, as well as additional results related to the grown-up turbulence.


\section{Growing of integrable turbulence}
\label{Sec:TheProblem}


\subsection{Formulation of the problem}
\label{Sec:TheProblem:A}

We examine the long-time statistics of solutions to the following problem,
\begin{eqnarray}
	\Psi|_{\tau=0} = C_{0}h(\xi),\quad \overline{|h|^{2}}=1, && \nonumber\\
	\left\{\begin{array}{rlllc} 
		i\Psi_{\tau} + \beta\Psi_{\xi\xi} + \gamma|\Psi|^{2}\Psi = i p\Psi, & \mbox{while} & \overline{|\Psi|^{2}}<C_{f}^{2},\\
		i\Psi_{\tau} + \beta\Psi_{\xi\xi} + \gamma|\Psi|^{2}\Psi = 0,       & \mbox{for}   & \overline{|\Psi|^{2}}=C_{f}^{2}. 
	\end{array}\right. \nonumber
\end{eqnarray}
Here $\tau$ is time, $\xi$ is spatial coordinate, $\Psi$ is wavefield, $C_{0}h(\xi)$ is the initial noise and $C_{0}$ is its average amplitude, the function $h(\xi)$ has unit average intensity $\overline{|h|^{2}}=1$ and characteristic spatial scale $\delta\xi = \ell$, $C_{f}$ is the final average amplitude, and the positive coefficients $\beta$, $\gamma$ and $p$ describe the dispersion, nonlinearity and pumping, respectively. 
The overline denotes spatial averaging,
$$
\overline{|h|^{2}} = \frac{1}{\Lambda}\int_{-\Lambda/2}^{\Lambda/2}|h|^{2}\,d\xi,
$$
over a wide simulation box $\xi\in[-\Lambda/2, \Lambda/2]$, $\Lambda\gg\ell$, which for the numerical study is taken with periodic boundaries. 
The average intensity $\overline{|\Psi|^{2}}$ changes from $C_{0}^{2}$ at $\tau=0$ to $C_{f}^{2}\gg C_{0}^{2}$ after turning off the pumping, and then remains constant, see Eq.~(\ref{wave-action}) below.

After renormalization $\tau = a\,t$, $\xi = b\,x$, $\Psi=C\psi$ with the coefficients $a = \ell^{2}/\beta$, $b = \ell$ and $C = (\beta/\gamma\ell^{2})^{1/2}$, we obtain the same problem in dimensionless form,
\begin{eqnarray}
	\psi_{t=0} = A_{0}f(x),\quad \overline{|f|^{2}}=1, && \label{NLSE-1}\\
	\left\{\begin{array}{rlllc}
		i\psi_{t} + \psi_{xx} + |\psi|^{2}\psi = ip_{0}\psi, & \mbox{while} & \overline{|\psi|^{2}}<A_{f}^{2},\\
		i\psi_{t} + \psi_{xx} + |\psi|^{2}\psi = 0, & \mbox{for} & \overline{|\psi|^{2}}=A_{f}^{2}, 
	\end{array}\right. \label{NLSE-2}
\end{eqnarray}
where the function $f(x)$ has unit average intensity and unit characteristic spatial scale $\delta x = 1$, while the coefficients $A_{0}=C_{0}/C$, $A_{f}=C_{f}/C$ and $p_{0}=p\ell^{2}/\beta$ denote the renormalized initial and final average amplitudes and pumping, respectively. 
The boundary conditions are periodic with a large period, $x\in[-L/2, L/2]$, where $L=\Lambda/\ell\gg 1$.


\subsection{Conditions for adiabatic growth}
\label{Sec:TheProblem:B}

In the present paper, we understand the adiabatic growth of turbulence as a quasistatic process, which runs successively through the statistically stationary states of the integrable 1D-NLSE. 
To engineer such a growth, we need to ensure that the dynamics is governed primarily by the terms of Eq.~(\ref{NLSE}), and also start evolution from statistical state which is sufficiently close to stationary. 
One way to achieve the latter is to start from small noise, since in this case the initial state is almost linear and, without the pumping, the linear turbulence would be stationary. 
To set the parameters in Eqs.~(\ref{NLSE-1})-(\ref{NLSE-2}) accordingly, we analyze the characteristic time scales associated with the pumping, dispersion and nonlinearity terms, and also consider Eq.~(\ref{NLSE-2}) in the Fourier space. 

When the pumping is on, the characteristic amplitude increases exponentially, see e.g. Eqs.~(\ref{wave-action}),~(\ref{wave-action-t}) below, that defines the characteristic pumping time as $t_{p}=1/p_{0}$. 
The characteristic dispersion time $t_{l}$ is connected with the characteristic length $l$ describing the function $\psi$ as $t_{l}=l^2/2$, see e.g.~\cite{agrawal2001nonlinear}. 
In our numerical experiments with not very large final average amplitudes $A_{f}\lesssim 1$, we observe that the Fourier spectrum of solution has the same characteristic width as the initial noise, meaning that $l$ does not change with time considerably and we may assume $l\simeq \delta x = 1$ at all times. 
Finally, the characteristic nonlinear time is inverse-proportional to the average intensity, $t_{nl}=1/\overline{|\psi|^{2}}$, and the latter changes from $\overline{|\psi|^{2}}=A_{0}^{2}$ at $t=0$ to $A_{f}^{2}$ at the final time. 

Hence, the turbulence should grow in adiabatic regime (i) when the characteristic pumping time is much larger than both the dispersion and nonlinear times, $t_{p}\gg t_{l}$ and $t_{p}\gg t_{nl}$, and (ii) when at the start of the growth stage the system is almost linear, $t_{l}\ll t_{nl}|_{t=0}$. 
These conditions are satisfied at all times if 
\begin{eqnarray}
	\frac{1}{2}\ll \frac{1}{A_{0}^{2}}\ll \frac{1}{p_{0}} \quad\leftrightarrow\quad p_{0}\ll A_{0}^{2} \ll 2. \label{puming-adiabatic-L-old}
\end{eqnarray}

The above set of inequalities can be relaxed, provided that the difference between $p_{0}$ and $2$ is large enough in order of magnitude.
Indeed, let us consider two experiments which differ only by the initial noise amplitudes $A_{1}$ and $A_{2}$, where $A_{1}$ satisfies the conditions~(\ref{puming-adiabatic-L-old}), $p_{0}\ll A_{1}^{2} \ll 2$, and $A_{2}$ is much smaller, $A_{2}^{2}\lesssim p_{0}$. 
Then, in the growth stage of the second experiment, there exists a certain moment of time $t^{*}$ when the average intensity $\overline{|\psi^{(2)}|^{2}}$ reaches level $A_{1}^{2}$ of the initial conditions of the first experiment. 
Meanwhile, the equation of motion~(\ref{NLSE-2}) on the time interval $[0, t^{*}]$ has dispersion as the dominant term, i.e., remains practically linear, so that the nonlinear correlation between the Fourier modes should be negligible and the statistical state of the second experiment at time $t^{*}$ should not differ from the initial conditions of the first one. 
These assumptions, confirmed in our numerical experiments, lead to the following conditions for adiabatic growth of turbulence, 
\begin{eqnarray}
	p_{0}\ll 2, \quad A_{0}^{2} \ll 2. \label{puming-adiabatic-L}
\end{eqnarray}
Effectively, these conditions mean that, as long as the dispersion term is dominant in Eq.~(\ref{NLSE-2}), the ratio between the nonlinearity and pumping is not important.

At the growth stage, it is also instructive to consider the equation of motion~(\ref{NLSE-2}) in the Fourier space, in which there exists an explicit dependency on the wavenumber,
\begin{eqnarray}
	i\frac{\partial\psi_{k}}{\partial t} - k^{2}\psi_{k} + (|\psi|^{2}\psi)_{k} = ip_{0}\psi_{k}. \label{NLSE-k-space}
\end{eqnarray}
Here $(|\psi|^{2}\psi)_{k}$ is the Fourier-transformed nonlinear term of the 1D-NLSE. 
For adiabatic growth of turbulence, we need to ensure that the pumping term is much smaller than all the other terms present in the equation, for all wavenumbers $k=2\pi\, m/L$, $m\in\mathbb{Z}$, including the smallest nonzero ones $k = \pm \Delta k = \pm 2\pi/L$, i.e., 
\begin{eqnarray}
	p_{0}\ll \bigg(\frac{2\pi}{L}\bigg)^{2} \quad\leftrightarrow\quad L\ll \frac{2\pi}{\sqrt{p_{0}}}, \label{puming-adiabatic-k}
\end{eqnarray}
where $\Delta k = 2\pi/L$ is distance between neighbor wavenumbers. 
Note that, for the zeroth harmonic $k=0$, the dispersion term in Eq.~(\ref{NLSE-k-space}) is absent and the condition similar to~(\ref{puming-adiabatic-k}) does not appear. 
For a wide basin length $L\gg 1$, the condition~(\ref{puming-adiabatic-k}) is much more strict for the pumping coefficient $p_{0}$ than Eq.~(\ref{puming-adiabatic-L}).

In the adiabatic regime, the turbulence runs successively through the statistically stationary states, and the result of such an evolution must not depend on its speed. 
Hence, we can use the independence of the wave statistics on the pumping coefficient $p_{0}$ -- and, as follows from Eq.~(\ref{puming-adiabatic-k}), on the basin length $L$ too -- as an additional test of adiabaticity. 
Conversely, if the statistics depends on $p_{0}$ and $L$, then the growth regime is non-adiabatic.

Note that the adiabatic regime of the turbulence growth could be realized if the growth stage begins not with small noise, but with some other initial state, including an essentially nonlinear one, provided that this state is sufficiently close to stationary and the condition~(\ref{puming-adiabatic-k}) is satisfied. 
The growth trajectory of such an adiabatic regime, i.e., the set of statistically stationary states through which the growth passes, must substantially depend on the initial state. 
In particular, in formulation of the problem~(\ref{NLSE-1})-(\ref{NLSE-2}), the noise amplitude might be small enough, so that the initial state is close to stationary and the adiabatic growth regime is realized, and at the same time large enough for the growth trajectory to be affected by the composition of noise, e.g., the presence of coherent structures such as solitons. 
In the present paper, we focus on the universal adiabatic regime, such that the decrease in the initial noise amplitude does not affect its trajectory at late growth stages.

As we discuss below, our simulations indicate that if the conditions~(\ref{puming-adiabatic-L}),~(\ref{puming-adiabatic-k}) are satisfied and the initial noise does not contain solitons, then the turbulence grows in this universal adiabatic regime. 
Moreover, if the conditions~(\ref{puming-adiabatic-L}),~(\ref{puming-adiabatic-k}) are violated or the initial noise contains solitons, then the turbulence grows either in a non-adiabatic regime, which depends on the pumping coefficient $p_{0}$ and basin length $L$, or in a ``non-universal'' adiabatic regime depending on the noise amplitude $A_{0}$.


\subsection{Invariants of the 1D-NLSE}
\label{Sec:TheProblem:C}

Without pumping, Eq.~(\ref{NLSE-2}) reduces to the 1D-NLSE~(\ref{NLSE}), which conserves an infinite series of invariants~\cite{zakharov1972exact,novikov1984theory}. 
In the present paper, we write them in a slightly modified form, 
\begin{eqnarray}
	\mathcal{I}_{j} &=& \frac{1}{L}\int_{-L/2}^{L/2}\psi\,\mathcal{A}_{j}\,dx, \label{integrals_rec1}\\
	\mathcal{A}_{j} &=& \frac{\partial\mathcal{A}_{j-1}}{\partial x} + \frac{1}{2}\psi\sum_{l+m=j-1}\mathcal{A}_{l}\mathcal{A}_{m}, \label{integrals_rec2}
\end{eqnarray}
where $\mathcal{A}_{1}=\psi^{*}$. 
The first three invariants are wave action (in our notations equals the average intensity),
\begin{equation}\label{wave-action}
	\mathcal{N} = \overline{|\psi|^{2}} = \frac{1}{L}\int_{-L/2}^{L/2}|\psi|^{2}\,dx = \sum_{k}|\psi_{k}|^{2},
\end{equation}
momentum
\begin{equation}\label{momentum}
	\mathcal{M} = \frac{i}{2L}\int_{-L/2}^{L/2}(\psi_{x}^{*}\psi-\psi_{x}\psi^{*})\,dx = \sum_{k}k|\psi_{k}|^{2},
\end{equation}
and total energy
\begin{eqnarray}
	&& \mathcal{E} = \mathcal{H}_{l} + \mathcal{H}_{nl}, \label{energy-1}\\
	&& \mathcal{H}_{l} = \overline{|\psi_{x}|^{2}} = \frac{1}{L}\int_{-L/2}^{L/2}|\psi_{x}|^{2}\,dx = \sum_{k}k^{2}|\psi_{k}|^{2}, \label{energy-2}\\
	&& \mathcal{H}_{nl} = -\frac{1}{2}\overline{|\psi|^{4}} = -\frac{1}{2L}\int_{-L/2}^{L/2}|\psi|^{4}\,dx. \label{energy-3}
\end{eqnarray}
Here $\mathcal{H}_{l}$ is the kinetic energy, $\mathcal{H}_{nl}$ is the potential energy and $\psi_{k}$ is the Fourier-transformed wavefield,
$$
	\psi_{k}(t) = \frac{1}{L}\int_{-L/2}^{L/2}\psi(t,x)\,e^{-ikx}\,dx.
$$

When the pumping is on, the functions~(\ref{integrals_rec1})-(\ref{integrals_rec2}) evolve with time. 
The integrands in~(\ref{integrals_rec1}) are polynomials of $\psi$, $\psi^{*}$ and their space derivatives. 
Differentiating these integrands over time, using the equation of motion~(\ref{NLSE-2}) for the time derivatives and taking into account that for $p_{0}=0$ the functions~(\ref{integrals_rec1}) are invariant, we get
\begin{eqnarray}
	d\mathcal{I}_{j}/dt = p_{0}\times\frac{1}{L}\int_{-L/2}^{L/2}\mathcal{B}_{j}\,dx, \label{integrals_rec1-t}
\end{eqnarray}
where $\mathcal{B}_{j}$ are polynomials of $\psi$, $\psi^{*}$ and their space derivatives, which do not depend on $p_{0}$ explicitly. 
In particular, for the wave action, momentum and total energy, the direct calculation yields
\begin{eqnarray}
	d\mathcal{N}/dt &=& 2p_{0}\,\mathcal{N},\label{wave-action-t}\\
	d\mathcal{M}/dt &=& 2p_{0}\,\mathcal{M},\label{momentum-t}\\
	d\mathcal{E}/dt &=& 2p_{0}\,(\mathcal{E} + \mathcal{H}_{nl}). \label{energy-1-t}
\end{eqnarray}
Thus, the wave action and momentum grow exponentially with time, and since $\mathcal{N}|_{t=0}=A_{0}^{2}$ and $\mathcal{N}|_{t=+\infty}=A_{f}^{2}$, the time $t_{0}$ for turning off the pumping is determined as $t_{0}=\frac{1}{p_{0}}\ln \frac{A_{f}}{A_{0}}$. 

Evolution of the total energy is less trivial. 
For the function $\psi$ having characteristic spacial scale $l\simeq 1$ and average intensity $\mathcal{N}=\overline{|\psi|^{2}}$, the kinetic and potential energies can be estimated as $\mathcal{H}_{l}\sim \mathcal{N}/l^{2}$ and $|\mathcal{H}_{nl}|\sim \mathcal{N}^{2}$. 
Then, at the early growth stage when the intensity is small $\mathcal{N}\ll 1$, the potential energy remains small compared to the kinetic one, $|\mathcal{H}_{nl}|\ll \mathcal{H}_{l}$, so that the total energy grows close to exponentially, $\mathcal{E}\propto e^{2p_{0}t}$. 
Later, when the kinetic and potential energies become comparable, evolution of the total energy depends strongly on their interplay. 
The behavior of the next-order integrals~(\ref{integrals_rec1}) is expected to be even more complex.

Note that if two different realizations of the initial noise have the same values of wave action $\mathcal{N}_{0}$ and the same values of momentum $\mathcal{M}_{0}$, then, after turning off the pumping, the two realizations of the grown-up wavefield will also have equal values of these integrals -- $\mathcal{N}_{0}e^{2p_{0}t_{0}}$ and $\mathcal{M}_{0}e^{2p_{0}t_{0}}$, respectively. 
The situation may be different for the next-order integrals, including the total energy, as their evolution relies also on some functions that may depend strongly on specific realization of the wavefield -- for instance, the potential energy in Eq.~(\ref{energy-1-t}). 
Thus, different realizations of the grown-up wavefield are expected to have different values of the total energy and the next-order integrals~(\ref{integrals_rec1}).


\section{Numerical methods}
\label{Sec:NumMethods}


\subsection{Numerical scheme}
\label{Sec:NumMethods:A}

We solve Eq.~(\ref{NLSE-2}) using the pseudo-spectral Runge-Kutta fourth-order method in adaptive grid, with the grid size $\Delta x$ set from analysis of the Fourier spectrum of the solution, see~\cite{agafontsev2015integrable} for detail. 
For more accurate simulation of the growth stage, we rewrite Eq.~(\ref{NLSE-2}) for the function $\upsilon=e^{-p_{0}t}\cdot\psi$, eliminating the right-hand side of the equation. 
After turning off the pumping, we have checked that the first ten integrals of motion~(\ref{integrals_rec1})-(\ref{integrals_rec2}) are conserved by our numerical scheme up to the relative errors from $10^{-10}$ (the first three invariants) to $10^{-6}$ (the tenth invariant) orders.

The function $f(x)$, that defines statistics of the initial noise in Eq.~(\ref{NLSE-1}) and has unit average intensity and unit characteristic spatial scale, is constructed as superposition of linear waves, 
\begin{eqnarray}
	f(x) = \sum_{k}F_{k}\,e^{ikx+i\phi_{k}},\quad F_{k} = G_{s}\,e^{-|k|^{s}}. \label{IC}
\end{eqnarray}
Here $s\in\mathbb{N}$ is the exponent defining shape of the Fourier spectrum $F_{k}$, $\phi_{k}$ are random phases for each $k$ and each realization of the initial noise, $G_{s}=[\pi\, 2^{1/s}/L\,\Gamma_{1+1/s}]^{1/2}$ is the normalization constant such that the average intensity is unity, $\overline{|f|^{2}}=1$, $L$ is the basin length, and $\Gamma$ is the Euler Gamma-function. 
We perform simulations for several profiles of the noise spectrum~(\ref{IC}), including exponential $s=1$, Gaussian $s=2$, and super-Gaussian $s=8$ and $s=32$, and also for one generic (non-symmetric) Fourier spectrum $F_{k}$ (referred to below as $NS$), which is constructed as discussed in~\cite{agafontsev2021extreme}. 
Specifically, the non-symmetric spectrum represents an arbitrary smooth function at small and moderate wavenumbers $|k|\lesssim 1$, decays as $\propto e^{-k^{2}}$ at $k\to -\infty$ and as $\propto e^{-k^{32}}$ at $k\to +\infty$, and has zero momentum~(\ref{momentum}). 

For each of the several numerical experiments presented in this paper, we perform simulations of time evolution in the formulation of Eqs.~(\ref{NLSE-1})-(\ref{NLSE-2}) for an ensemble of several hundred realizations of the initial noise, which differ only in the set of random Fourier phases $\phi_{k}$ in Eq.~(\ref{IC}), and then average the statistical results over these realizations. 
We have checked that larger ensemble sizes lead to the same statistical results.

As a ``base'' numerical experiment, we consider the one with parameters $L=128\pi$, $s=2$, $A_{0}=10^{-2}$, $A_{f}=1$ and $p_{0}=10^{-5}$, so that the pumping is turned off at $t_{0}=\frac{1}{p_{0}}\ln \frac{A_{f}}{A_{0}}\approx 4.6\times 10^{5}$; the statistical results are averaged over $200$ random realizations of the initial noise. 
As can be easily verified, these parameters satisfy the conditions~(\ref{puming-adiabatic-L}),~(\ref{puming-adiabatic-k}) needed for the adiabatic growth of turbulence. 
After turning off the pumping, the kinetic and potential energies should be of unity order for this experiment, $\mathcal{H}_{l}\sim \mathcal{N}/l^{2}\sim 1$ and $|\mathcal{H}_{nl}|\sim \mathcal{N}^{2} = 1$, so that the grown-up turbulence is expected to be strongly nonlinear, $\mathcal{H}_{l}\sim |\mathcal{H}_{nl}|$. 
In our experiments, we study how the wave statistics depends on the pumping coefficient $p_{0}$, initial and final amplitudes $A_{0}$ and $A_{f}$, basin length $L$ and shape of the Fourier spectrum of the initial noise.


\subsection{Measurement of statistical functions}
\label{Sec:NumMethods:B}

After turning off the pumping, we start measurement of the basic statistical functions, averaging them over the ensemble of random realizations of the initial noise. 
We examine the ensemble-averaged kinetic $\langle\mathcal{H}_{l}(t)\rangle$ and potential $\langle\mathcal{H}_{nl}(t)\rangle$ energies, the fourth-order moment of amplitude $\kappa_{4}=\langle\overline{|\psi|^{4}}\rangle/\langle\overline{|\psi|^{2}}\rangle^{2}$, the probability density function (PDF) $\mathcal{P}(I,t)$ of relative wave intensity $I=|\psi|^{2}/\langle\overline{|\psi|^{2}}\rangle$ where $\langle\overline{|\psi|^{2}}\rangle$ is the average intensity, the wave-action spectrum,
\begin{equation}\label{wave-action-spectrum}
	S_{k}(t) = \langle|\psi_{k}|^{2}\rangle/\Delta k,
\end{equation}
where $\Delta k = 2\pi/L$ is distance between neighbor wavenumbers, and the autocorrelation of the intensity,
\begin{equation}\label{g2}
	g_{2}(x,t) = \frac{\langle \overline{|\psi(y+x,t)|^{2}\cdot |\psi(y,t)|^{2}}\rangle}{\langle \overline{|\psi(y,t)|^{2}}\rangle^{2}}.
\end{equation}
In the latter relation, the overline denotes spatial averaging over the $y$ coordinate. 
Note that, at $x=0$, the autocorrelation equals the fourth-order moment, $g_{2}(0,t)=\kappa_{4}(t)$, and at $x\to\infty$ it must approach unity, $g_{2}(x,t)\to 1$.
For the wave-action spectrum and the PDF, we use normalization conditions $\int S_{k}\,dk = \mathcal{N}$ and $\int \mathcal{P}(I)\,dI = 1$, respectively.


\begin{figure*}[t]\centering
    \includegraphics[width=8.0cm]{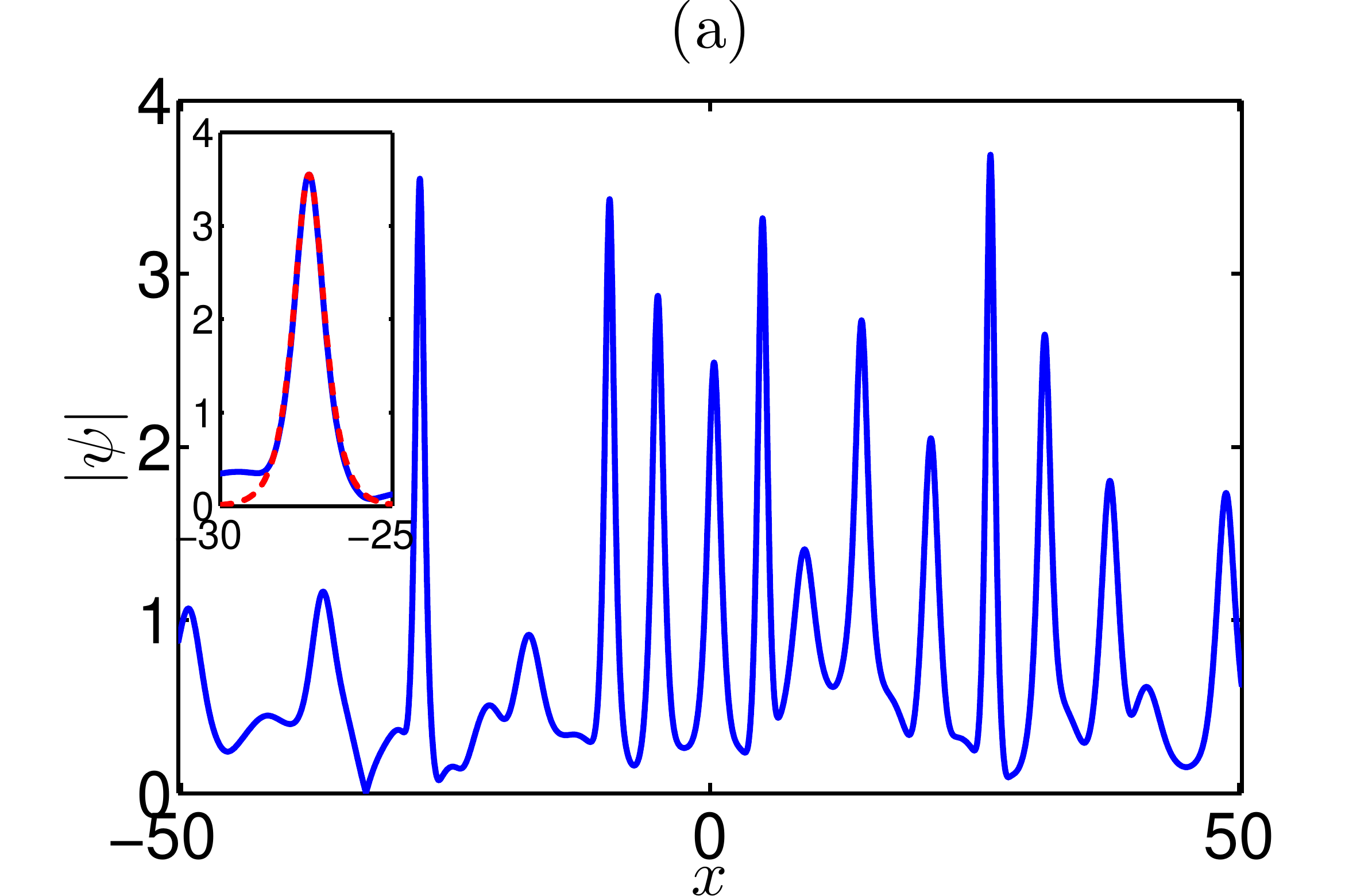}
	\includegraphics[width=9.8cm]{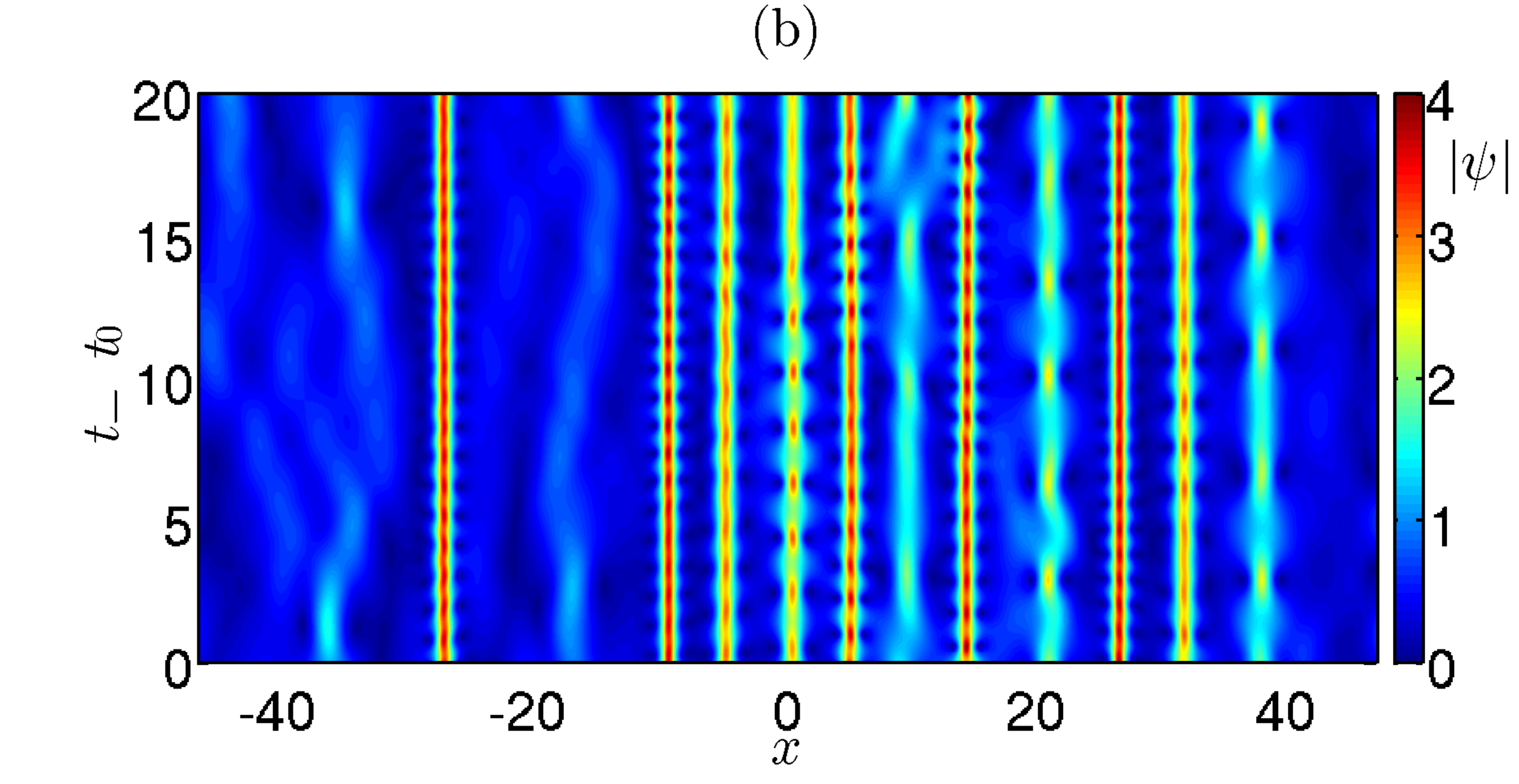}
	
	\caption{\small {\it (Color on-line)} 
	(a) Amplitude $|\psi|$ for a single realization of wavefield at the end of the growth stage $t=t_{0}$ and (b) its subsequent space-time evolution after turning off the pumping, $t\ge t_{0}$. 
	The parameters correspond to the base numerical experiment: $L=128\pi$, $s=2$, $A_{0}=10^{-2}$, $A_{f}=1$ and $p_{0}=10^{-5}$. 
	The inset in panel (a) shows fit for one of the pulses with the one-soliton solution~(\ref{1-SS}). 
	Note that both panels demonstrate only the central quarter of the basin length. 
	}
\label{fig:fig1}
\end{figure*}

\begin{figure*}[t]\centering
	\includegraphics[width=8.9cm]{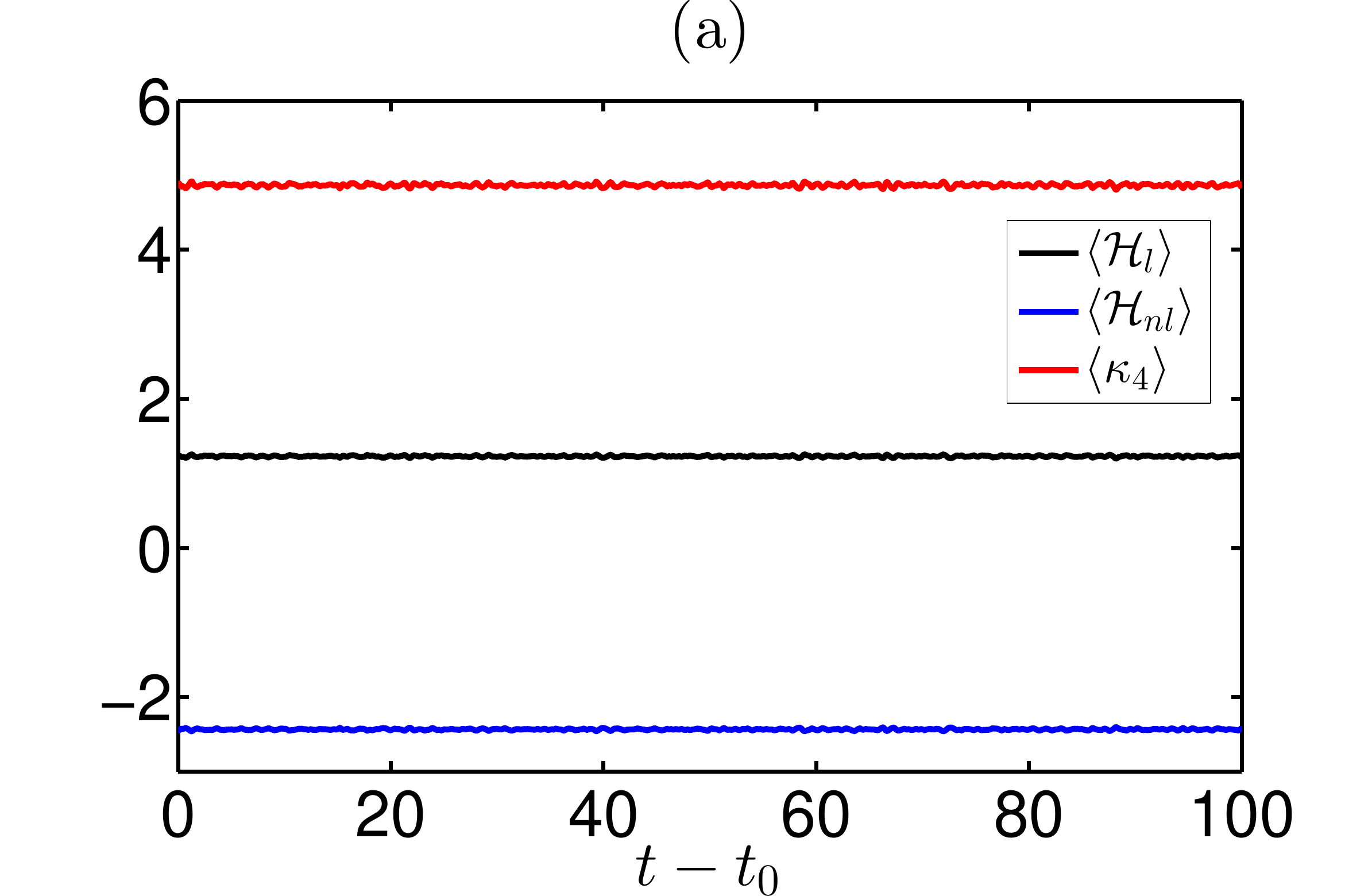}
	\includegraphics[width=8.9cm]{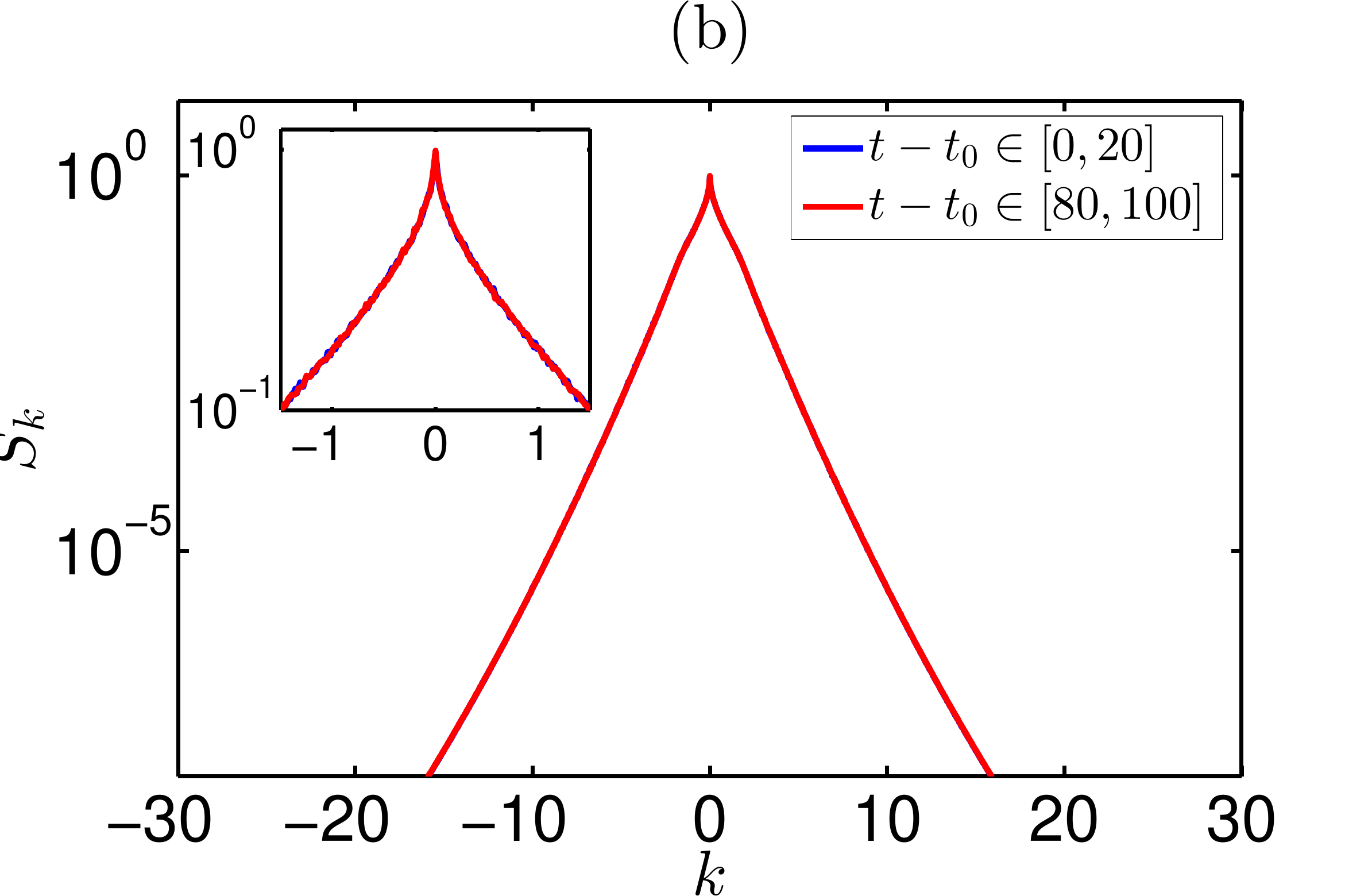}\\
	\includegraphics[width=8.9cm]{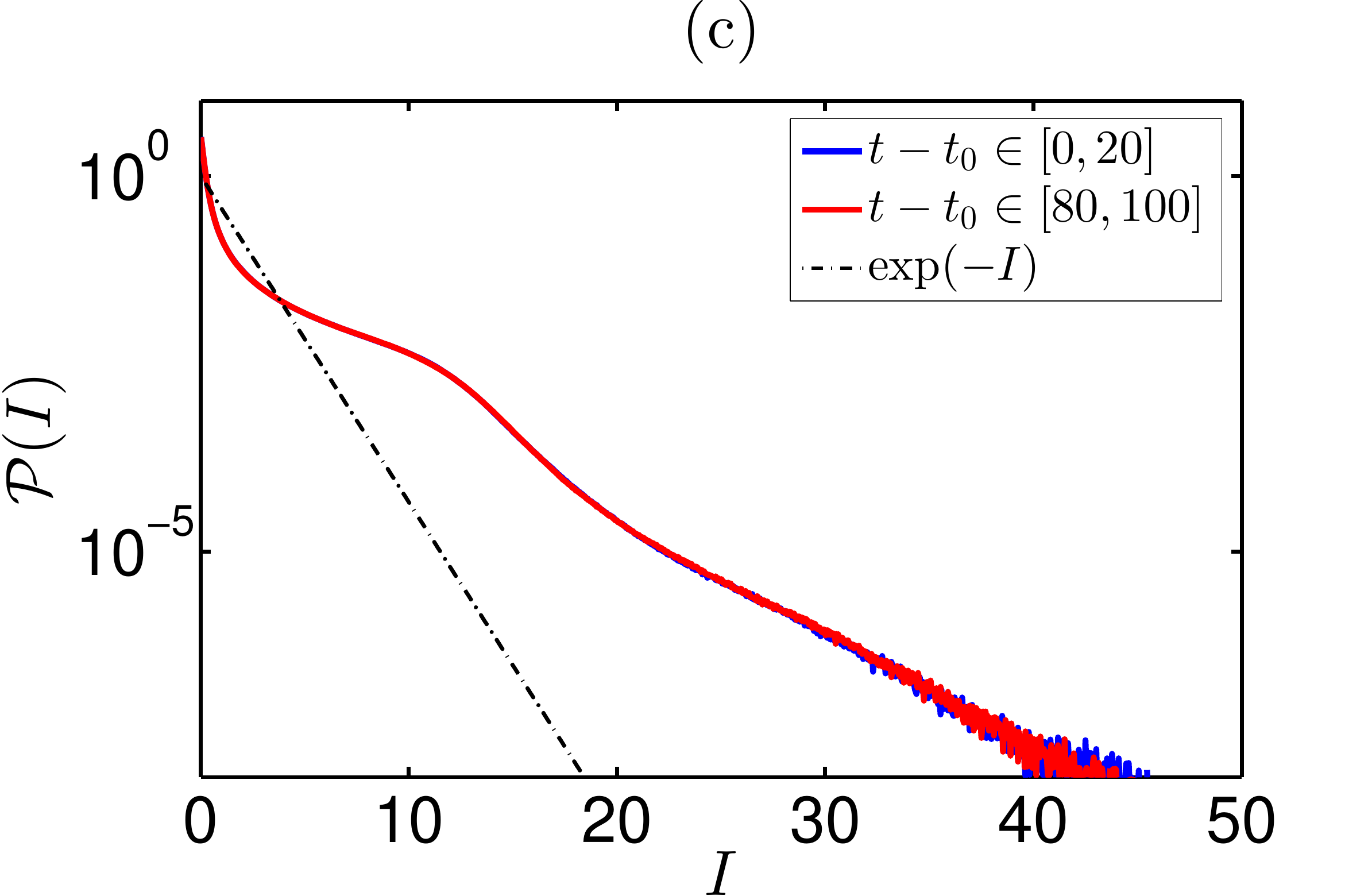}
	\includegraphics[width=8.9cm]{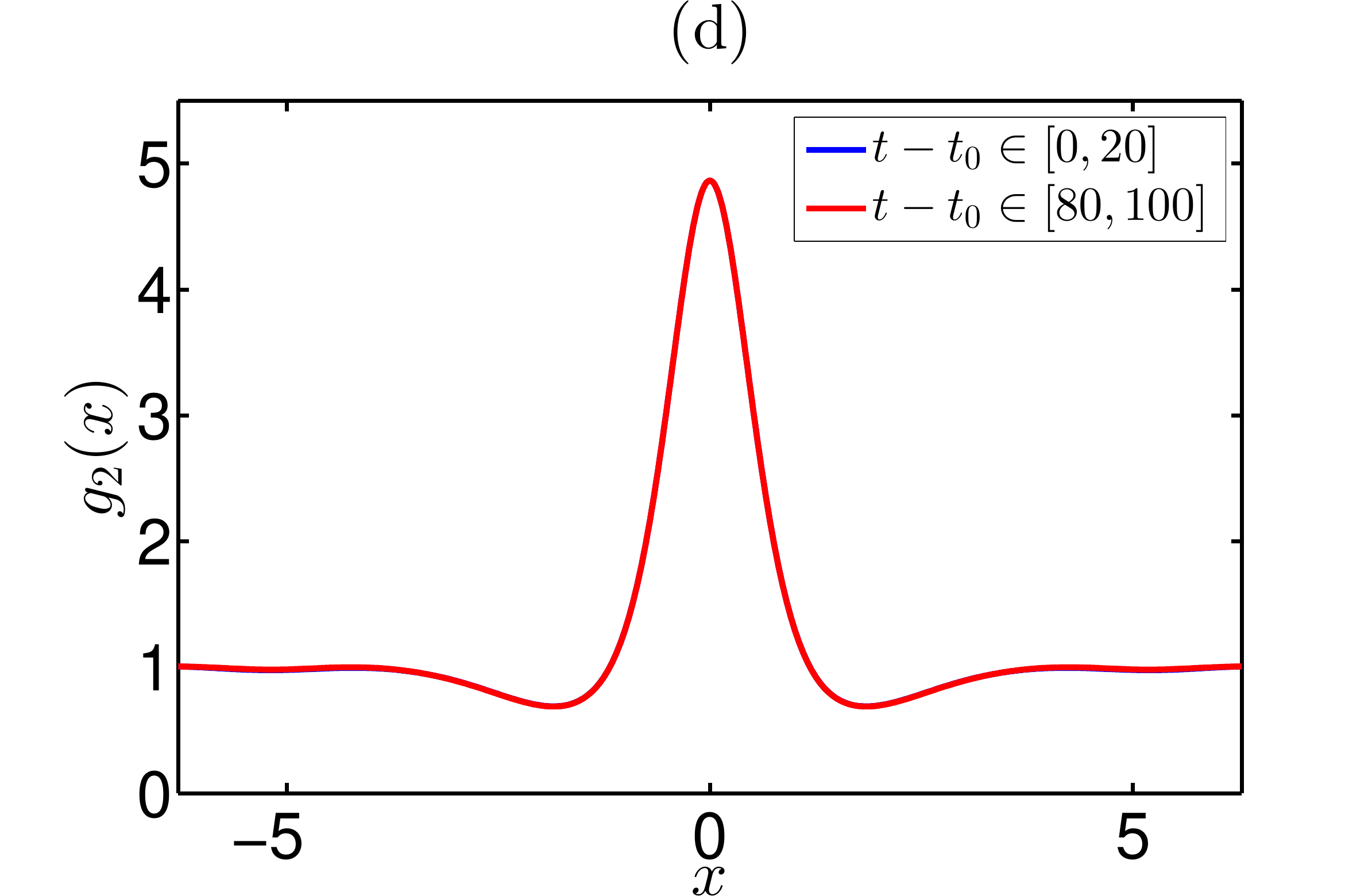}
	
	\caption{\small {\it (Color on-line)} 
	(a) Evolution of the ensemble-averaged kinetic energy $\langle\mathcal{H}_{l}\rangle$, potential energy $\langle\mathcal{H}_{nl}\rangle$ and fourth-order moment of amplitude $\kappa_{4}$ after turning off the pumping, $t\ge t_{0}$. 
	(b-d) Statistical functions averaged over the ensemble and different time intervals $t-t_{0}\in[0,20]$ and $t-t_{0}\in[80,100]$: (b) the wave-action spectrum $S_{k}$, (c) the PDF $\mathcal{P}(I)$ of relative wave intensity $I=|\psi|^{2}/\langle\overline{|\psi|^{2}}\rangle$ and (d) the autocorrelation of intensity $g_{2}(x)$. 
	The parameters correspond to the base numerical experiment: $L=128\pi$, $s=2$, $A_{0}=10^{-2}$, $A_{f}=1$ and $p_{0}=10^{-5}$. 
	The inset in panel (b) shows wave-action spectrum at smaller wavenumbers, and the black dash-dot line in panel (c) indicates the exponential PDF~(\ref{Rayleigh}). 
	}
\label{fig:fig2}
\end{figure*}

\begin{figure*}[t]\centering
	\includegraphics[width=5.9cm]{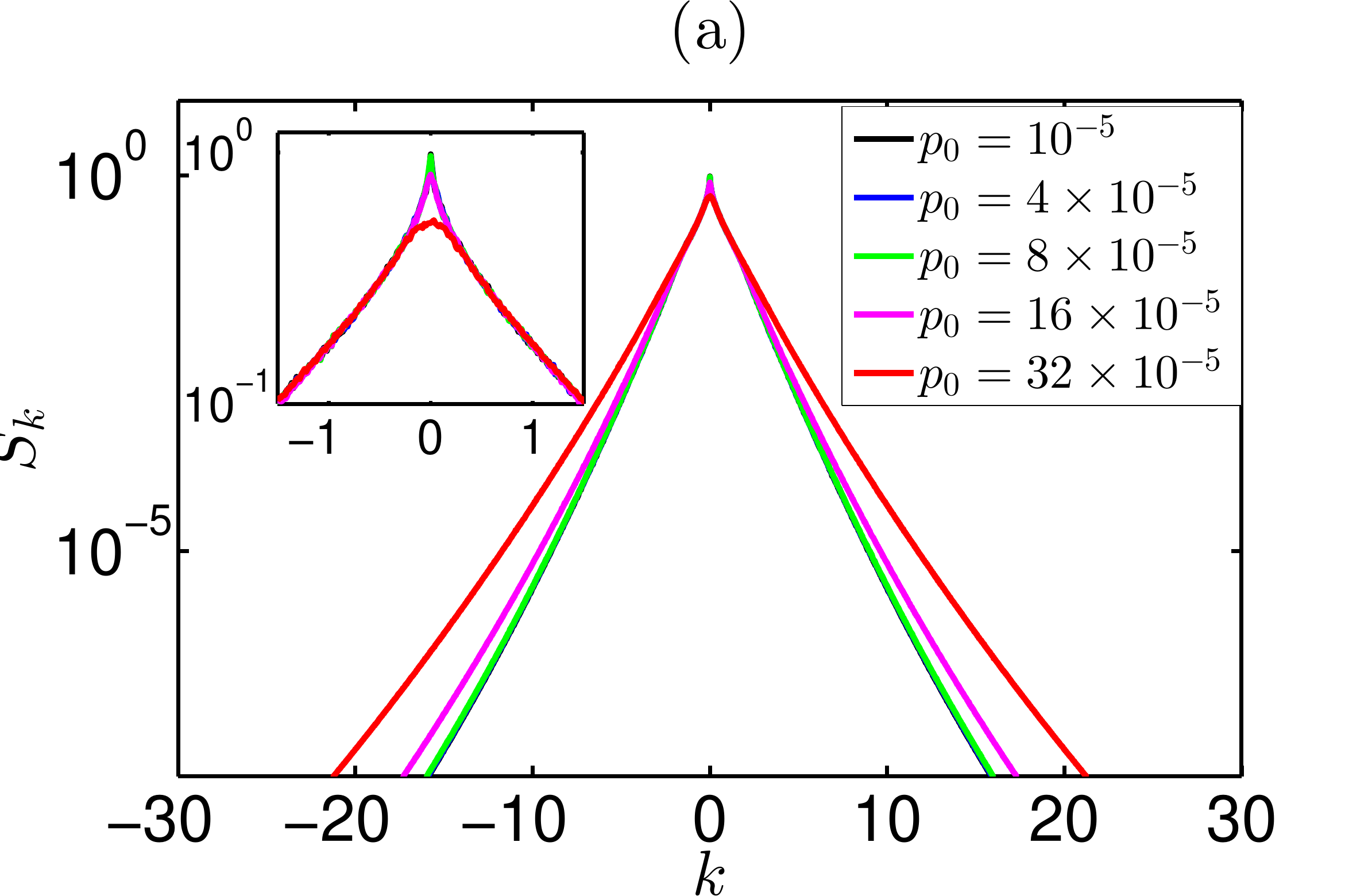}
	\includegraphics[width=5.9cm]{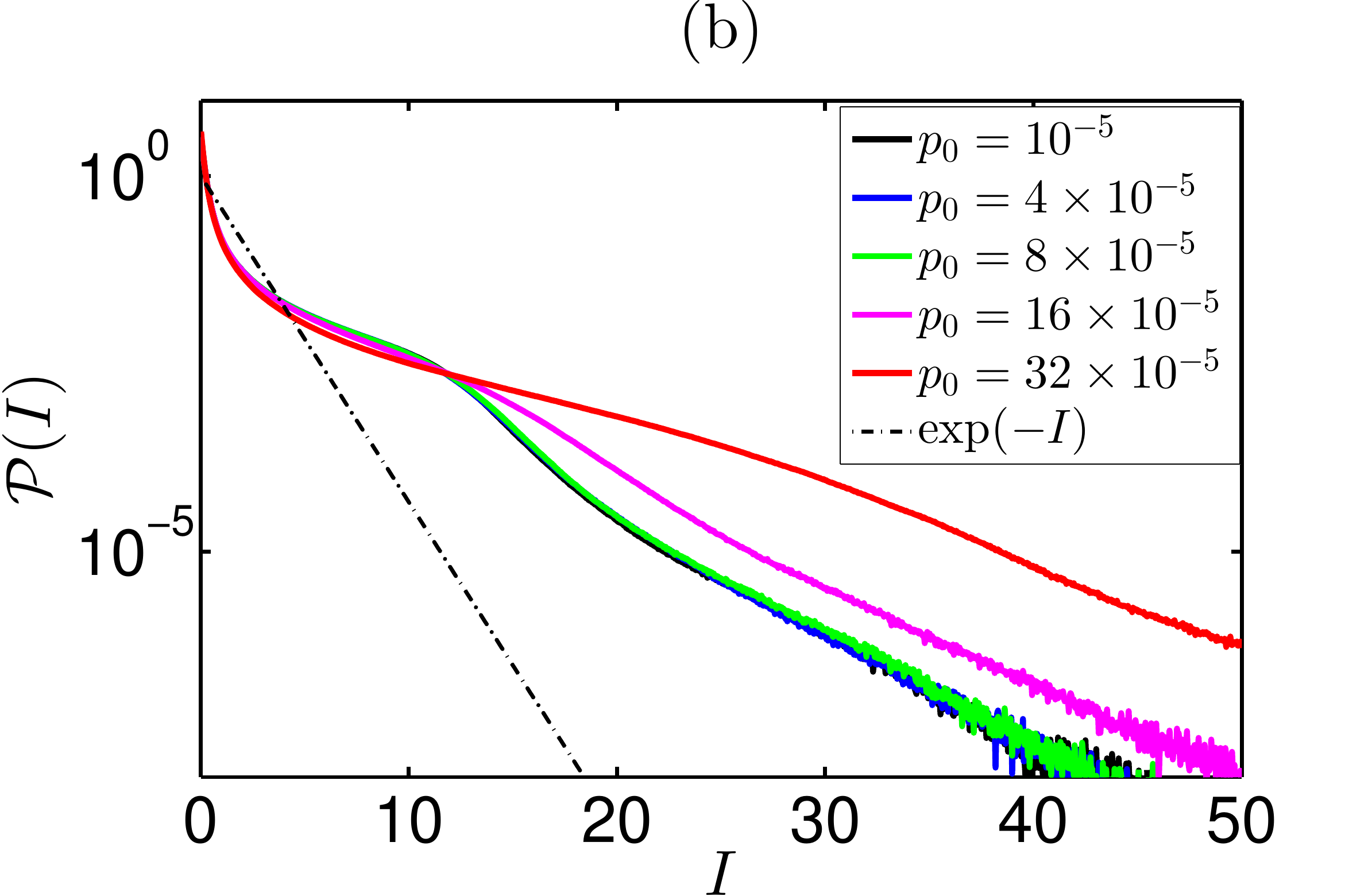}
	\includegraphics[width=5.9cm]{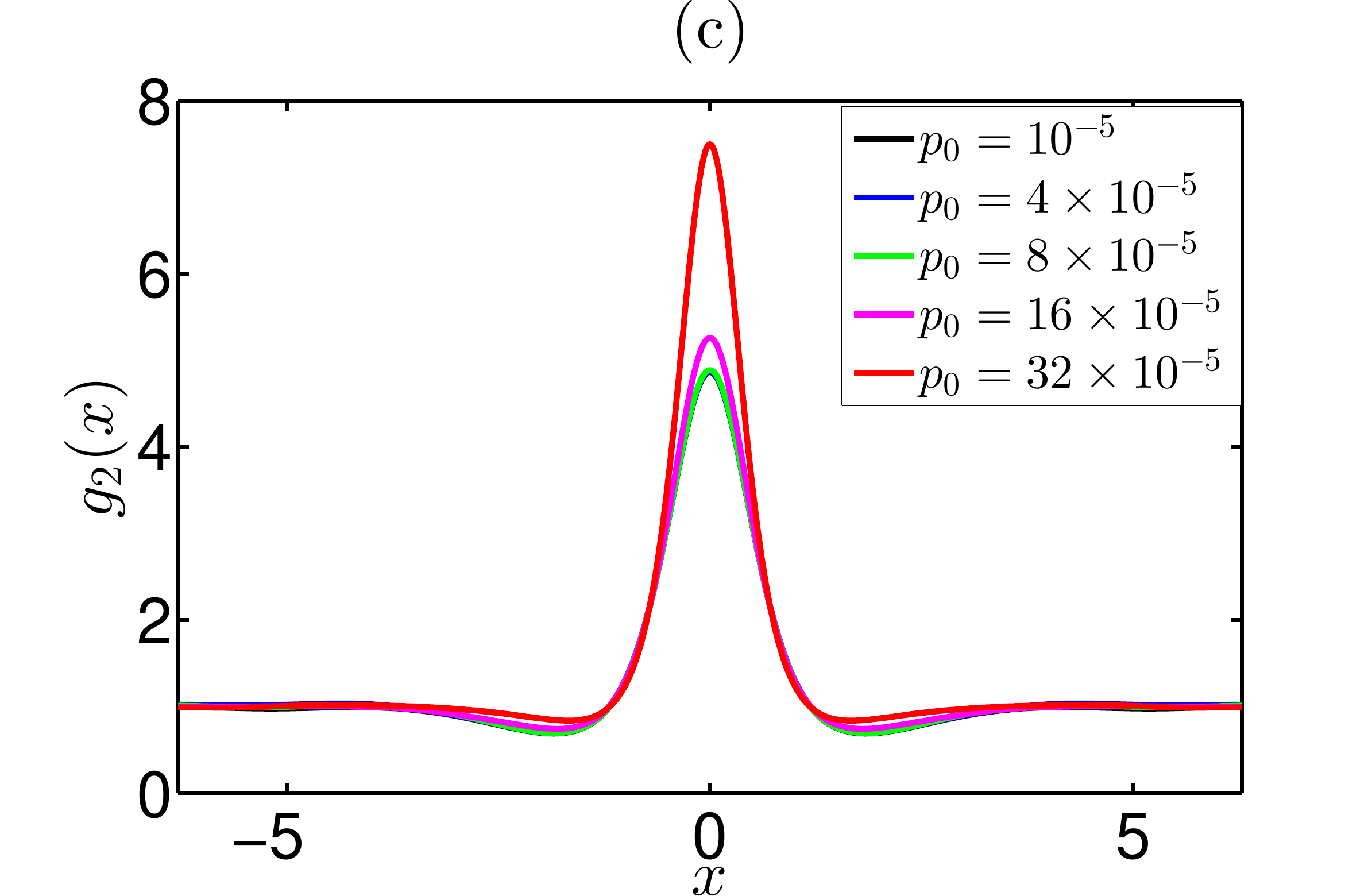}\\
	\includegraphics[width=5.9cm]{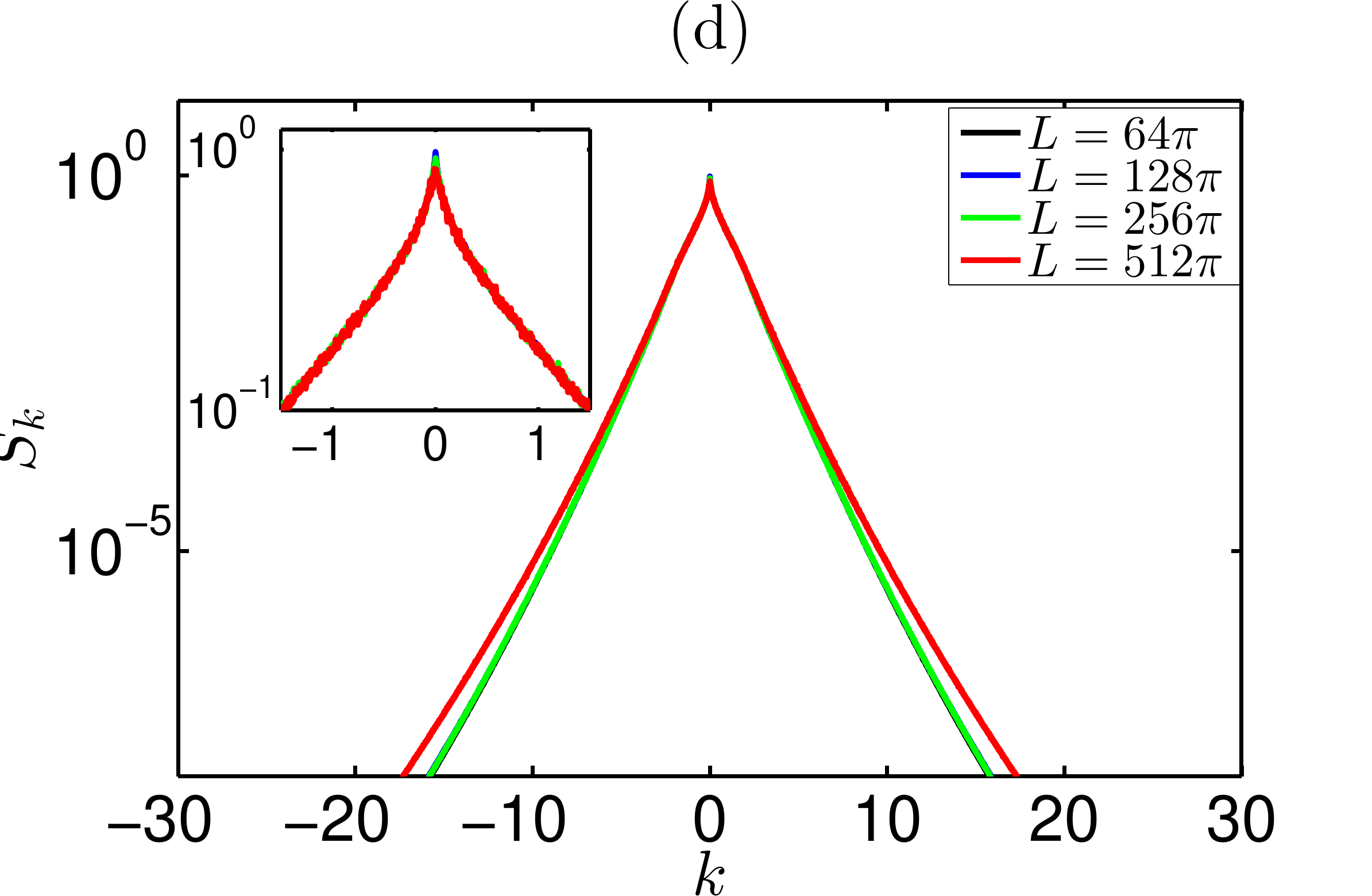}
	\includegraphics[width=5.9cm]{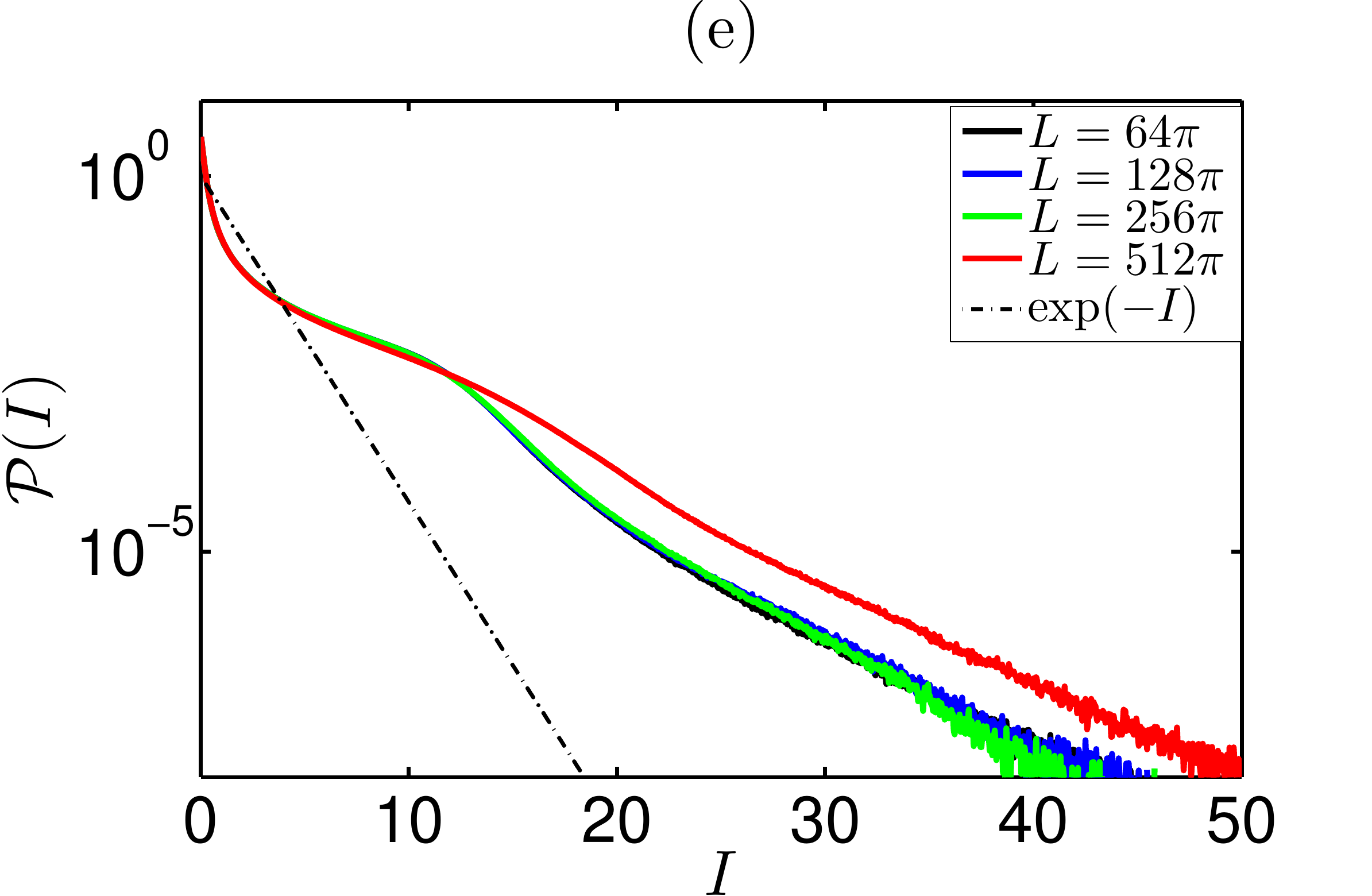}
	\includegraphics[width=5.9cm]{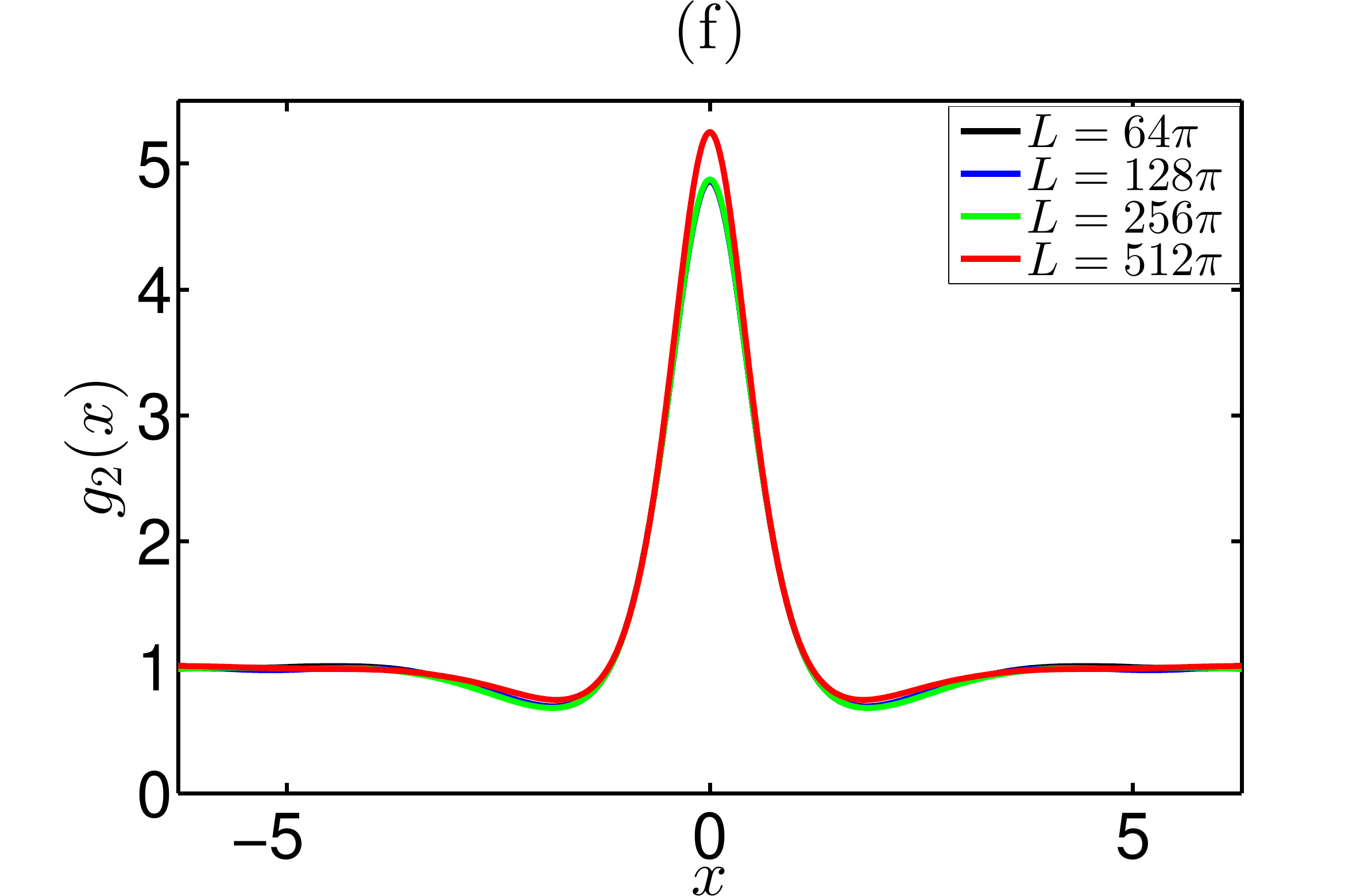}\\
	\includegraphics[width=5.9cm]{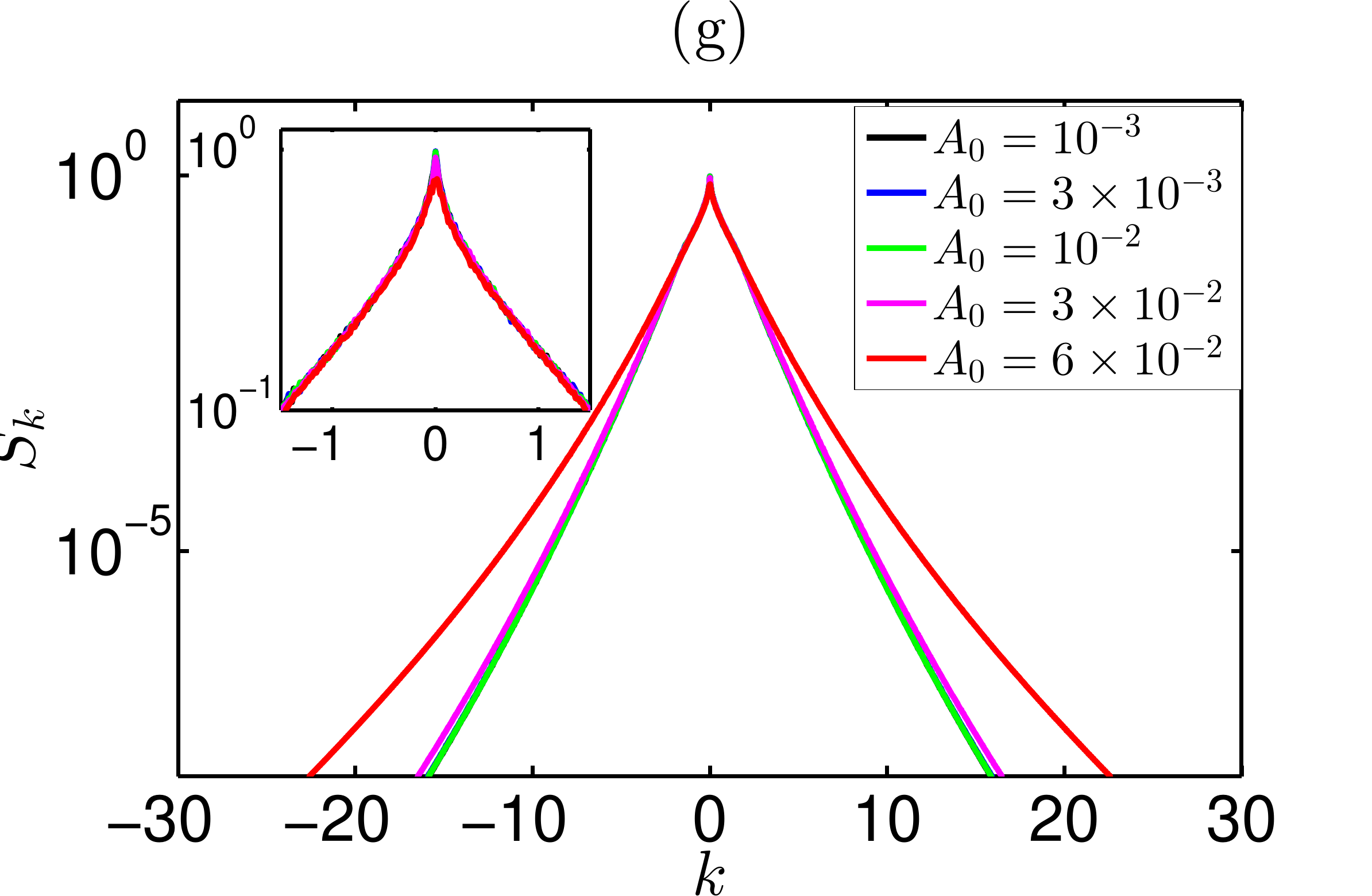}
	\includegraphics[width=5.9cm]{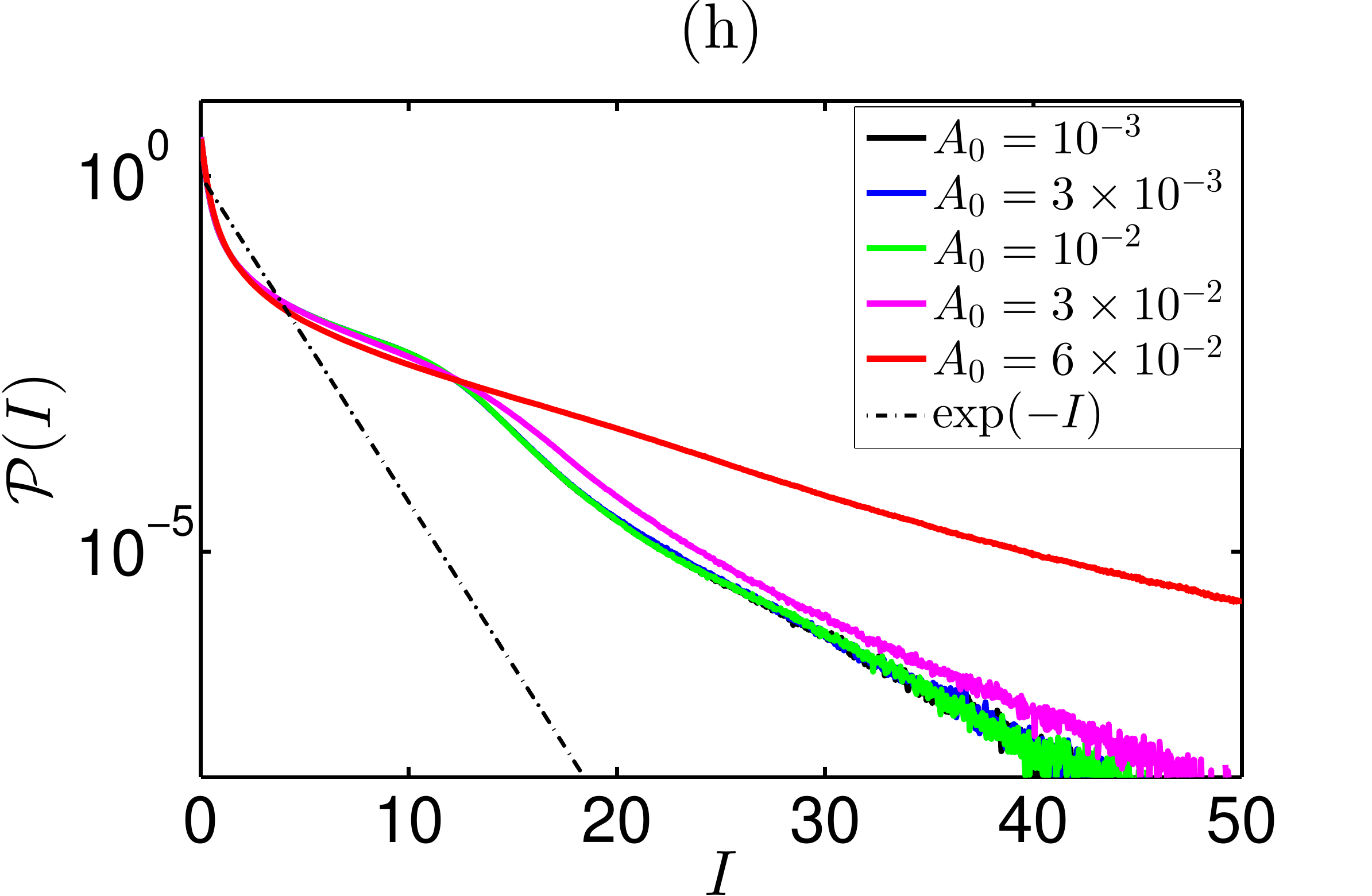}
	\includegraphics[width=5.9cm]{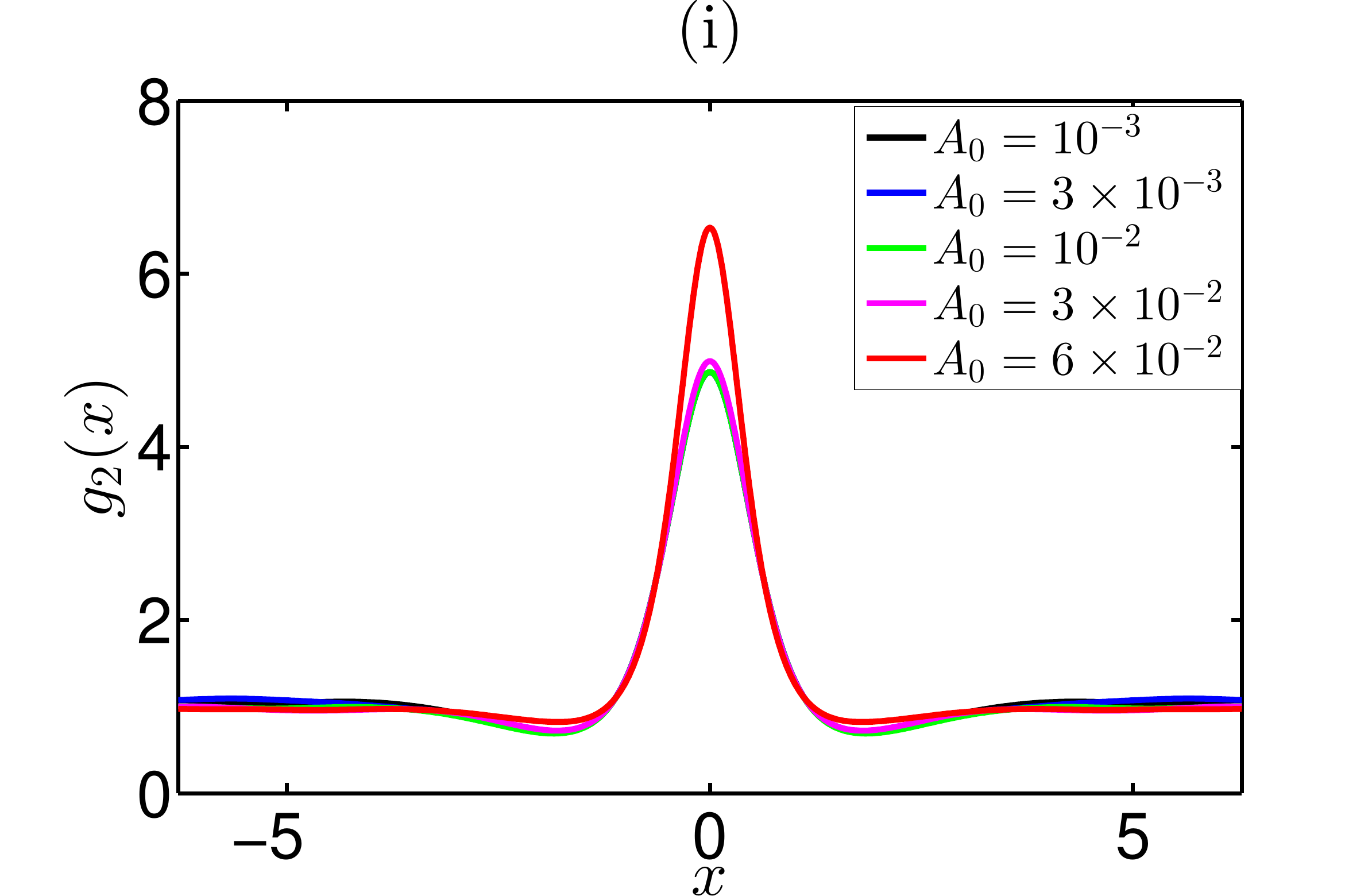}
	
	\caption{\small {\it (Color on-line)} 
	Statistical functions averaged over the ensemble and time interval $t-t_{0}\in[0,20]$ for different sets of numerical experiments: (a,d,g) the wave-action spectrum $S_{k}$, (b,e,h) the PDF $\mathcal{P}(I)$ of relative wave intensity $I=|\psi|^{2}/\langle\overline{|\psi|^{2}}\rangle$ and (c,f,i) the autocorrelation of intensity $g_{2}(x)$. 
	For all the experiments, the initial spectrum is Gaussian, $s=2$, and the final average amplitude equals unity, $A_{f}=1$. 
	Panels (a-c) illustrate experiments with different pumping coefficients $p_{0}$ and fixed $A_{0}=10^{-2}$ and $L=128\pi$. 
	Panels (d-f) show experiments with different basin lengths $L$ and fixed $p_{0}=10^{-5}$ and $A_{0}=10^{-2}$. 
	Panels (g-i) demonstrate experiments with different initial amplitudes $A_{0}$ and fixed $p_{0}=10^{-5}$ and $L=128\pi$. 
	The insets in panels (a,d,g) show wave-action spectrum at smaller wavenumbers, the black dash-dot lines in panels (b,e,h) indicate the exponential PDF~(\ref{Rayleigh}). 
	}
\label{fig:fig3}
\end{figure*}

\begin{figure*}[t]\centering
	\includegraphics[width=5.9cm]{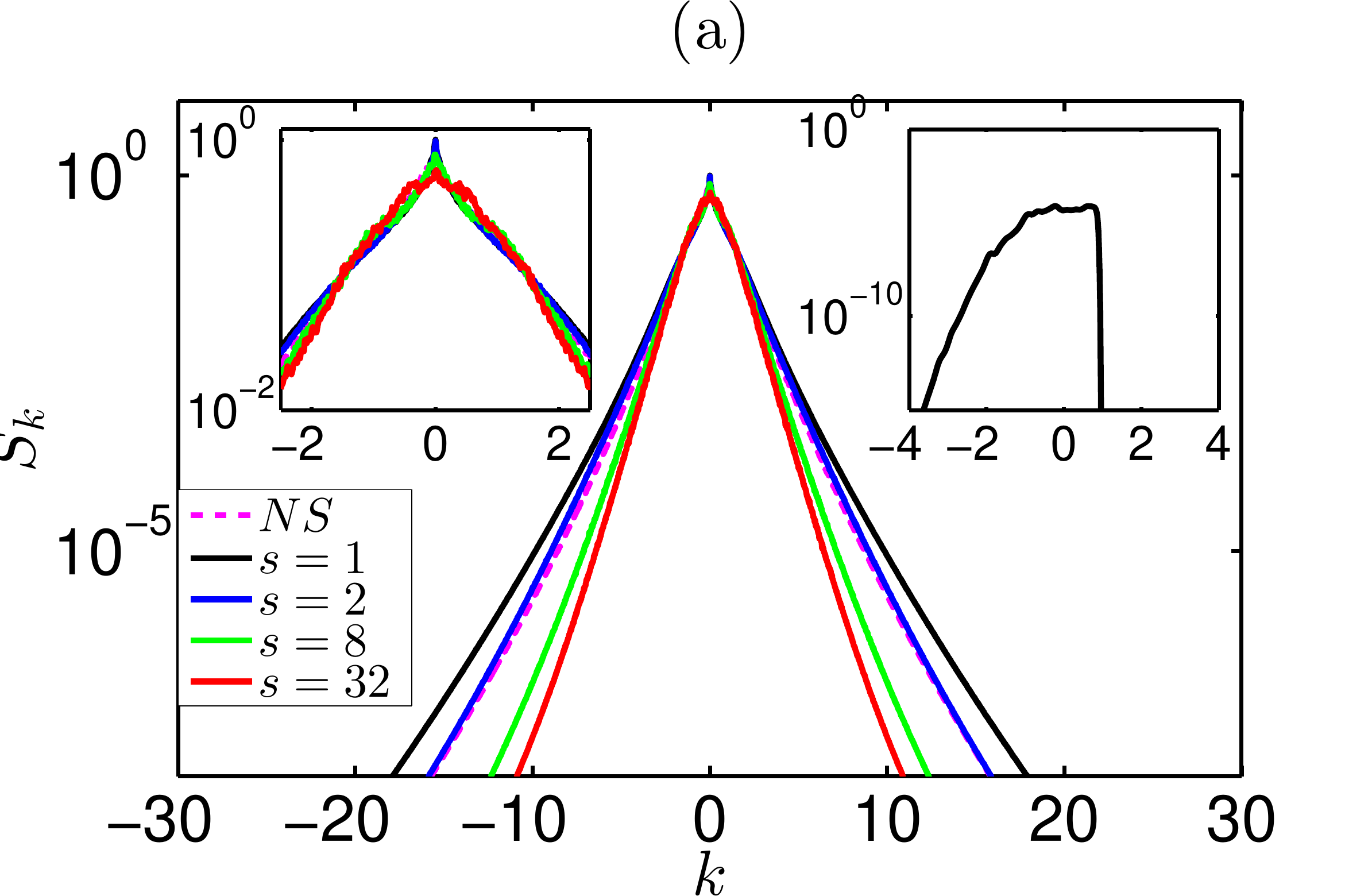}
	\includegraphics[width=5.9cm]{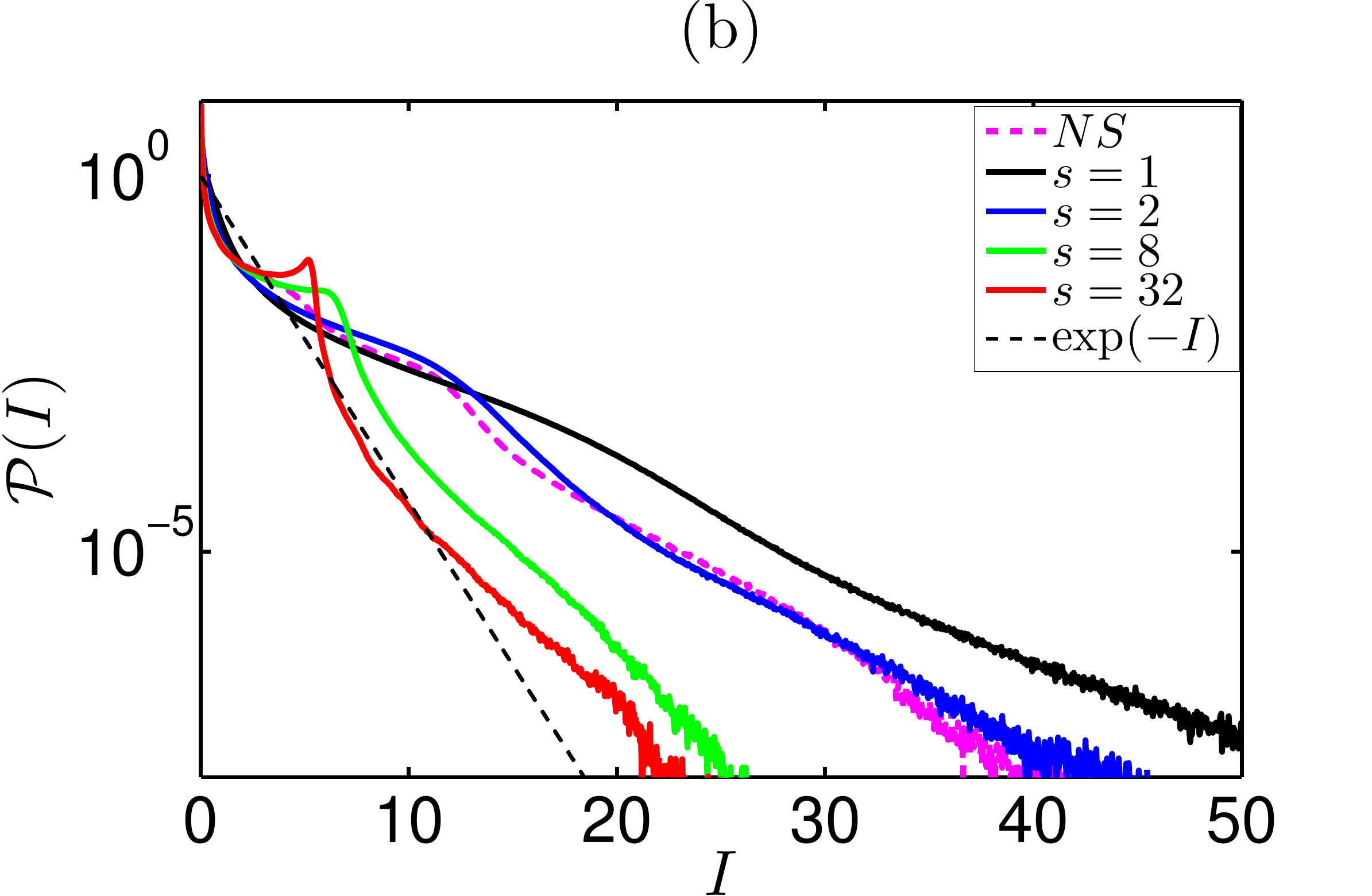}
	\includegraphics[width=5.9cm]{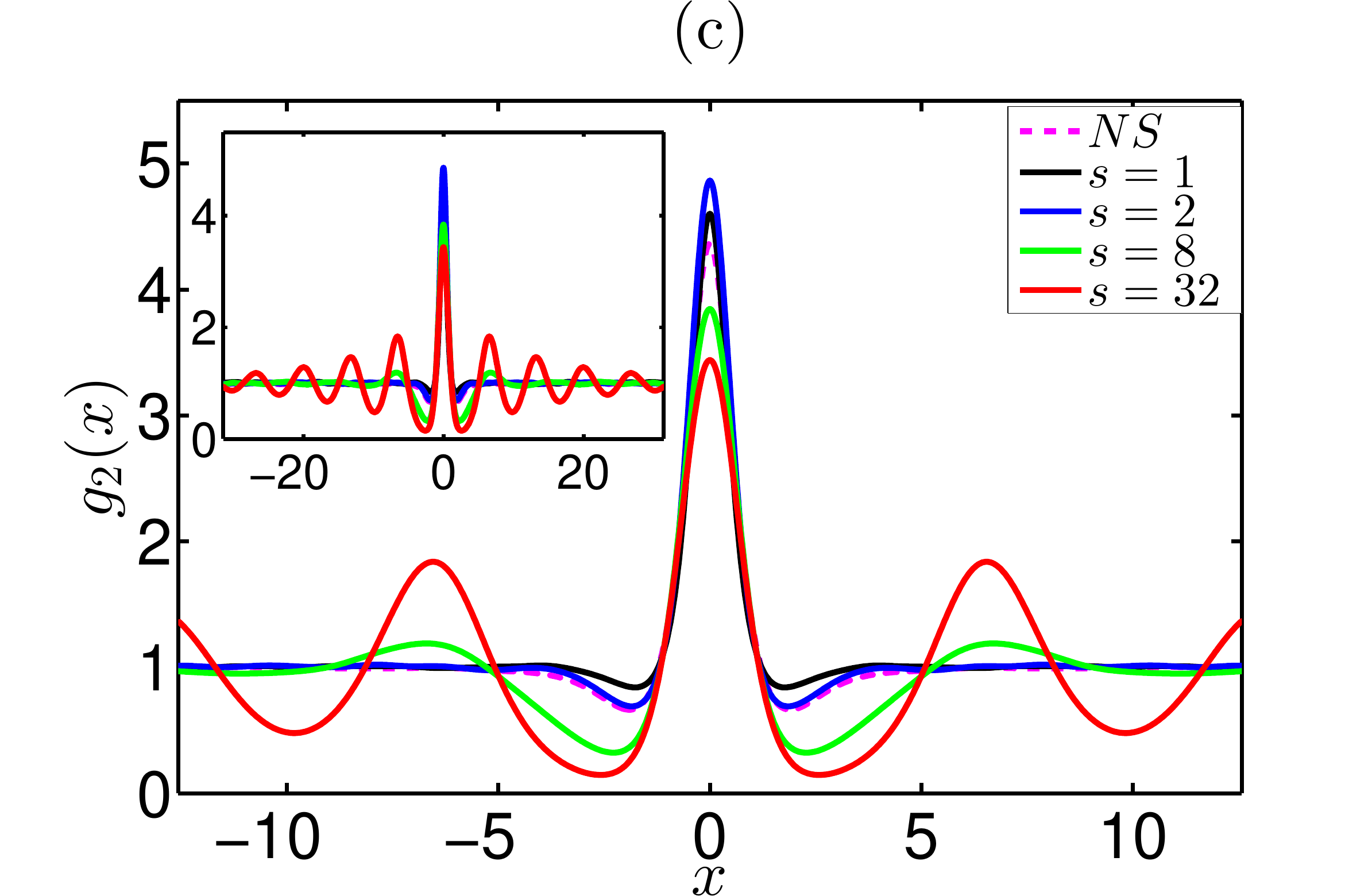}
	
	\caption{\small {\it (Color on-line)} 
	Statistical functions averaged over the ensemble and time interval $t-t_{0}\in[0,20]$ for four super-Gaussian initial spectra with the exponents $s=1$, $2$, $8$ and $32$, and one non-symmetric initial spectrum $NS$, see Section~\ref{Sec:NumMethods:A}: (a) the wave-action spectrum $S_{k}$, (b) the PDF $\mathcal{P}(I)$ of relative wave intensity $I=|\psi|^{2}/\langle\overline{|\psi|^{2}}\rangle$ and (c) the autocorrelation of intensity $g_{2}(x)$. 
	The other parameters are the same as for the base experiment: $L=128\pi$, $A_{0}=10^{-2}$, $A_{f}=1$ and $p_{0}=10^{-5}$. 
	The insets in panel (a) show (left) the wave-action spectrum at smaller wavenumbers and (right) the non-symmetric initial spectrum $NS$, $F_{k}^{2}/\Delta k$, see Eqs.~(\ref{IC}),~(\ref{wave-action-spectrum}). 
	The inset in panel (c) shows autocorrelation of intensity at larger distances. 
	The black dash-dot line in panel (b) indicates the exponential PDF~(\ref{Rayleigh}).
	}
\label{fig:fig4}
\end{figure*}

\begin{figure*}[t]\centering
	\includegraphics[width=5.9cm]{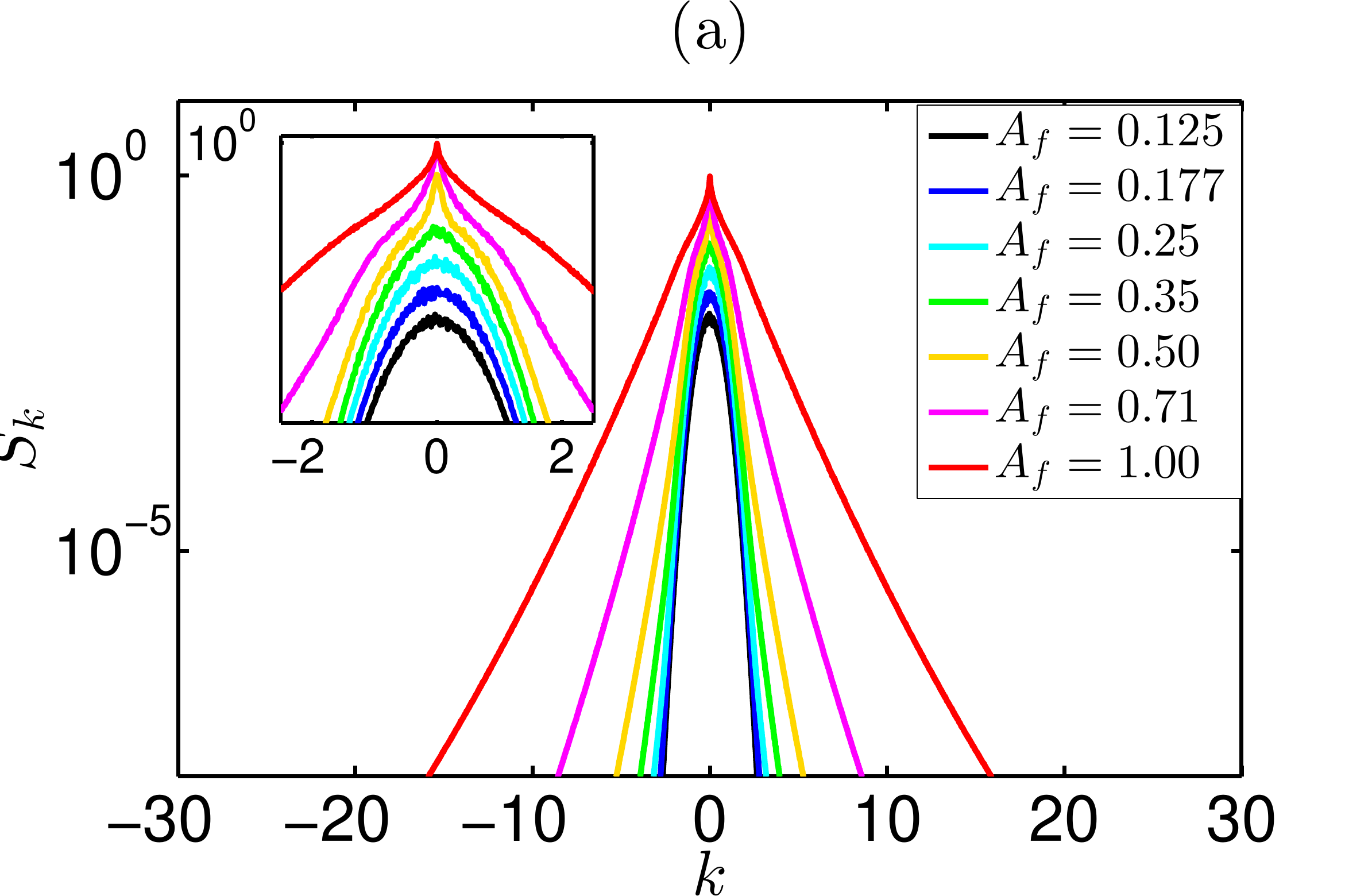}
	\includegraphics[width=5.9cm]{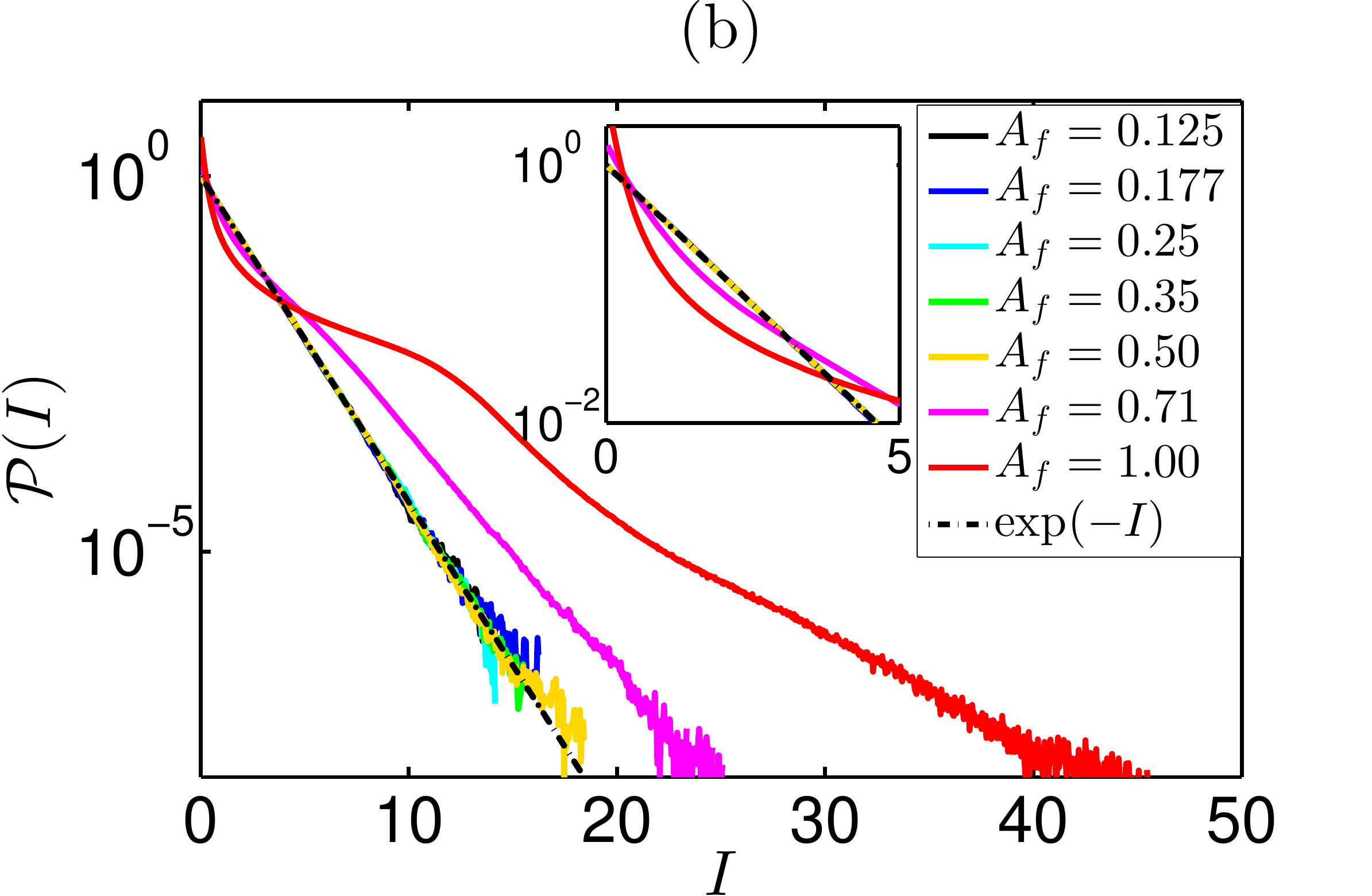}
	\includegraphics[width=5.9cm]{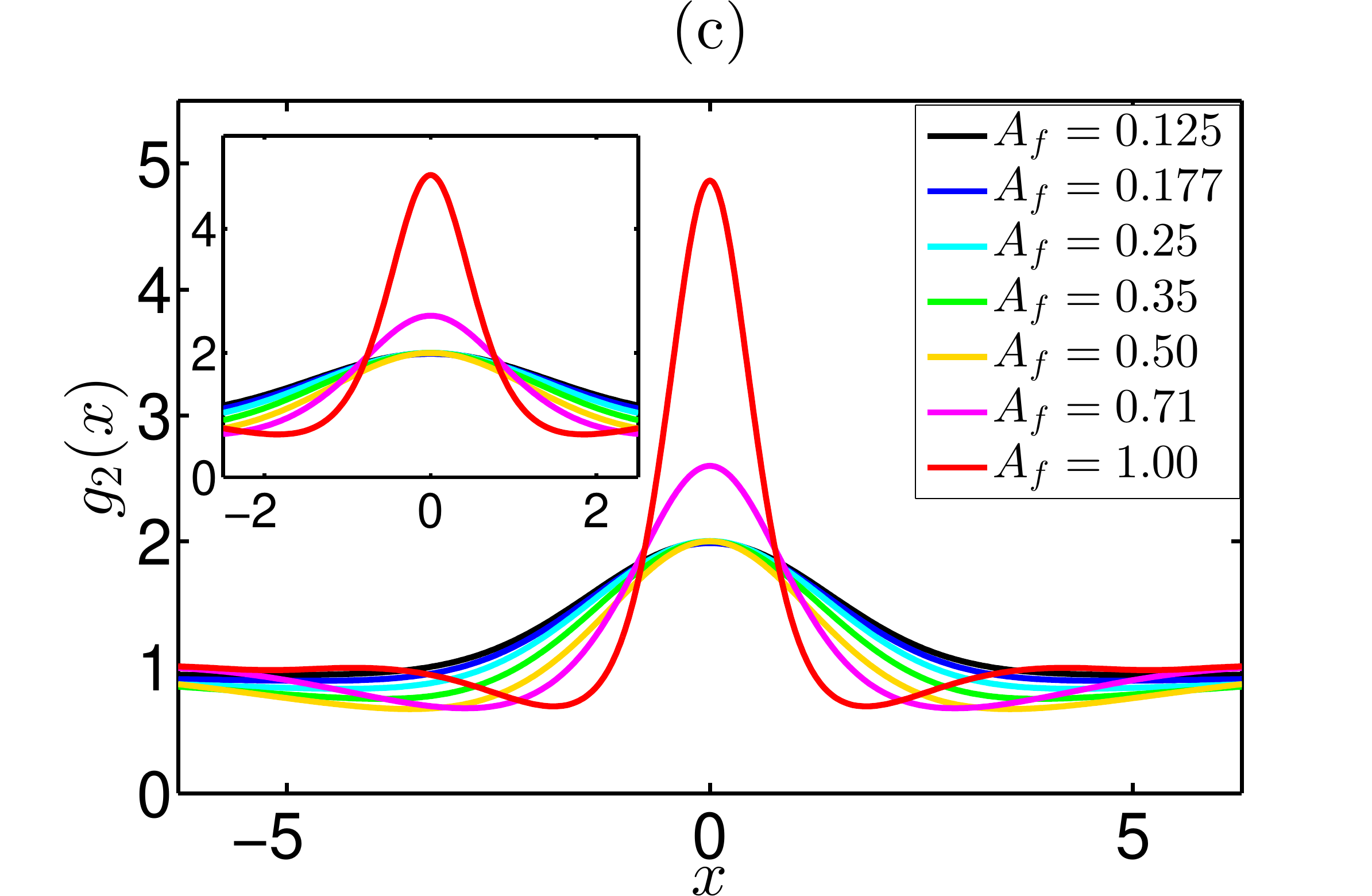}
	
	\caption{\small {\it (Color on-line)} 
	Statistical functions averaged over the ensemble and time interval $t-t_{0}\in[0,20]$ at different average amplitudes $A_{f}=0.125$, $0.177$, $0.25$, $0.35$, $0.5$, $0.71$ and $1$: (a) the wave-action spectrum $S_{k}$, (b) the PDF $\mathcal{P}(I)$ of relative wave intensity $I=|\psi|^{2}/\langle\overline{|\psi|^{2}}\rangle$ and (c) the autocorrelation of intensity $g_{2}(x)$. 
	All the other parameters correspond to the base experiment: $L=128\pi$, $s=2$, $A_{0}=10^{-2}$ and $p_{0}=10^{-5}$. 
	The insets in panels show the same functions at smaller scales. 
	The black dash-dot line in panel (b) indicates the exponential PDF~(\ref{Rayleigh}). 
	}
\label{fig:fig5}
\end{figure*}

\begin{figure*}[t]\centering
	\includegraphics[width=5.9cm]{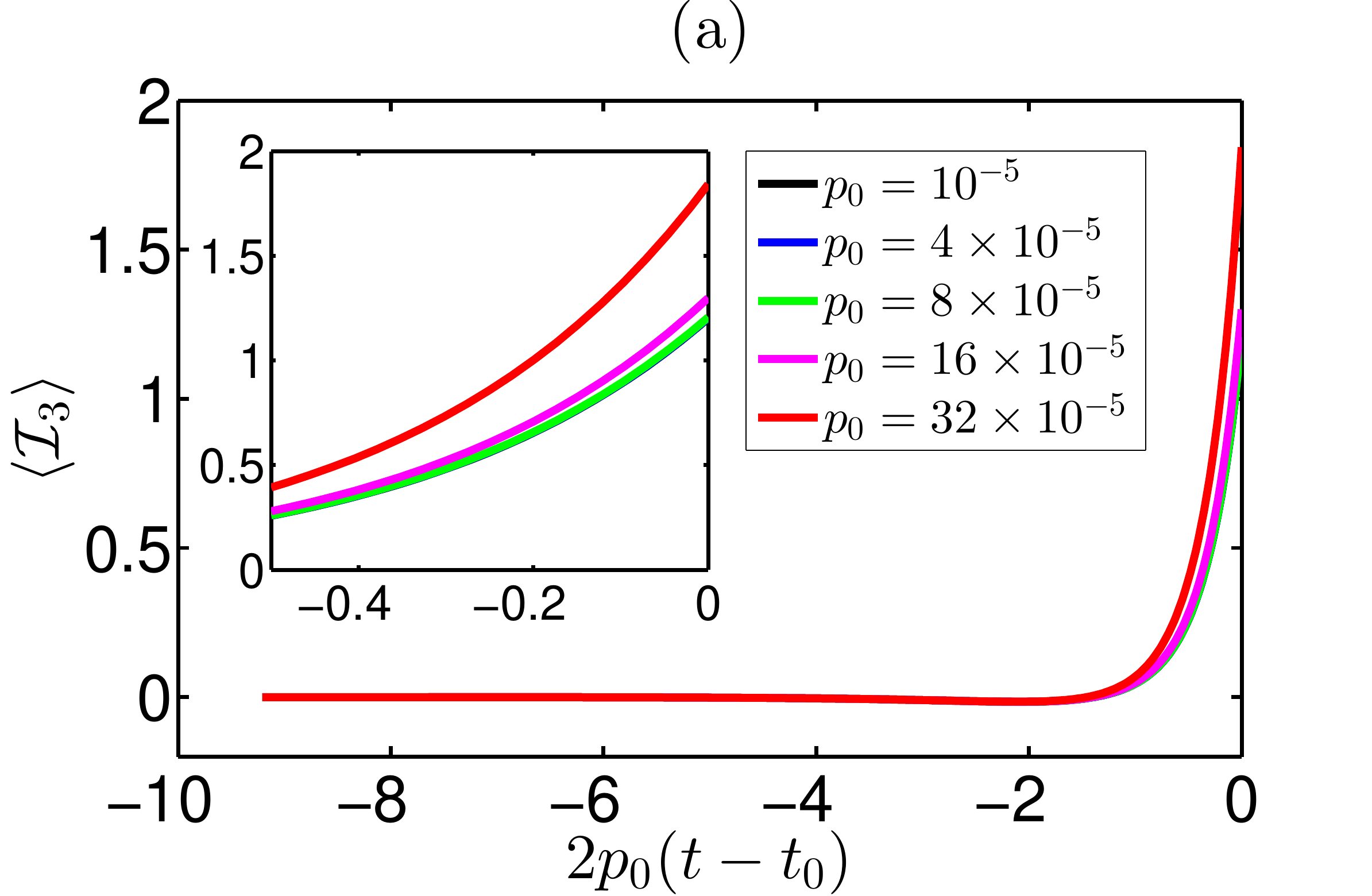}
	\includegraphics[width=5.9cm]{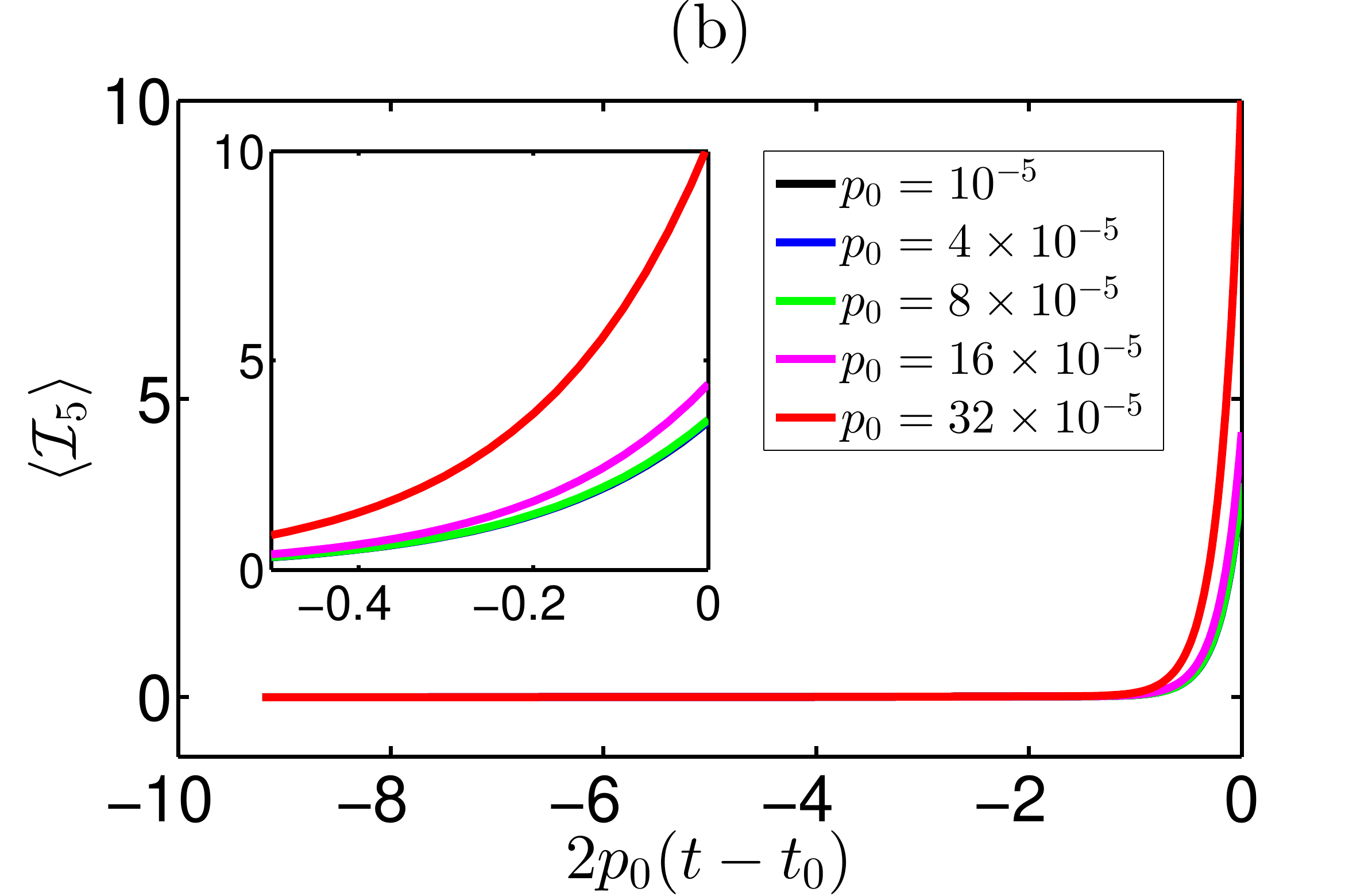}
	\includegraphics[width=5.9cm]{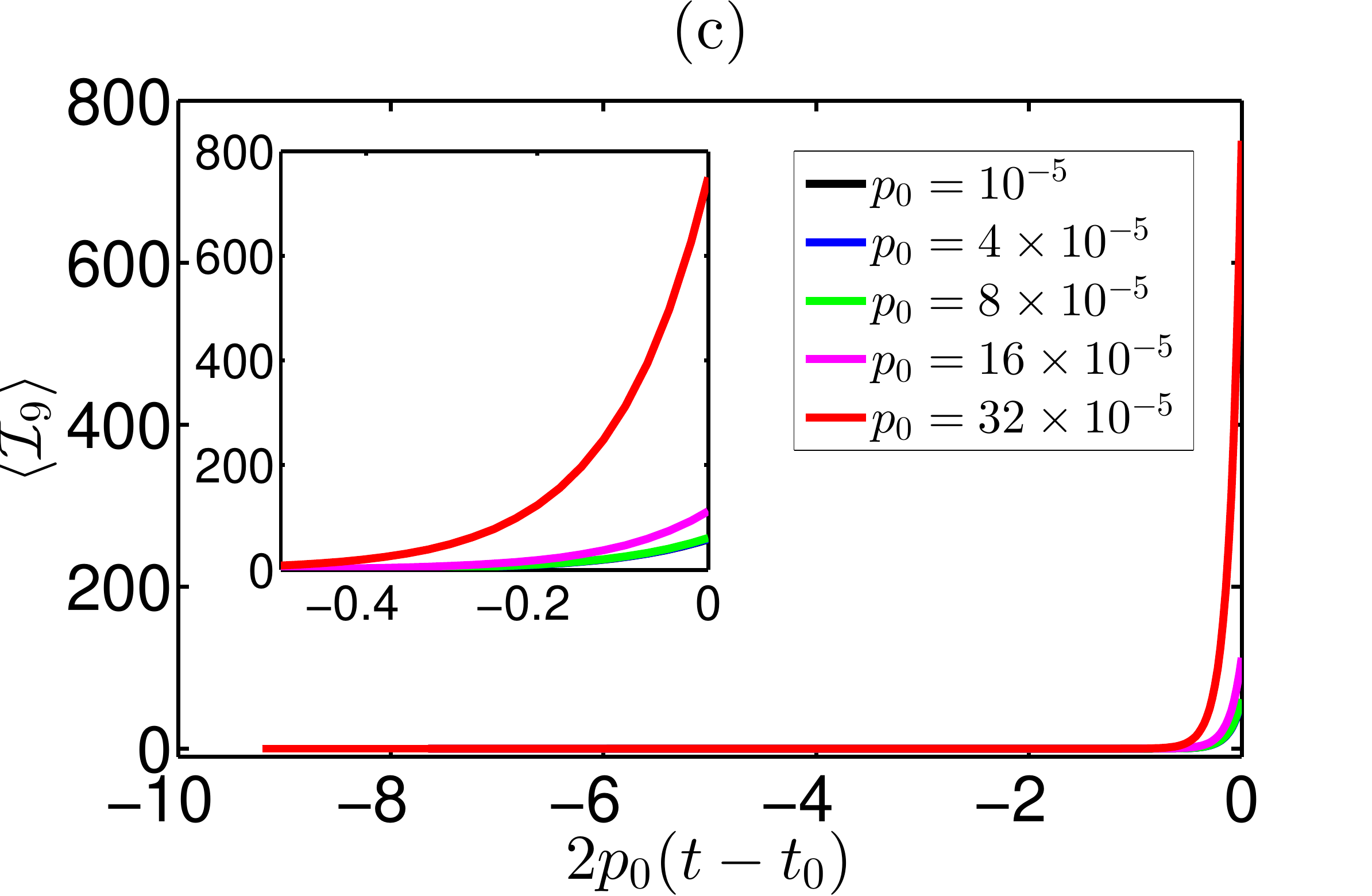}
	
	\caption{\small {\it (Color on-line)} 
	Evolution of the ensemble-averaged invariants~(\ref{integrals_rec1})-(\ref{integrals_rec2}) of the 1D-NLSE during the growth stage versus renormalized time $2p_{0}(t-t_{0})$ for the experiments with different pumping coefficients: (a) $\langle\mathcal{I}_{3}\rangle$, (b) $\langle\mathcal{I}_{5}\rangle$ and (c) $\langle\mathcal{I}_{9}\rangle$. 
	All the other parameters correspond to the base experiment: $L=128\pi$, $s=2$, $A_{0}=10^{-2}$ and $A_{f}=1$; the final states for these experiments are shown in Fig.~\ref{fig:fig3}(a-c). 
	Note that the third-order invariant equals minus total energy~(\ref{energy-1}), $\mathcal{I}_{3} = -\mathcal{E}$. 
	The insets in panels show the same functions near the end of the growth stage. 
	}
\label{fig:fig6}
\end{figure*}

\begin{figure*}[t]\centering
	\includegraphics[width=5.9cm]{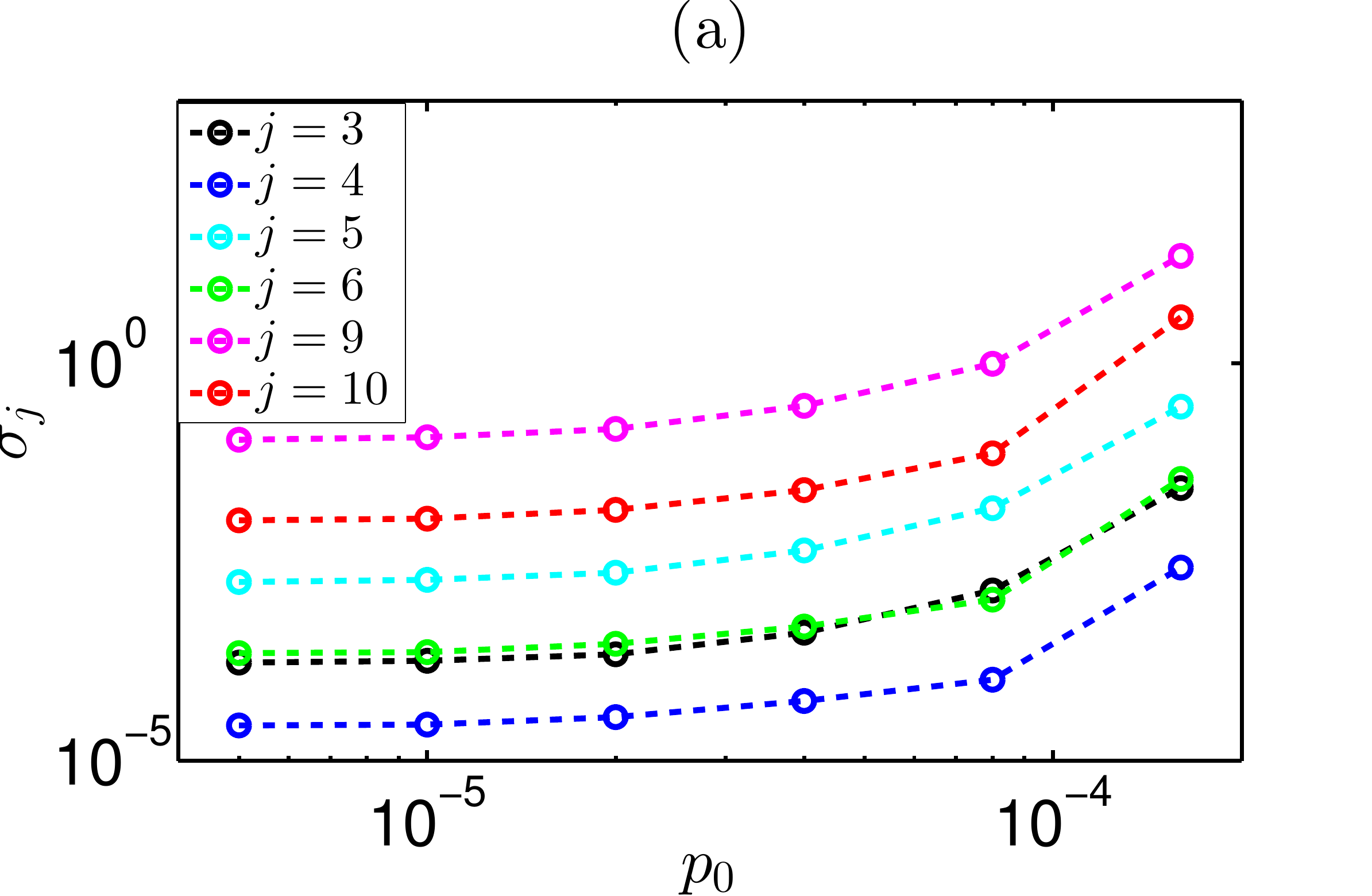}
	\includegraphics[width=5.9cm]{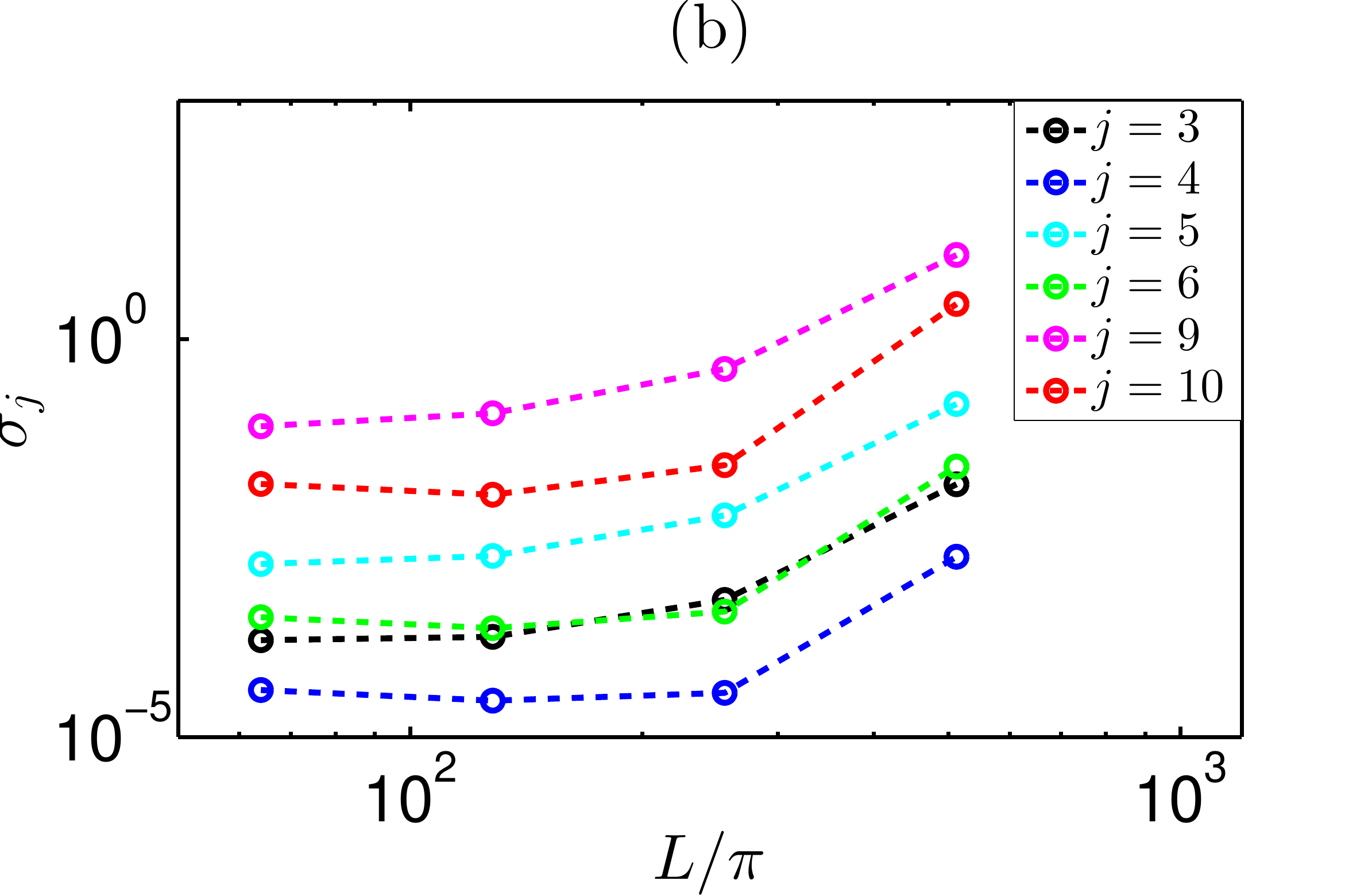}
	\includegraphics[width=5.9cm]{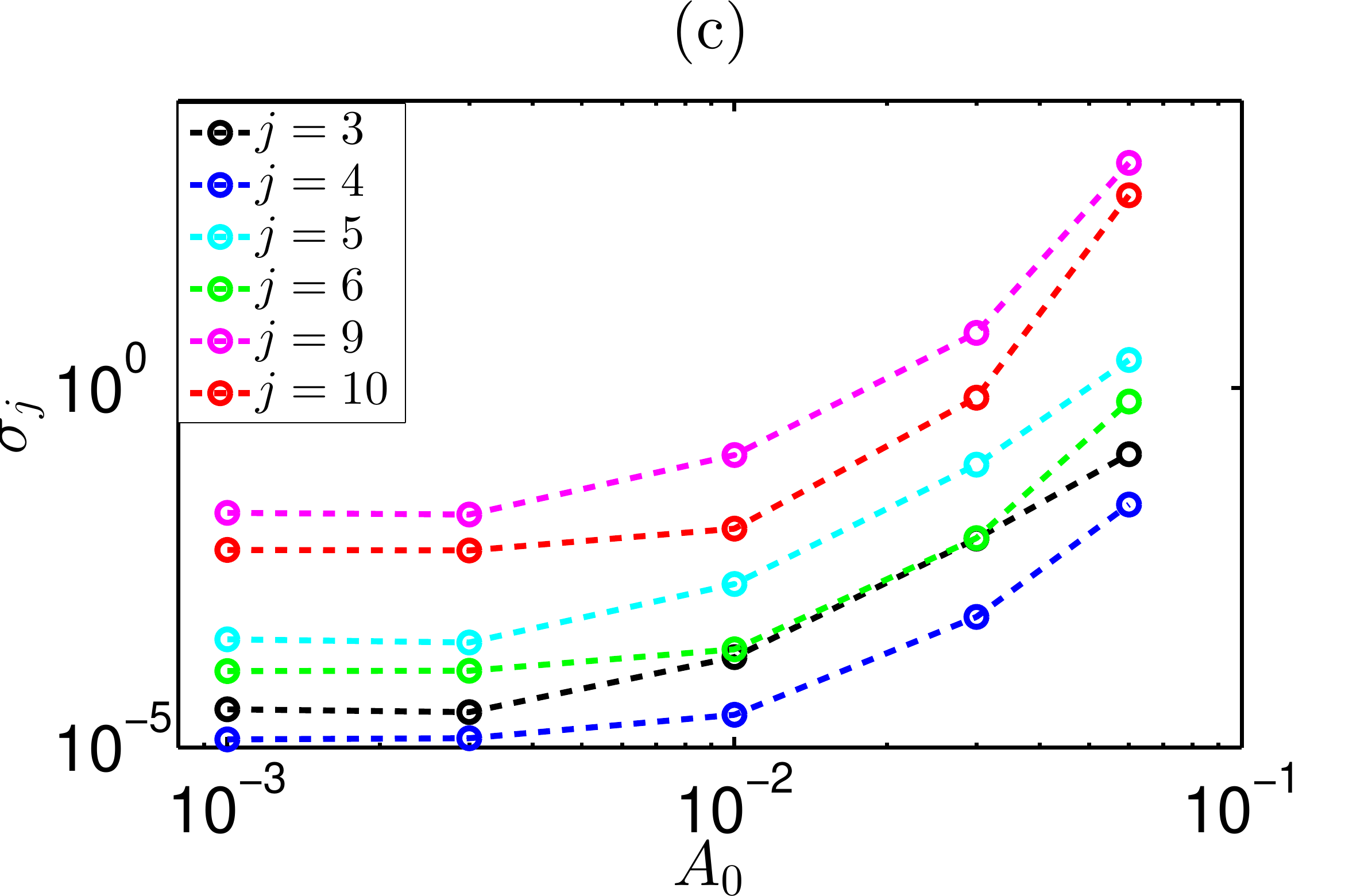}
	
	\caption{\small {\it (Color on-line)} 
	Standard deviations $\sigma_{j}$ of the ensemble-averaged 1D-NLSE invariants~(\ref{integrals_rec1})-(\ref{integrals_rec2}) $\langle\mathcal{I}_{j}\rangle$ of orders $j=3$, $4$, $5$, $6$, $9$ and $10$, at the end of the growth stage $A_{f}=1$ for three sets of numerical experiments shown in Fig.~\ref{fig:fig3}; the initial noise has Gaussian Fourier spectrum $s=2$. 
	Panel (a) illustrates experiments with different pumping coefficients $p_{0}$ and fixed $L=128\pi$ and $A_{0}=10^{-2}$, see Fig.~\ref{fig:fig3}(a-c). 
	Panel (b) shows experiments with different basin lengths $L$ and fixed $A_{0}=10^{-2}$ and $p_{0}=10^{-5}$, see Fig.~\ref{fig:fig3}(d-f). 
	Panel (c) demonstrates experiments with different initial amplitudes $A_{0}$ and fixed $L=128\pi$ and $p_{0}=10^{-5}$, see Fig.~\ref{fig:fig3}(g-i).
	}
\label{fig:fig7}
\end{figure*}


\section{Grown-up turbulence}
\label{Sec:Results1}

\subsection{Universal adiabatic regime}
\label{Sec:Results1:A}

We start this Section with a discussion of our results for the base numerical experiment with parameters $L=128\pi$, $s=2$, $A_{0}=10^{-2}$, $A_{f}=1$ and $p_{0}=10^{-5}$. 

Figure~\ref{fig:fig1} illustrates a typical realization of wavefield at the end of the growth stage $t=t_{0}$, as well as its subsequent space-time evolution after turning off the pumping, $t\ge t_{0}$. 
Note that the two panels of the figure show only the central quarter of the basin length. 
The wavefield in Fig.~\ref{fig:fig1}(a) contains several peaks of amplitude $|\psi|\simeq 3.5$ having characteristic width $\delta x\simeq 1$. 
The spatio-temporal dynamics demonstrated in Fig.~\ref{fig:fig1}(b) reveals that these peaks represent persistent structures that do not change their positions, suggesting that they are in fact solitons having zero velocities. 
The one-soliton solution of the 1D-NLSE~(\ref{NLSE}) can be written as, 
\begin{eqnarray}
	&&\psi_{\mathrm{s}}(x,t) = \chi \frac{\exp\left[i V (x-x') + i\theta'\right]}{\cosh \chi \left(x-x'\right)}, \label{1-SS}\\
	&& x' = x_{0}' + Vt, \quad
	\theta' = \theta_{0}' + \frac{1}{2} (\chi^2+V^2) t, \nonumber
\end{eqnarray}
where $\chi>0$, $V$, $x'$ and $\theta'$ are real-valued soliton amplitude, velocity and current position and phase, while the constants $x'_{0}$ and $\theta_{0}'$ stand for the position and phase at $t=0$. 
The function (\ref{1-SS}) indeed fits the observed peaks very well, as demonstrated in the inset of Fig.~\ref{fig:fig1}(a) for one of the pulses. 
The figure contains $8$ pulses satisfying the rogue wave criterion $I = |\psi|^{2}/\langle\overline{|\psi|^{2}}\rangle > 8$, see e.g.~\cite{kharif2003physical,dysthe2008oceanic,onorato2013rogue,dudley2019rogue}, where $\langle\overline{|\psi|^{2}}\rangle = A_{f}^{2} = 1$.  
This means that rogue waves emerge very frequently in the grown-up turbulence, and many of these waves, apparently, are standing solitons.

As shown in Fig.~\ref{fig:fig2}(a), after turning off the pumping $t\ge t_{0}$, the ensemble-averaged kinetic energy $\langle\mathcal{H}_{l}\rangle$, potential energy $\langle\mathcal{H}_{nl}\rangle$ and fourth-order moment of amplitude $\kappa_{4}$ do not change with time. 
The other statistical functions do not change with time either, meaning that in addition to averaging over ensemble of initial noise we can use time-averaging as well, in order to improve accuracy of the measurement. 
Averaging over time intervals $t-t_{0}\in[0,20]$ and $t-t_{0}\in[80,100]$ does not lead to a difference in the results for the wave-action spectrum, the PDF of relative wave intensity and the autocorrelation of intensity, see Fig.~\ref{fig:fig2}(b-d). 
Turning off the pumping at intermediate states -- before the average intensity $\overline{|\psi|^{2}}$ reaches $A_{f}^{2}=1$, we observe the same effect, i.e., that the statistical functions do not show dependency on time. 
Hence, we conclude that, after turning off the pumping at intermediate or final states, the resulting integrable turbulence turns out to be stationary for the given set of experimental parameters. 

The statistically stationary state, demonstrated in Fig.~\ref{fig:fig2}, is characterized by the kinetic energy $\langle\mathcal{H}_{l}\rangle=1.23$, potential energy $\langle\mathcal{H}_{nl}\rangle=-2.43$ and the fourth-order moment $\kappa_{4}=4.9$; here and below all the statistical functions are averaged over both the ensemble of initial conditions and time interval $t-t_{0}\in[0,20]$. 
Hence, this state is strongly nonlinear: the potential energy related to the nonlinear term of the 1D-NLSE turns out to be almost twice larger than the kinetic one related to (linear) dispersion, and the fourth-order moment exceeds significantly the value $\kappa_{4}=2$ characterizing a random superposition of linear waves, see e.g.~\cite[chapter~5]{nazarenko2011wave}. 

The wave-action spectrum shown in Fig.~\ref{fig:fig2}(b) is a continuous function, which decays slightly slower than exponentially at large wavenumbers $|k|\gg 1$ and has a sharp triangular shape at small and moderate wavenumbers $|k|\lesssim 1$. 
Compared to the exponential PDF, 
\begin{equation}\label{Rayleigh}
	\mathcal{P}_{R}(I) = e^{-I},
\end{equation}
which describes the intensity distribution for a random superposition of linear waves, see e.g.~\cite[chapter~5]{nazarenko2011wave} and~\cite[chapter~3]{mandel1995optical}, the PDF in Fig.~\ref{fig:fig2}(c) is smaller at moderate intensities $I\lesssim 4$ and larger by orders of magnitude at large intensities $I\gtrsim 10$. 
The probability to meet intensity above the rogue wave threshold $I>8$ can be calculated as the corresponding integral of the PDF, 
\begin{equation}\label{Probability-RW}
	P_{RW} = \int_{8}^{+\infty}\mathcal{P}(I)\,dI.
\end{equation}
For the PDF shown in Fig.~\ref{fig:fig2}(c), the result $P_{RW}=2.2\times 10^{-2}$ is almost two orders of magnitude larger than for a random superposition of linear waves $P_{RW}^{(R)}=3.4\times 10^{-4}$. 
The autocorrelation of intensity has a bell shape at moderate distances $|x|\lesssim 1.5$, which is characterized by the full width at half maximum $\Delta_{FWHM}=1.1$ and the maximum $g_{2}(0)=\kappa_{4}=4.9$, and at larger distances it practically reaches unity, $g_{2}(x)\approx 1$ at $|x|\gtrsim 4$. 

Repeating the experiment for different sets of parameters, we observe the same statistically stationary state after turning off the pumping, provided that the conditions~(\ref{puming-adiabatic-L}),~(\ref{puming-adiabatic-k}) for adiabatic growth of turbulence are satisfied. 
In particular, Fig.~\ref{fig:fig3}(a-c) shows the results for experiments with different pumping coefficients from $p_{0}=10^{-5}$ to $32\times 10^{-5}$, with all the other parameters the same as for the base experiment. 
As demonstrated in the figure, the wave-action spectrum, the PDF and the autocorrelation of intensity are the same for all experiments with $p_{0}\le 8\times 10^{-5}$. 
Note that we have done experiments with $p_{0} < 10^{-5}$ not shown in Fig.~\ref{fig:fig3}(a-c) and came to the same results. 
Also note that the ensemble-averaged kinetic energy, potential energy and fourth-order moment are uniquely defined by the wave-action spectrum, the PDF and the autocorrelation of intensity, respectively; see Eqs.~(\ref{energy-2}),~(\ref{energy-3}),~(\ref{wave-action-spectrum}) and~(\ref{g2}). 

Experiments with different basin lengths from $L=64\pi$ to $512\pi$ and all the other parameters the same as for the base experiment demonstrate coinciding results for $L\le 256\pi$, see Fig.~\ref{fig:fig3}(d-f). 
For these experiments, we have chosen the ensemble size inverse-proportional to the basin length $L$ in order to make the ``total length'' of all the realizations the same: $400$ for $L=64\pi$, $200$ for $L=128\pi$, $100$ for $L=256\pi$ and $50$ for $L=512\pi$. 

Note that, for the case $p_{0} = 8\times 10^{-5}$ in Fig.~\ref{fig:fig3}(a-c), which leads to the same statistical state as the base experiment, the criterion~(\ref{puming-adiabatic-k}) for the adiabatic growth of turbulence is satisfied poorly, as $p_{0}/\Delta k^{2}\approx 0.33$. 
The same applies to the case $L = 256\pi$ in Fig.~\ref{fig:fig3}(d-f), as $p_{0}/\Delta k^{2}\approx 0.16$. 
Increasing the pumping coefficient to $p_{0} = 16\times 10^{-5}$ in Fig.~\ref{fig:fig3}(a-c) or the basing length to $L = 512\pi$ in Fig.~\ref{fig:fig3}(d-f), we amplify the ratio to $p_{0}/\Delta k^{2}\approx 0.66$, so that the criterion~(\ref{puming-adiabatic-k}) becomes violated. 
Experiments with the corresponding parameters show coinciding statistical results -- compare the magenta lines in Fig.~\ref{fig:fig3}(a-c) and red lines in Fig.~\ref{fig:fig3}(d-f) -- which differ from those of the base experiment.

For the experiments with different initial noise amplitudes from $A_{0}=10^{-3}$ to $6\times 10^{-2}$ and all other parameters the same as for the base experiment, the results coincide for $A_{0}\le 10^{-2}$, see Fig.~\ref{fig:fig3}(g-i). 
For $A_{0}\ge 3\times 10^{-2}$, the wave statistics becomes dependent on $A_{0}$. 
By checking intermediate and final states of the pumping for the experiments with $A_{0} = 3\times 10^{-2}$ and $6\times 10^{-2}$ in the same way as shown in Fig.~\ref{fig:fig2}, we observe that they are stationary, so that the adiabatic growth regime is realized. 
As an additional test, we have performed several more experiments for the same noise amplitudes and increased pumping coefficient (i.e., increased speed of the turbulence growth), and came to the same statistical results for $p_{0}\le 8\times 10^{-5}$; see Section~\ref{Sec:TheProblem:B} for the discussion of such an adiabaticity test. 
Hence, the experiments in Fig.~\ref{fig:fig3}(g-i) with $A_{0} = 3\times 10^{-2}$ and $6\times 10^{-2}$ represent the ``non-universal'' adiabatic regime depending on the initial noise amplitude $A_{0}$. 
Apparently, this means that for $A_{0}\ge 3\times 10^{-2}$ the composition of noise changes. 
As we discuss below in Section~\ref{Sec:Results2}, this is indeed the case, since the initial noise contains solitons for $A_{0}\gtrsim 3\times 10^{-2}$ and does not contain them for $A_{0}\lesssim 10^{-2}$. 

Thus, for the two sets of experiments with (i) different pumping coefficients, Fig.~\ref{fig:fig3}(a-c), and (ii) different basin lengths, Fig.~\ref{fig:fig3}(d-f), the non-adiabatic regime is realized for the parameters $p_{0}\gtrsim 16\times 10^{-5}$ and $L\gtrsim 512\pi$, respectively. 
For the third set of experiments with different initial noise amplitudes, Fig.~\ref{fig:fig3}(g-i), the turbulence grows adiabatically, but this growth becomes dependent on $A_{0}$ for $A_{0}\ge 3\times 10^{-2}$. 
The universal adiabatic regime is realized in Fig.~\ref{fig:fig3}(a-c) for $p_{0}\le 8\times 10^{-5}$, in Fig.~\ref{fig:fig3}(d-f) for $L\le 256\pi$ and in Fig.~\ref{fig:fig3}(g-i) for $A_{0}\le 10^{-2}$, when the wave statistics does not depend on the pumping coefficient, basin length or noise amplitude. 
Note that, in Fig.~\ref{fig:fig3}, both (i) the non-adiabatic regime and (ii) the non-universal adiabatic regime lead to statistical states characterized by substantially higher far tail of the PDF, reflecting a much more frequent appearance of very large waves compared to the universal adiabatic regime. 

All the experiments discussed so far have fixed parameters $s=2$ and $A_{f}=1$ describing shape of the initial Fourier spectrum and final average amplitude respectively. 
We have done other sets of experiments with different $s$ and $A_{f}$, and came to the same conclusions on the adiabatic and non-adiabatic regimes of the turbulence growth, as discussed above.


\subsection{Dependency on the noise spectrum}
\label{Sec:Results1:B}

The linear pumping term in Eq.~(\ref{NLSE-k-space}) pumps in the same spectrum, which is already present in the system. 
Then, the resulting statistical state should depend on the shape of the initial noise spectrum. 
To test this hypothesis, in addition to the experiment with the Gaussian $s=2$ noise spectrum discussed above, we have done three more experiments with the exponential $s=1$ and super-Gaussian $s=8$ and $s=32$ spectra, and also one experiment with the non-symmetric spectrum $NS$, see Section~\ref{Sec:NumMethods:A}. 
In all these experiments, the characteristic noise spectral width is the same, $\delta k = 1$, and all the other parameters coincide with those of the base experiment: $L=128\pi$, $A_{0}=10^{-2}$, $A_{f}=1$ and $p_{0}=10^{-5}$. 

The statistical functions shown in Fig.~\ref{fig:fig4} demonstrate a clear dependency on the noise spectrum. 
For the symmetric initial spectra, with increasing exponent $s$, the wave-action spectrum at small and moderate wavenumbers $|k|\lesssim 1$ changes its shape from a sharp triangle into a bell-like, while at large wavenumbers $|k|\gg 1$ the decay of the spectrum with $k$ becomes much faster, see Fig.~\ref{fig:fig4}(a). 
The similar effect is observed for the far tail of the PDF at $I\gtrsim 20$, where the tail is lower for larger $s$, see Fig.~\ref{fig:fig4}(b). 
At $I\gtrsim 20$, the tail of the PDF for $s=1$ is by several orders of magnitude higher than for $s=32$, meaning that emergence of very large waves with $I\gtrsim 20$ is by orders of magnitude more frequent for the $s=1$ case. 
Note that, for $s=32$, the PDF is very similar to the PDF for the long-time statistically stationary state of the modulational instability of cnoidal waves~\cite{agafontsev2016integrable}; we will discuss this similarity in more detail in Appendix~\ref{Sec:Annex-super-Gaussian}. 

For the experiments with $s=1$, $2$, $8$ and $32$, the fourth-order moment of amplitude, which determines the maximum of the autocorrelation of intensity $g_{2}(0)=\kappa_{4}$, takes values $\kappa_{4}=4.6$, $4.9$, $3.8$ and $3.4$ respectively, see Fig.~\ref{fig:fig4}(c). 
Meanwhile, the central peak of the autocorrelation function widens moderately with increasing $s$. 
For the cases $s=1$ and $2$, the central peak at $|x|\lesssim 1$ is followed by a short trough at $|x|\simeq 1.5$, after which the function becomes practically indistinguishable from unity at $|x|\gtrsim 4$. 
For $s=32$, the autocorrelation tends to unity in the form of damped oscillations with a period of about $2\pi$, which are visible even at large enough distances $|x|\sim 10\pi$. 
Thus, the case $s=32$ is distinguished by a significantly more distant correlation of intensity compared to $s=1$ and $2$. 
The experiment $s=8$ looks as a transitional one between $s=1$ and $2$ on the one hand and $s=32$ on the other hand. 

Interesting results have been obtained for the experiment with non-symmetric noise spectrum $NS$. 
In particular, the wave-action spectrum turns out to be practically symmetric, almost coinciding with that for the symmetric $s=2$ case, see Fig.~\ref{fig:fig4}(a). 
The ``symmetrization'' of spectrum can only occur due to strong nonlinearity, since the nonlinear term $(|\psi|^{2}\psi)_{k}$ in Eq.~(\ref{NLSE-k-space}) is the only one capable to it, while at small average intensities $\overline{|\psi|^{2}}\ll 1$ the spectrum remains non-symmetric. 
The PDF turns out to be very similar to the $s=2$ case also, Fig.~\ref{fig:fig4}(b), and the autocorrelation function almost coincides with that for $s=2$ except at very small distances $|x|\ll 1$, where the non-symmetric noise spectrum leads to smaller fourth-order moment $\kappa_{4}=4.4$ versus $4.9$ for $s=2$, see Fig.~\ref{fig:fig4}(c). 
We remind that at $k\to -\infty$ the decay of the non-symmetric spectrum is the same as for the $s=2$ case, $\propto e^{-k^{2}}$, while at $k\to +\infty$ it is much faster, $\propto e^{-k^{32}}$.


\subsection{Intermediate stages of the pumping}
\label{Sec:Results1:C}

Turning off the pumping at intermediate states, corresponding to the average amplitude $A_{f}=0.125$, $0.177$, $0.25$, $0.35$, $0.5$ and $0.71$, we observe in Fig.~\ref{fig:fig5} evolution of the wave statistics for the base experiment with parameters $L=128\pi$, $s=2$, $A_{0}=10^{-2}$ and $p_{0}=10^{-5}$. 
At these $A_{f}$, the ratio of the potential energy to the kinetic one equals $\langle|\mathcal{H}_{nl}|\rangle/\langle\mathcal{H}_{l}\rangle = 0.0625$, $0.125$, $0.25$, $0.5$, $1$ and $1.74$, respectively, reaching $1.98$ for $A_{f}=1$. 
Thus, in line with discussion in Section~\ref{Sec:TheProblem:C}, at the early growth stage the potential-to-kinetic energy ratio increases proportionally to the average intensity, $\langle|\mathcal{H}_{nl}|\rangle/\langle\mathcal{H}_{l}\rangle \propto A_{f}^{2}$, and for $A_{f}\ge 0.5$ the statistical state becomes essentially nonlinear, $\langle|\mathcal{H}_{nl}|\rangle/\langle\mathcal{H}_{l}\rangle\ge 1$. 
At late growth stages, the energy ratio saturates at $\langle|\mathcal{H}_{nl}|\rangle/\langle\mathcal{H}_{l}\rangle \approx 2$; this value is known empirically as the maximum possible for statistically stationary states of the 1D-NLSE, see e.g.~\cite{agafontsev2015integrable,agafontsev2016integrable,agafontsev2021extreme}, which has yet to be explained theoretically.

With increasing average amplitude $A_{f}$, the wave-action spectrum at small and moderate wavenumbers $|k|\lesssim 1$ changes its shape from a bell-like, corresponding to the initial Gaussian spectrum, to a sharp triangular at $A_{f}=1$, see Fig.~\ref{fig:fig5}(a). 
Simultaneously, the tails of the spectrum at $|k|\gg 1$ grow noticeably. 
These two processes accelerate sharply in the region $0.35\lesssim A_{f}\lesssim 0.5$, when the statistical state becomes essentially nonlinear. 

Until $A_{f}\le 0.5$, the PDF coincides with the exponential distribution~(\ref{Rayleigh}), see Fig.~\ref{fig:fig5}(b), even despite the significant nonlinearity, $\langle|\mathcal{H}_{nl}|\rangle/\langle\mathcal{H}_{l}\rangle = 1$, at $A_{f}=0.5$.
With a further increase in the average amplitude, the PDF decreases at moderate intensities $I\simeq 2$ and increases greatly at large intensities $I\gtrsim 10$. 

With increasing $A_{f}$, the autocorrelation of intensity does not change until $A_{f}\simeq 0.1$; as discussed in Section~\ref{Sec:Results2}, we detect solitons in the wavefield for the first time staring from approximately this average amplitude. 
At larger average amplitudes, the autocorrelation function gradually shrinks, see Fig.~\ref{fig:fig5}(c), while its maximum stays at $g_{2}(0) = \kappa_{4} = 2$ until $A_{f}\le 0.5$ and only then starts to grow, reaching $\kappa_{4}=4.9$ at $A_{f}=1$. 


\subsection{Invariants of the 1D-NLSE during the growth stage}
\label{Sec:Results1:D}

As we have noted in Section~\ref{Sec:TheProblem:C}, the invariants of the 1D-NLSE~(\ref{integrals_rec1})-(\ref{integrals_rec2}) evolve with time during the growth stage; in particular, the wave action and momentum grow exponentially, $\mathcal{N}=\mathcal{N}_{0}\,e^{2p_{0}t}$ and $\mathcal{M}=\mathcal{M}_{0}\,e^{2p_{0}t}$. 
For the initial noise~(\ref{IC}), we have $\mathcal{N}_{0}=A_{0}^{2}$ and $\mathcal{M}_{0}=0$, with the latter relation calculated easily in the Fourier space, see Eq.~(\ref{momentum}). 
Thus, at the end of the growth stage $t_{0}=\frac{1}{p_{0}}\ln\frac{A_{f}}{A_{0}}$, all realizations of wavefield have the same wave action $\mathcal{N}=A_{f}^{2}$ and zero momentum $\mathcal{M}=0$. 
Evolution of the next-order invariants is nontrivial, see e.g. Eq.~(\ref{energy-1-t}) for the total energy, so that different realizations may have different values of these invariants. 

However, surprisingly, our experiments indicate very close values of the next-order invariants for the universal adiabatic regime of the turbulence growth. 
For instance, at the end of the growth stage of the base experiment, the third-, fifth- and ninth-order invariants averaged over ensemble of $200$ realizations equal $\langle\mathcal{I}_{3}\rangle = (1.2022\pm 0.0002)$, $\langle\mathcal{I}_{5}\rangle = (3.541\pm 0.002)$ and $\langle\mathcal{I}_{9}\rangle = (58.1\pm 0.1)$. 
Note that the third invariant equals minus total energy, $\mathcal{I}_{3}=-\mathcal{E}$. 
The standard deviations for the invariants of even orders are similar, but the invariants themselves practically equal zero; for instance, $\langle\mathcal{I}_{4}\rangle = i(0.1\pm 2.9)\times 10^{-5}$, $\langle\mathcal{I}_{6}\rangle = i(0.0\pm 2.3)\times 10^{-4}$ and $\langle\mathcal{I}_{10}\rangle = i(0.0\pm 1.1)\times 10^{-2}$. 
We have checked that, at intermediate states of the pumping, different realizations of the grown-up wavefield also have very close values of the higher-order invariants and invariants of even orders practically equal zero as well. 
In the future, we are going to check whether these conclusions can be proved analytically using the perturbation theory for the IST spectrum~\cite{KivsharRMP1989,mullyadzhanov2021solitons}.

Evolution of the ensemble-averaged third-, fifth- and ninth-order invariants $\langle\mathcal{I}_{3}\rangle$, $\langle\mathcal{I}_{5}\rangle$ and $\langle\mathcal{I}_{9}\rangle$ during the growth stage is shown in Fig.~\ref{fig:fig6} for the same set of experiments with different pumping coefficients $p_{0}$, which has been illustrated in Fig.~\ref{fig:fig3}(a-c). 
As follows from Fig.~\ref{fig:fig6}, for sufficiently small pumping coefficient $p_{0}\le 8\times 10^{-5}$, the invariants evolve self-similarly versus the renormalized time $2p_{0}(t-t_{0})$; the time renormalization is applied according to Eq.~(\ref{integrals_rec1-t}). 
For larger pumping coefficient, evolution of the invariants becomes non-self-similar and their values become dependent on $p_{0}$. 
This result corroborates Fig.~\ref{fig:fig3}(a-c), where the statistical functions also become dependent on the pumping coefficient for $p_{0}\gtrsim 16\times 10^{-5}$. 

We have checked that the same self-similarity of the invariants versus time is valid for the other two sets of experiments with different basin lengths $L$ and different noise amplitudes $A_{0}$, shown in Figs.~\ref{fig:fig3}(d-f) and~\ref{fig:fig3}(g-i) respectively, if only the turbulence grows in adiabatic regime, regardless of whether this regime is universal (i.e., does not depend on $A_{0}$), or not. 
For the non-adiabatic regime, the self-similarity is absent. 

The standard deviations of the invariants increase with increasing pumping coefficient, see Fig.~\ref{fig:fig7}(a), or basin length, Fig.~\ref{fig:fig7}(b), or initial noise amplitude, Fig.~\ref{fig:fig7}(c). 
For the non-adiabatic or non-universal adiabatic regimes, realized in Fig.~\ref{fig:fig7} at $p_{0}\ge 16\times 10^{-5}$, $L\ge 512\pi$, or $A_{0}\ge 3\times 10^{-2}$, the standard deviations become fairly noticeable, especially for the higher-order invariants. 
Thus, another distinction of the universal adiabatic regime versus the non-universal adiabatic and non-adiabatic regimes is the very small differences in the invariant values between different realizations of the grown-up wavefield. 


\section{IST analysis}
\label{Sec:Results2}

\subsection{Methods}
\label{Sec:Results2:A}

Following the modern theoretical-numerical approach to the studies of nonlinear waves~\cite{OsborneBook2010,Randoux2018nonlinear,chekhovskoy2019nonlinear,slunyaev2021persistence,bruhl2022comparative}, we find the scattering data of the grown-up wavefields by solving the direct scattering problem for the auxiliary Zakharov-Shabat (ZS) system~\cite{zakharov1972exact}, in which these wavefields play the role of the potential. 
As we demonstrate below, already at $A_{f} = 0.25$, when the statistics of waves does not differ much from that for a random superposition of linear waves, the fraction of nonlinear dispersive waves (the continuous spectrum) in the total wave action $\mathcal{N}$ turns out to be small, meaning that the corresponding wavefields represent almost pure solitonic states. 
For this reason, we focus on the solitonic part of the IST spectrum and do not consider the continuous part.

Using the efficient numerical tools developed by us recently in~\cite{mullyadzhanov2019direct,gelash2020anomalous,mullyadzhanov2021magnus}, we accurately identify the complete set of soliton parameters -- their amplitudes $\chi$, velocities $V$, positions $x'$ and phases $\theta'$ -- that represents a challenging numerical problem. 
In particular, to cope with computational instabilities when computing soliton positions and phases, we use a fine numerical grid and perform arithmetic operations with high precision. 
The corresponding theoretical formalism and numerical methods are outlined in Appendixes~\ref{Sec:Annex-IST-theory} and~\ref{Sec:Annex-IST-numerics}, where the IST method is discussed for localized wavefields. 

Note that we grow turbulence under periodic boundary conditions, so that one formally needs another IST technique for finite-band scattering data~\cite{novikov1984theory,belokolos1994algebro,bobenko2011computational}. 
However, the characteristic width of coherent structures in our wavefields turns out to be small compared to the basin length and we neglect the periodicity effects. 
A similar idea was suggested in~\cite{el2001soliton}, where a soliton gas was considered as a limit of finite-band solutions; later, we have used this approach in~\cite{agafontsev2021rogue} for analysis of rogue waves in a bound-state soliton gas. 
Below we demonstrate that the grown-up wavefields are very well approximated by exact multi-soliton solutions reconstructed from the solitonic part of the scattering data, meaning that the impact of periodicity effects on our analysis is indeed small.

Before discussion of the results, we confirm that the initial noise does not contain solitons and consists entirely of nonlinear dispersive waves, provided that its amplitude is sufficiently small. 
In particular, for the basin length $L=128\pi$ and noise amplitude $A_{0}=10^{-2}$, we do not detect solitons, as expected for low-amplitude oscillating wavefield~\cite{turitsyn2008soliton,derevyanko2008soliton}. 
For $A_{0}=3\times 10^{-2}$, some realizations of noise contain one or two solitons immersed in the dispersive background. 
These solitons have small characteristic amplitude $\chi\simeq 0.015 \sim A_{0}$, so that their width $\chi^{-1}\simeq 70$ approaches the whole basin length $L=128\pi$ in order of magnitude. 
For larger noise amplitude, e.g. $A_{0}=10^{-1}$, we observe that all noise realizations contain several solitons. 
Thus, we can suppose that solitons are present in the initial noise if $1/A_{0} \ll L$ and absent if
\begin{equation}
	A_{0}L \lesssim 1. \label{no-solitons-in-noise}
\end{equation}
Assuming that solitons and nonlinear dispersive waves may react differently to the pumping, see e.g.~\cite{KivsharRMP1989,YangBook2010,slunyaev2015wave}, we think that the presence of solitons may explain dependency of the statistical results on the noise amplitude for $A_{0}\ge 3\times 10^{-2}$, which has been discussed in Section~\ref{Sec:Results1:A}. 
Then, for the universal adiabatic growth, the conditions~(\ref{puming-adiabatic-L}),~(\ref{puming-adiabatic-k}) should be supplemented with the condition~(\ref{no-solitons-in-noise}) of the absence of solitons in the initial noise.


\begin{figure*}[!t]\centering
	\includegraphics[width=17.8cm]{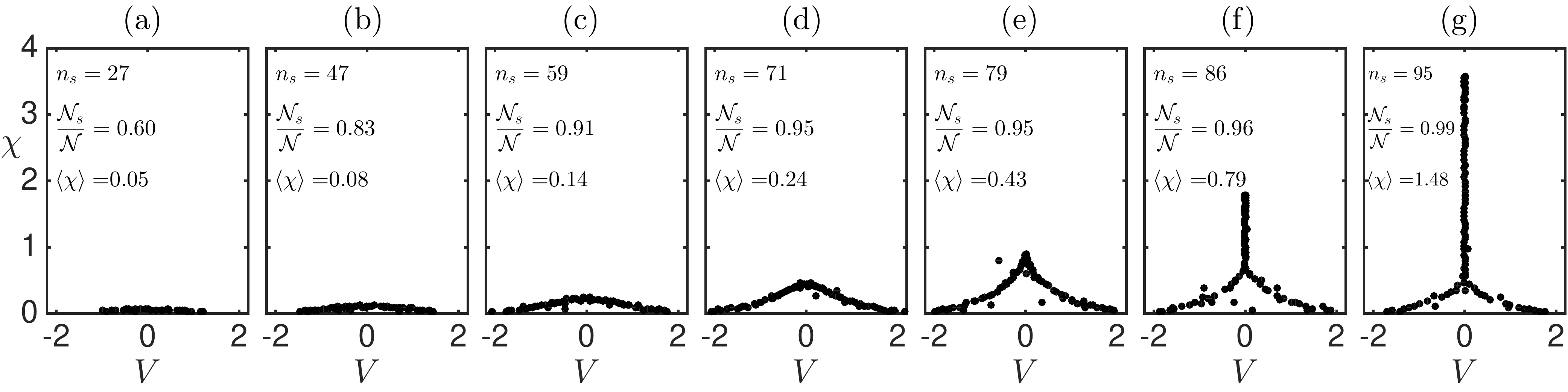}
	
	\caption{\small Amplitude-velocity dot diagrams of solitons for simulation from a single realization of initial conditions at different stages of the pumping: (a) $A_{f}=0.125$, (b) $A_{f}=0.177$, (c) $A_{f}=0.25$, (d) $A_{f}=0.35$, (e) $A_{f}=0.5$, (f) $A_{f}=0.71$ and (g) $A_{f}=1$. 
	The parameters correspond to the base experiment: $L=128\pi$, $s=2$, $A_{0}=10^{-2}$ and $p_{0}=10^{-5}$; panel (g) refers to the wavefield from Fig.~\ref{fig:fig1}(a). 
	Each dot represents a soliton within the studied wavefield, $n_{s}$ shows total number of solitons detected, $\mathcal{N}_{s}/\mathcal{N}$ demonstrates ratio between the wave action of the solitonic part $\mathcal{N}_{s}$, see Eq.~(\ref{Is1}), and the total wave action $\mathcal{N}$, and $\langle\chi\rangle$ indicates the mean soliton amplitude.
	}
\label{fig:fig08}
\end{figure*}

\begin{figure*}[!t]\centering
	\includegraphics[width=17.8cm]{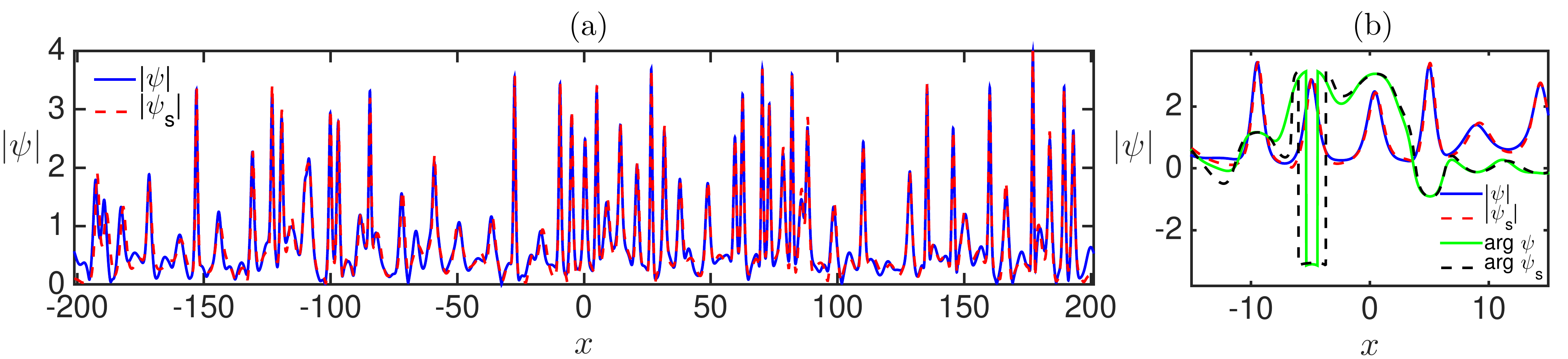}
	
	\caption{\small Typical grown-up wavefield and its approximation with the multi-soliton solution $\psi_{s}$ generated from the discrete part of the scattering data, for the base numerical experiment with parameters $L=128\pi$, $s=2$, $A_{0}=10^{-2}$, $A_{f}=1$ and $p_{0}=10^{-5}$. 
	Panel (a) shows absolute value $|\psi|$ of the grown-up wavefield (solid blue) and multi-soliton solution (dashed red), while panel (b) represents zoom of panel (a) at the central region. 
	Also, panel (b) illustrates complex phase $\mathrm{arg}\,\psi$ of the grown-up wavefield (solid green) and multi-soliton solution (dashed black). 
	Note that panel (a) shows the same grown-up wavefield as in Fig.~\ref{fig:fig1}(a), but over the entire basin length.
	}
\label{fig:fig09}
\end{figure*}

\begin{figure}[!t]
	\includegraphics[width=8.8cm]{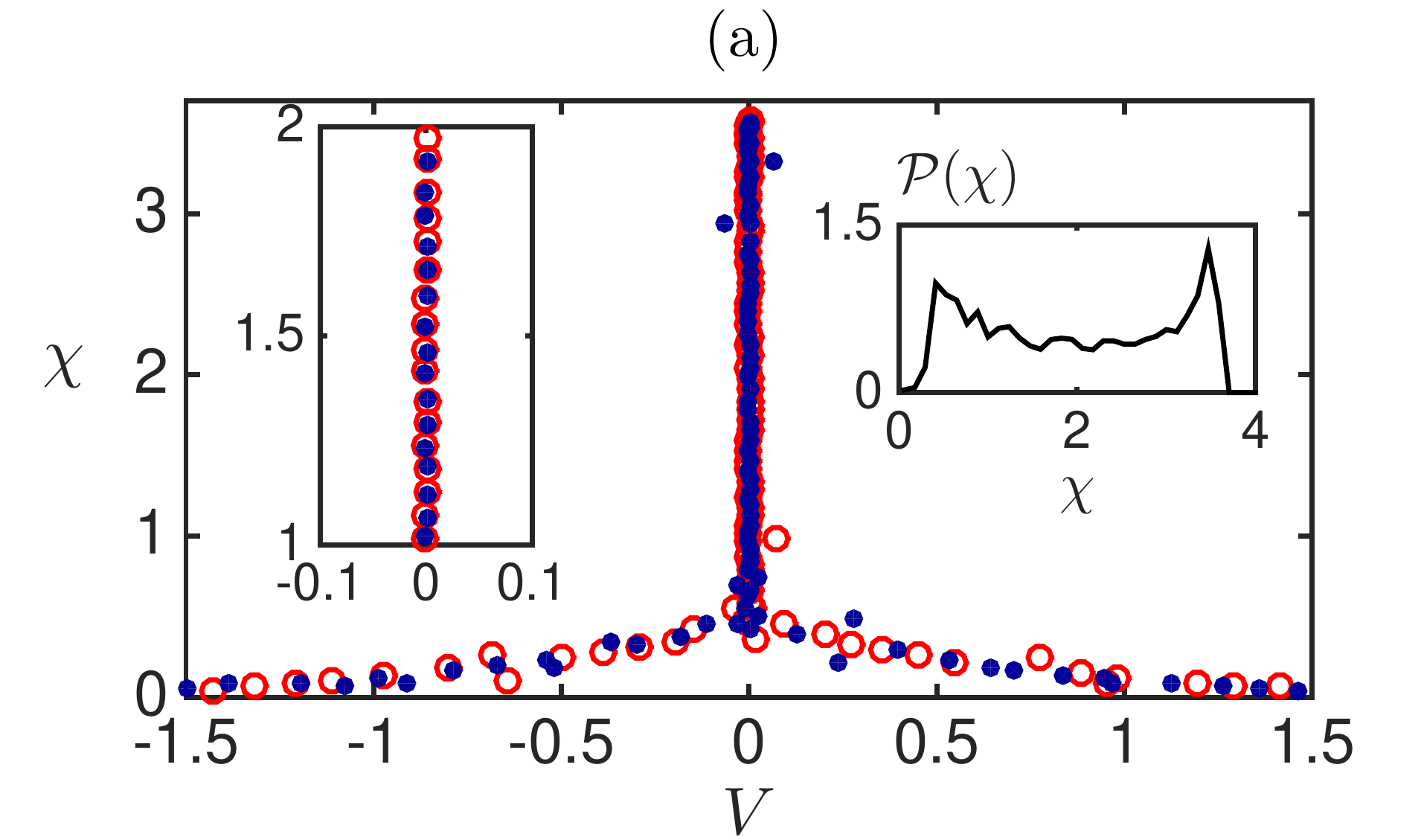}
	\includegraphics[width=8.9cm]{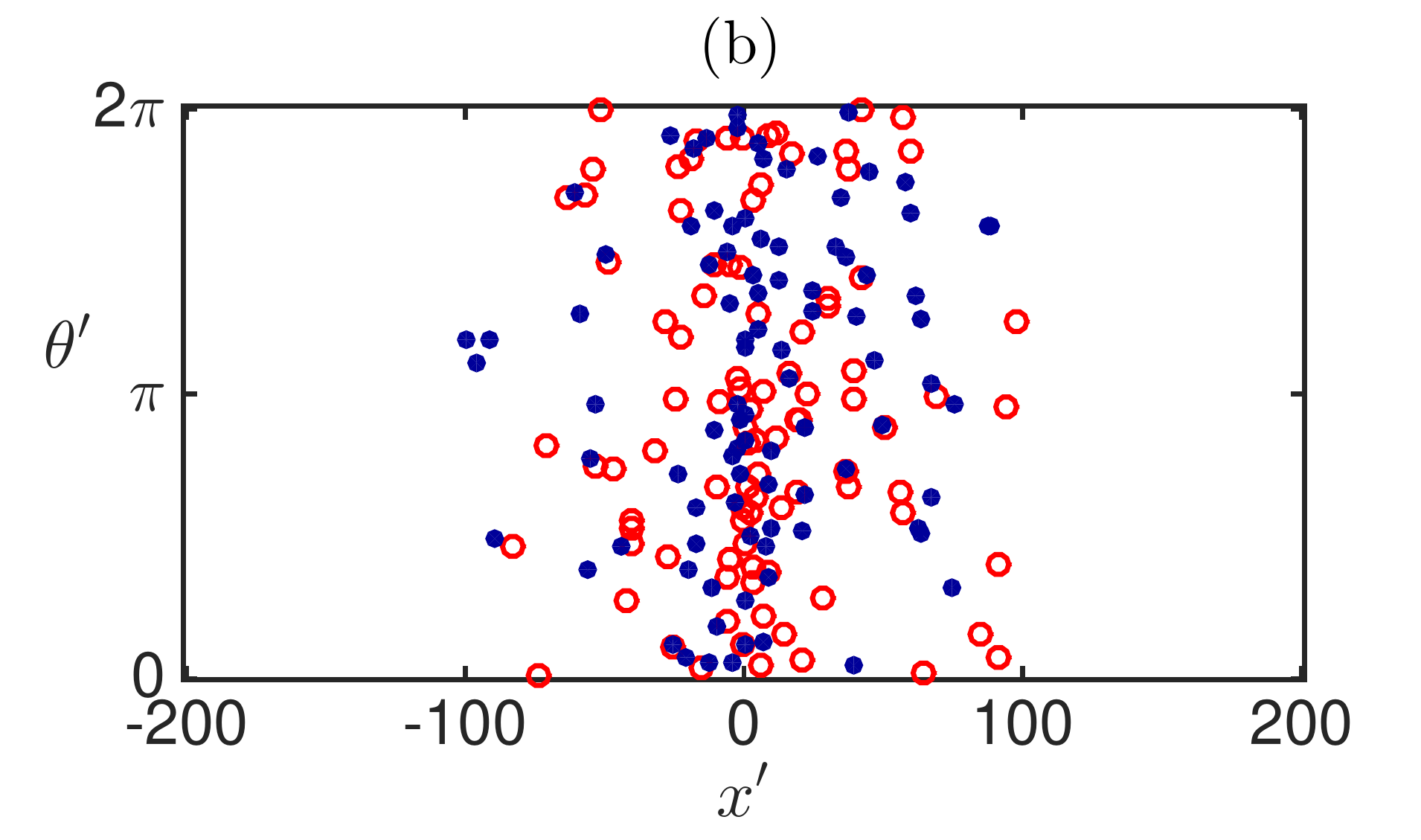}
	
	\caption{\small Dot diagrams of solitons for the base numerical experiment with parameters $L=128\pi$, $s=2$, $A_{0}=10^{-2}$, $A_{f}=1$ and $p_{0}=10^{-5}$: (a) for amplitudes $\chi$ and velocities $V$, and (b) for positions $x'$ and phases $\theta'$. 
	Blue dots and red circles represent solitons for two different realizations of the grown-up wavefield; red circles correspond to the same realization as in Figs.~\ref{fig:fig1}(a),~\ref{fig:fig08}(g),~\ref{fig:fig09}. 
	The left inset in panel (a) shows zoom of the main figure, indicating that sufficiently large solitons have very small velocities, while their amplitudes are strongly correlated between different realizations. 
	The right inset in panel (a) demonstrates the ensemble-averaged PDF of soliton amplitudes $\mathcal{P}(\chi)$, which looks rough since different realizations have very close sets of soliton amplitudes. 
	}
\label{fig:fig10}
\end{figure}

\begin{figure}[!t]\centering
	\includegraphics[width=8.9cm]{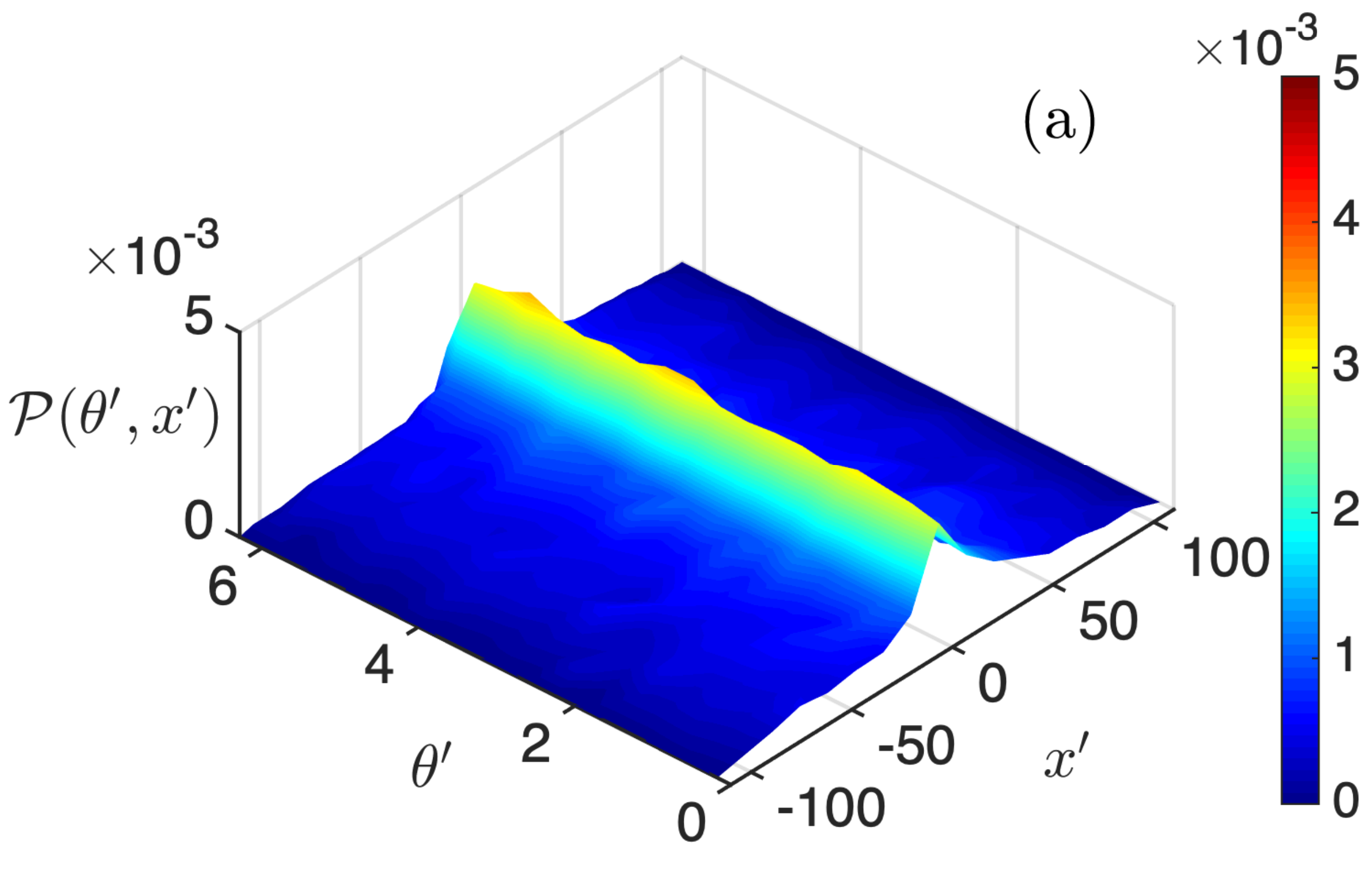}
	\includegraphics[width=8.9cm]{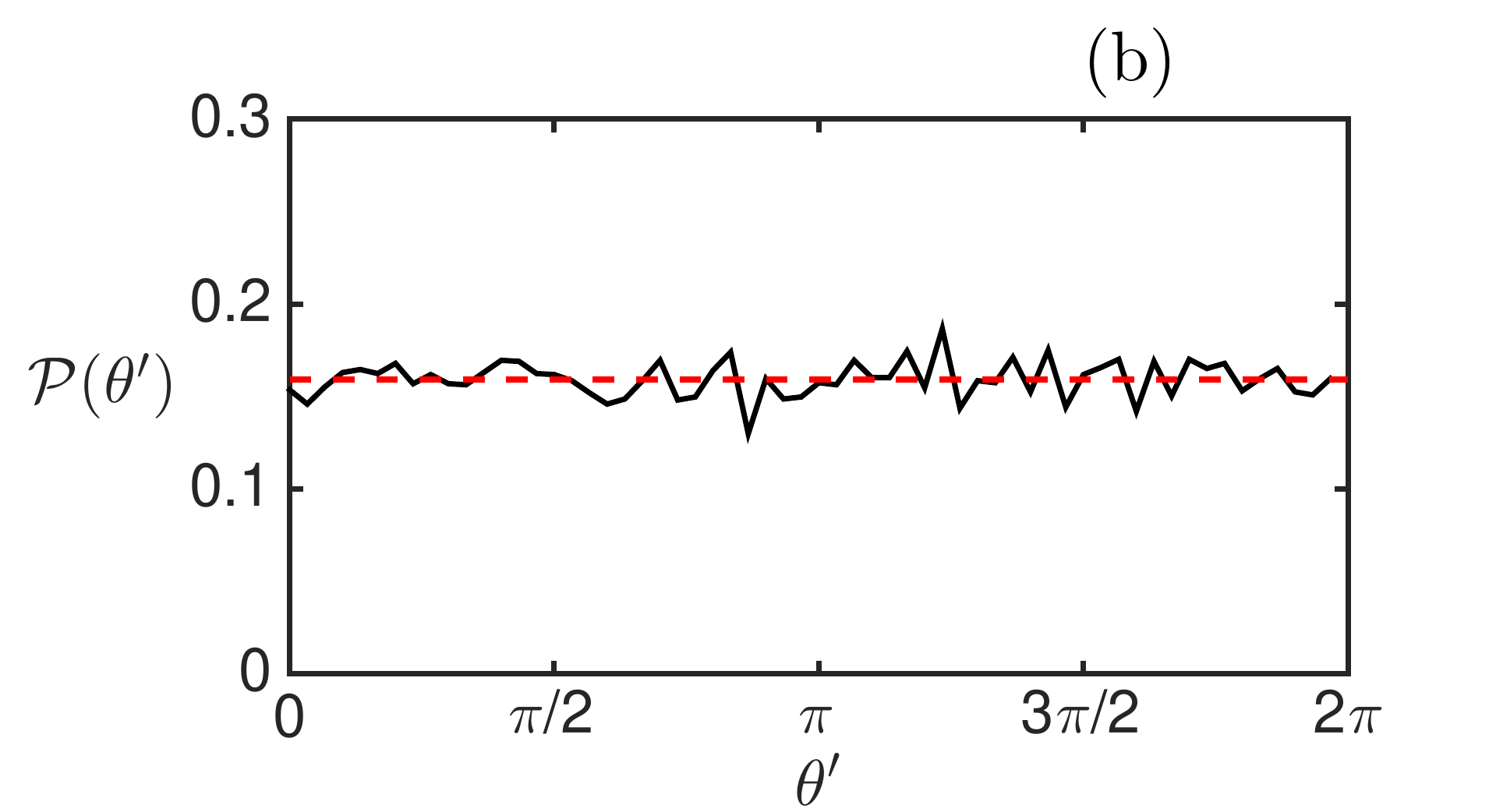}
	\includegraphics[width=8.9cm]{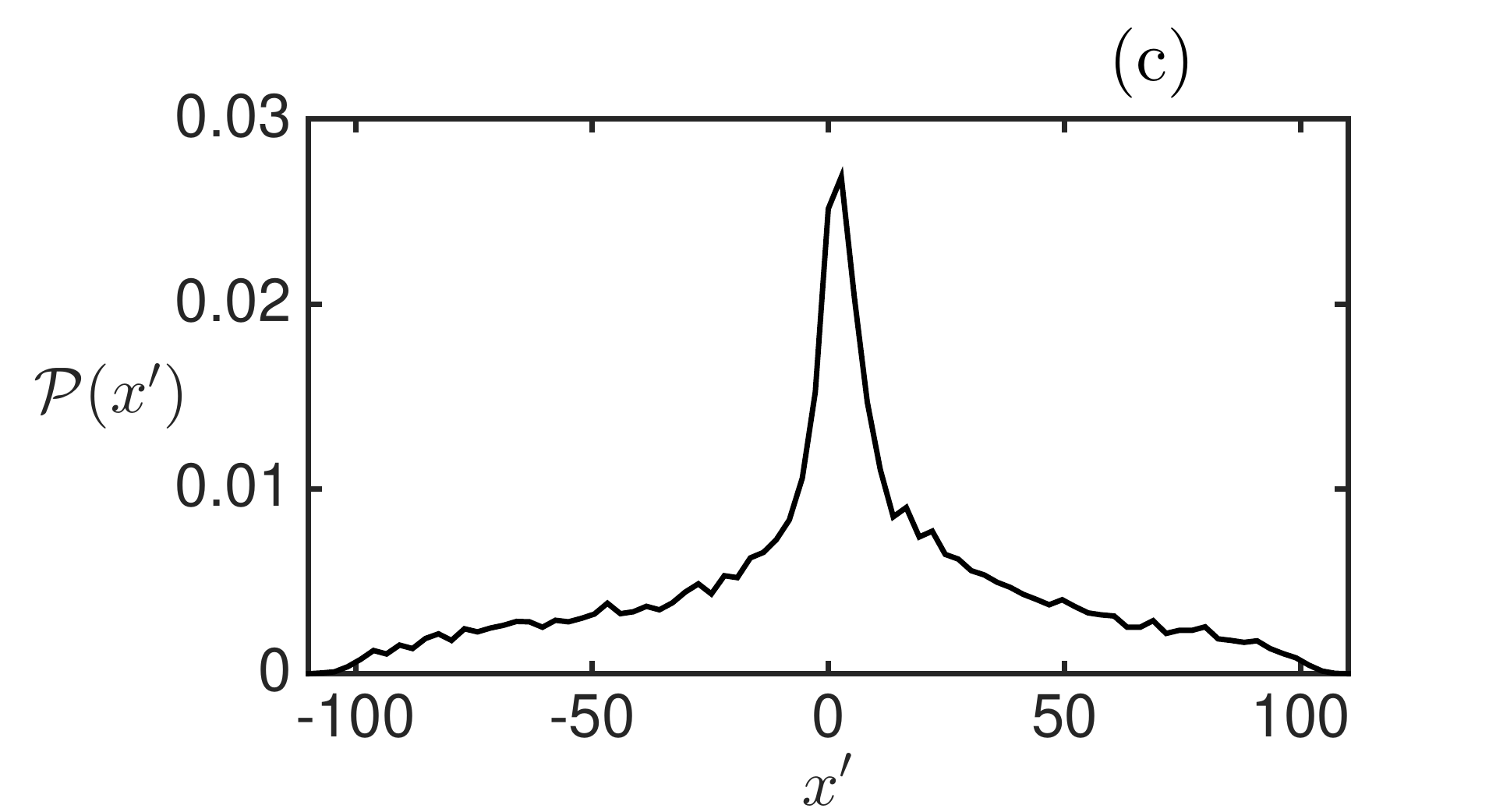}
	
	\caption{\small Ensemble-averaged PDFs of soliton phases $\theta'$ and positions $x'$ at the end of the growth stage for the base numerical experiment with parameters $L=128\pi$, $s=2$, $A_{0}=10^{-2}$, $A_{f}=1$ and $p_{0}=10^{-5}$: (a) joint PDF $\mathcal{P}(\theta',x')$ of soliton phases and positions, (b) PDF $\mathcal{P}(\theta')$ of soliton phases and (c) PDF $\mathcal{P}(x')$ of soliton positions. 
	The dashed red line in panel (b) indicates the uniform PDF $\mathcal{P}(\theta')=1/2\pi$. 
	}
\label{fig:fig11}
\end{figure}


\subsection{Universal adiabatic regime}
\label{Sec:Results2:B}

We now discuss the IST analysis for the base numerical experiment with parameters $L=128\pi$, $s=2$, $A_{0}=10^{-2}$ and $p_{0}=10^{-5}$, which demonstrates the universal adiabatic regime of the turbulence growth. 

Figure~\ref{fig:fig08} shows amplitude-velocity dot diagrams of solitons found in the scattering data for simulation from a single realization of initial noise at different stages of the pumping $A_{f}=0.125$, $0.177$, $0.25$, $0.35$, $0.5$, $0.71$ and $1$. 
Note that at $A_{f} = 0.06$ we do not detect solitons for all realizations of the grown-up wavefield (i.e., this state already differs from the initial noise of the same amplitude $A_{0}=0.06$, which typically contains a few solitons), while at $A_{f}=0.09$ we find only $3$-$4$ solitons per a realization. 

For the first stage $A_{f}=0.125$ shown in Fig.~\ref{fig:fig08}(a), in total $n_{s}=27$ solitons are detected; these solitons have mean amplitude $\langle\chi\rangle = 0.07$ and a broad distribution of velocities within the interval $|V|\le 1$. 
Their collective impact on the wavefield can be estimated through the wave action of the solitonic part, which can be found analytically by known soliton amplitudes, 
\begin{equation}
\label{Is1}
    \mathcal{N}_{s} = \frac{2}{L}\sum_{n=1}^{n_{s}}\chi_{n},
\end{equation}
see~\cite{zakharov1972exact} and Appendix~\ref{Sec:Annex-IST-theory}. 
The ratio to the total wave action yields $\mathcal{N}_{s}/\mathcal{N} = 0.6$, meaning that this wavefield is already essentially influenced by solitons. 
Note that, for the corresponding statistical state, the ratio of the potential energy to the kinetic one is small, $\langle|\mathcal{H}_{nl}|\rangle/\langle\mathcal{H}_{l}\rangle = 0.0625$, see Section~\ref{Sec:Results1:C}, so that the nonlinearity is weak; also, the PDF of relative wave intensity is exponential, see Fig.~\ref{fig:fig5}(b). 
This situation is possible since solitons have relatively large velocities, so that their kinetic energy is much larger than the potential one; see~\cite{gelash2018strongly}, where the same effect has been observed for a dense soliton gas.

With increasing average amplitude $A_{f}$, the number of solitons together with their mean amplitude and fraction in the total wave action increase, see Fig.~\ref{fig:fig08}, reaching $n_{s} = 95$, $\langle\chi\rangle = 1.5$ and $\mathcal{N}_{s}/\mathcal{N}=0.99$ at $A_{f}=1$. 
Solitons dominate the wavefield, $\mathcal{N}_{s}/\mathcal{N}\ge 0.9$, already starting from $A_{f}=0.25$, see Fig.~\ref{fig:fig08}(c); in the corresponding statistical state, the potential-to-kinetic energy ratio is still relatively small, $\langle|\mathcal{H}_{nl}|\rangle/\langle\mathcal{H}_{l}\rangle = 0.25$, and the PDF of relative wave intensity is still exponential, see Section~\ref{Sec:Results1:C}.

During the growth stage, small solitons continuously emerge with nonzero velocities. 
Between average amplitudes $A_{f}=0.125$ and $0.35$, the interval of soliton velocities increases, as if a wedge-shaped formation of solitons emerges from the lower half-plane of the spectral parameter; see the corresponding discussion in~\cite{mullyadzhanov2021solitons}. 
Having emerged, solitons gain amplitude and loose velocity; at $A_{f}\gtrsim 0.5$, a bound state of the largest solitons with practically zero velocities is starting to form. 
At the end of the growth stage in Fig.~\ref{fig:fig08}(g), $64$ solitons ($67$\%) containing $95$\% of all wave action $\mathcal{N}$ belong to this bound state. 
Hence, with increasing average amplitude, the grown-up wavefield approaches a \textit{bound-state} soliton gas.

Using the complete set of soliton parameters determined at the end of the growth stage $A_{f}=1$, see Fig.~\ref{fig:fig08}(g), we compute wavefield for a pure multi-soliton solution $\psi_{s}$ constructed from these solitons with methods discussed in Appendix~\ref{Sec:Annex-IST-theory}, see also~\cite{gelash2018strongly,gelash2019bound}, and observe that it approximates the original grown-up wavefield remarkably well. 
As shown in Fig.~\ref{fig:fig09}, the approximation is especially good in amplitude $|\psi|$ at its peaks and lacks accuracy near the troughs of $|\psi|$ and in the complex phase $\mathrm{arg}\,\psi$. 
Similar results are obtained for all experimental realizations of the grown-up wavefield. 
Note that, at the edges of the simulation box, we cannot approximate the wavefield with a multi-soliton solution properly, since the former is periodic and the latter is localized; see the corresponding discussion in Appendix~\ref{Sec:Annex-IST-theory}.

Amplitude-velocity dot diagrams of solitons turn out to be strongly correlated between different realizations of the grown-up wavefield -- see Fig.~\ref{fig:fig10}(a), where two different realizations are shown at the end of the growth stage. 
We observe the same correlation at intermediate stages of the pumping as well, meaning that the presented in Fig.~\ref{fig:fig08} soliton diagrams for a single realization of initial noise fully illustrate the general picture. 
This observation means, in essence, that despite the different random sets of Fourier phases $\phi_{k}$ which determine the specific realization of the initial noise~(\ref{IC}), the adiabatic regime of turbulence growth always leads to the same collection of solitons as the average intensity increases.

For a pure solitonic state, the amplitude-velocity dot diagram determines all the invariants~(\ref{integrals_rec1})-(\ref{integrals_rec2}) of the 1D-NLSE, see Appendix~\ref{Sec:Annex-IST-theory}. 
Since at the end of the growth stage $A_{f}=1$ the wavefield is almost purely solitonic, the observed in Fig.~\ref{fig:fig10}(a) correlation of the amplitude-velocity diagrams explains why different realizations have very close values of the higher-order invariants, see Section~\ref{Sec:Results1:D}. 
Then, close values of the invariants at intermediate stages of the pumping should be explained both by the correlation of the amplitude-velocity dot diagrams of solitons (which we observe) and by the correlation of the properties of nonlinear dispersive waves between different realizations of the grown-up wavefield.

Unlike amplitudes and velocities, soliton positions $x'$ and phases $\theta'$ represent different random sets of values for each experimental realization of wavefield, see the examples in Fig.~\ref{fig:fig10}(b).
However, these sets of values indicate very similar statistical distributions of positions and phases for different realizations. 
Note that the computed with the direct scattering transform position $x'$ and phase $\theta'$ coincide with those observed in the physical space only for the one-soliton solution~(\ref{1-SS}). 
In presence of other solitons or dispersive waves, the observed position and phase may differ considerably from $x'$ and $\theta'$. 

The ensemble-averaged joint PDF $\mathcal{P}(\theta', x')$ of soliton phases and positions, shown at the end of the growth stage in Fig.~\ref{fig:fig11}(a), indicates a uniform profile over phase $\theta'$ at each position $x'$. 
Similarly, the ensemble-averaged PDF of soliton phases on the interval $\theta'\in[0,2\pi)$ is very close to uniform $\mathcal{P}(\theta')\simeq 1/2\pi$, see Fig.~\ref{fig:fig11}(b). 
The same properties of these PDFs are observed at intermediate stages of the pumping as well. 

The adiabatic regime of the turbulence growth moves successively through the statistically stationary states, and for this reason we think that there is no correlation between soliton phases within the wavefield. 
Indeed, if the phases are correlated, then turning off the pumping should lead to their randomization due to different rates of their change with time for different solitons, see Eqs.~(\ref{Ck_param})-(\ref{DM-norming-constants-evolution}) in Appendix~\ref{Sec:Annex-IST-theory}; this randomization should change the statistical state, so that the correlation of phases should mean a statistical state which is not yet stationary. 
Note that in~\cite{gelash2019bound} this hypothesis has been used in a different way, demonstrating that a bound-state soliton gas with random phases of solitons rests in a statistically stationary state. 
In Fig.~\ref{fig:fig10}(b), one can see a few groups of solitons having simultaneously close positions and phases; without further study of their temporal evolution, it is difficult to conclude whether this is a phase-locking or a coincidence. 
We will study this question in more detail in a separate publication.

The ensemble-averaged PDF $\mathcal{P}(x')$ of soliton positions, shown at the end of the growth stage in Fig.~\ref{fig:fig11}(c), is close to symmetric and has a sharp triangular shape near the center of the simulation box $x=0$. 
We believe that the slight asymmetry observed in the figure is related to the finite ensemble size and will vanish if the latter increases. 
The full width at half maximum of $\mathcal{P}(x')$ equals $L_{0}=13$, that is much smaller than the basin length, $L_{0}\ll L$. 
This means a high soliton density, which can be defined as a number of solitons per unit length, 
\begin{equation}\label{soliton-density}
	\rho = \frac{n_{s}}{L}.
\end{equation}

Indeed, physical positions of solitons in a multi-soliton solution coincide with parameters $x'$ up to mutual soliton space shifts if and only if distances between solitons are large enough, see~\cite{novikov1984theory}, i.e., when the soliton density is low, $\rho\ll 1$. 
In this case, characteristic width $L_{0}$ of the distribution of positions will be of the same order as the total width of the multi-soliton solution, which equals $L$ in our case, i.e., $L_{0}\sim L$. 
With decreasing $L_0$, the multi-soliton solution shrinks and the soliton density increases, reaching maximum as $L_0$ approaches zero, see~\cite{gelash2018strongly,gelash2019bound}. 

For instance, the solitonic part of the plain wave solution of unit intensity $\psi=1$ has all the solitons located at a single point (i.e., $L_{0}=0$) and spatial density $\rho\approx 1/\sqrt{2}\pi\approx 0.23$, see~\cite{gelash2021solitonic} (note that in that paper a different normalization of the 1D-NLSE was used). 
At the end of the growth stage of the base experiment, the ensemble-averaged soliton density reaches $\rho=0.24$. 
In~\cite{gelash2018strongly}, we have used a slightly different definition of density, which accounts for the mean soliton amplitude, 
$$
	\tilde{\rho} = \frac{2n_{s}}{L\langle\chi\rangle}.
$$
With this definition, the solitonic content of the plain wave has density $\tilde{\rho}\approx 2\sqrt{2}/\pi^{2}\approx 0.29$, while the grown-up turbulence -- $\tilde{\rho}=0.32$. 
Thus, in our experiments we observe a fairly dense soliton gases.

As has been discussed in Section~\ref{Sec:Results1:B}, the grown-up turbulence depends significantly on the Fourier spectrum of initial noise. 
Repeating our IST analysis for the experiments with other noise spectra, we come to all the same conclusions as discussed above with the exception that the amplitude-velocity dot diagrams indicate a different distribution of solitons in amplitude and velocity. 
In particular, for the super-Gaussian initial spectrum $s=32$, the majority of solitons have practically equal amplitudes, so that, with increasing average amplitude $A_{f}$, the wavefield tends to a dense bound-state soliton gas which consists of identical solitons; see Appendix~\ref{Sec:Annex-super-Gaussian}. 


\begin{figure}[!t]
	\includegraphics[width=8.9cm]{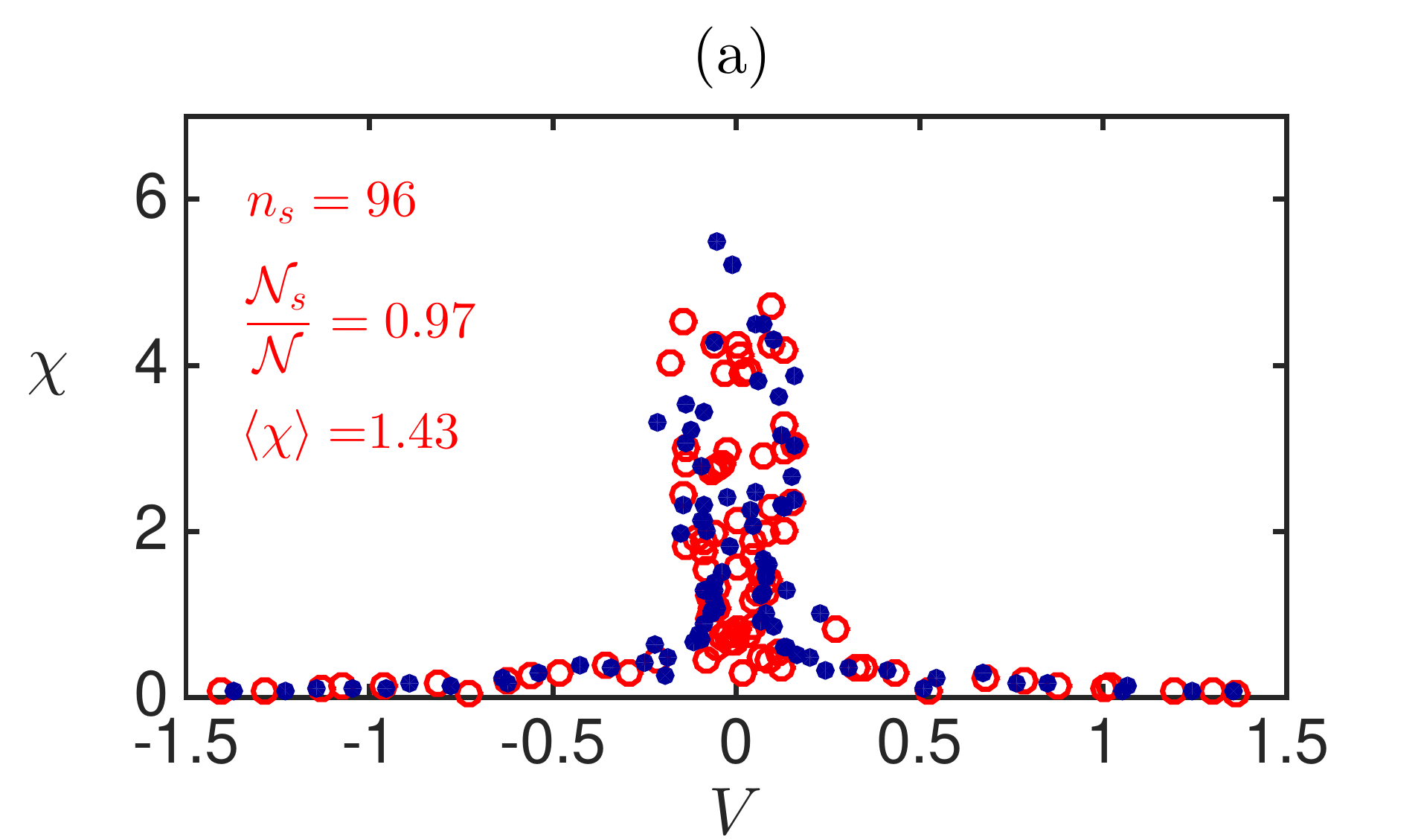}
	\includegraphics[width=8.9cm]{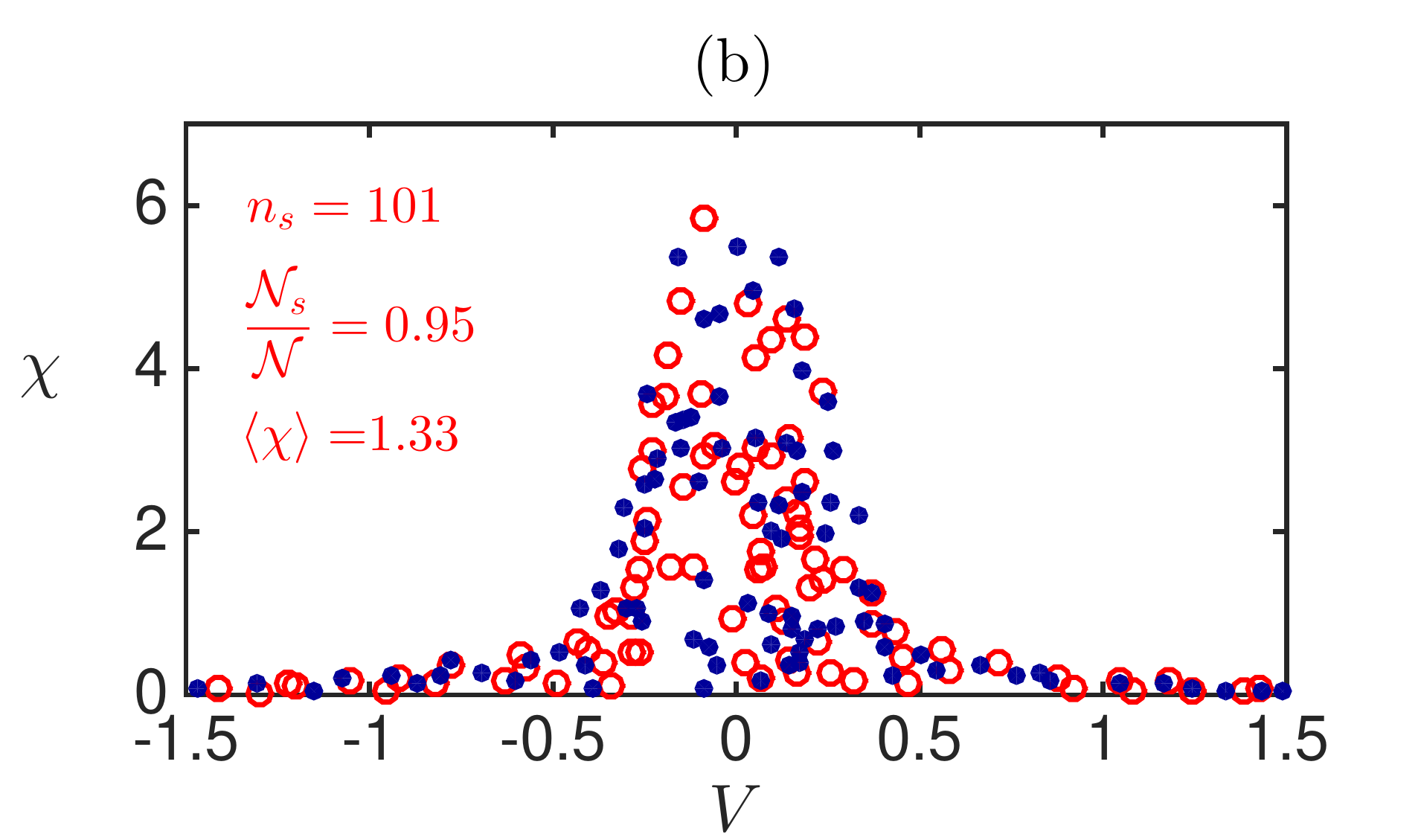}	
	
	\caption{\small Amplitude-velocity dot diagrams of solitons at the end of the growth stage for the experiments with (a) adiabatic regime with increased amplitude of the initial noise $A_{0}=6\times 10^{-2}$, see red lines in Fig.~\ref{fig:fig3}(g-i), and (b) non-adiabatic regime with increased pumping coefficient $p_{0}=32\times 10^{-5}$, see red lines in Fig.~\ref{fig:fig3}(a-c) and Fig.~\ref{fig:fig6}. 
	Blue dots and red circles represent solitons for two different realizations of the grown-up wavefield. 
	For the realizations illustrated with red circles, $n_{s}$ shows total number of solitons detected, $\mathcal{N}_{s}/\mathcal{N}$ demonstrates ratio between the wave action of the solitonic part $\mathcal{N}_{s}$, see Eq.~(\ref{Is1}), and the total wave action $\mathcal{N}$, and $\langle\chi\rangle$ indicates the mean soliton amplitude.	 
	}
\label{fig:fig12}
\end{figure}


\subsection{Deviations from the universal adiabatic regime}
\label{Sec:Results2:C}

Although in the present paper we focus on the universal adiabatic regime of the turbulence growth, whose trajectory through the statistically stationary states of the integrable 1D-NLSE does not change when starting from a smaller initial noise amplitude $A_{0}$, in this subsection we briefly outline the main distinctions that we observe analyzing the scattering data for the experiments deviating from this universal adiabatic regime.

First, while the wavefield still tends to a soliton gas in these experiments and the fraction of solitons reaches $95$-$99$\% of the total wave action at the end of the growth stage $A_{f}=1$, the distribution of soliton amplitudes becomes significantly wider. 
In particular, for the adiabatic experiment with increased initial noise amplitude $A_{0}=6\times 10^{-2}$ and the non-adiabatic experiment with increased pumping coefficient $p_{0}=32\times 10^{-5}$ (all the other parameters are the same as for the base experiment), the maximum soliton amplitude at the end of the growth stage reaches $\max\chi \simeq 6$, see Fig.~\ref{fig:fig12}, versus $3.5$ for the base experiment with the universal adiabatic regime, see Fig.~\ref{fig:fig10}(a). 
The total number of solitons in Fig.~\ref{fig:fig12} stays roughly the same at $95$-$105$, so that the soliton density and mean amplitude practically do not change compared to the base experiment. 
For the (non-universal) adiabatic regime depending on $A_{0}$, see Fig.~\ref{fig:fig12}(a), the increased deviations between soliton amplitudes should be the result of the presence of solitons in the initial noise, since these solitons are affected by pumping from the very beginning of the growth stage, in contrast to the rest of the solitons, which have not yet emerged. 
Concerning the non-adiabatic regime in Fig.~\ref{fig:fig12}(b), we think that solitons are affected by self-organization, which is known to occur in non-integrable and dissipative systems~\cite{jordan2000self,dudley2006supercontinuum,laurie2012one}, when solitons collide inelastically and the larger solitons grow while the smaller ones decay.

Second, even at the end of the growth stage $A_{f}=1$ of the two experiments shown in Fig.~\ref{fig:fig12}, large solitons still have noticeable velocities; hence, the wavefields for these experiments are almost purely solitonic states, which are still relatively far from being bound. 
Note that, at the same average amplitude $A_{f}$, the universal adiabatic regime produces practically a bound state of solitons: see Figs.~\ref{fig:fig08}(g),~\ref{fig:fig10} and also Fig.~\ref{fig:fig14} in Appendix~\ref{Sec:Annex-super-Gaussian} for the super-Gaussian initial spectrum $s=32$. 

Third, for both experiments shown in Fig.~\ref{fig:fig12}, different realizations of the grown-up wavefield exhibit rather different amplitude-velocity dot diagrams of their solitonic content. 
This difference explains large deviations in values of the higher-order invariants~(\ref{integrals_rec1})-(\ref{integrals_rec2}) of the 1D-NLSE between different realizations of the grown-up wavefield, which have been discussed in Section~\ref{Sec:Results1:D}.

Note that, in terms of the presented IST analysis, the non-universal adiabatic regime, Fig.~\ref{fig:fig12}(a), and the non-adiabatic regime, Fig.~\ref{fig:fig12}(b), do not differ qualitatively from each other. 
The difference may appear if we consider correlation between soliton phases within the wavefield, which we think is absent for the adiabatic regime and present for the non-adiabatic regime; we will study this question in more detail in a separate publication. 


\section{Conclusions and discussions}
\label{Sec:Conclusions}

We have studied numerically the integrable turbulence in the framework of the 1D-NLSE model using a new approach called ``growing of turbulence''. 
In this approach, a small linear pumping term is added to the equation and the evolution is started from statistically homogeneous Gaussian noise of small amplitude. 
After reaching a certain level of average intensity, the pumping is switched off and the resulting integrable turbulence is examined. 

For sufficiently small initial noise and pumping coefficient, and also for not very wide simulation box (basin length), we have found that the turbulence grows in a universal adiabatic regime. 
In this regime, when we switch off the pumping, the resulting integrable turbulence turns out to be stationary and does not depend on the pumping coefficient, amplitude of the initial noise or basing length. 
Focusing on this regime and turning off the pumping at different moments of time, we have studied the whole family of novel statistically stationary states, which lie on the trajectory of adiabatic growth from weak nonlinearity through the states of intermediate nonlinearity to strongly nonlinear states. 
Analyzing these states, we have observed how the wave statistics changes from Gaussian when the nonlinearity is weak towards essentially non-Gaussian with a strong presence of rogue waves when the nonlinearity becomes strong. 
Using the IST method to monitor this evolution, we have found that the solitonic part of the wavefield becomes dominant even when the (linear) dispersion effects are still leading in the dynamics and with increasing average intensity the wavefield approaches a dense bound-state soliton gas, whose properties are defined by the Fourier spectrum of the initial noise. 
Regimes deviating from the universal adiabatic growth also lead to solitonic states, but solitons in these states have noticeably different velocities and a significantly wider distribution by amplitude, while the wave statistics indicates a much more frequent appearance of very large waves.

Note that, in our formulation of the problem~(\ref{NLSE-1})-(\ref{NLSE-2}), we have suggested conditions~(\ref{puming-adiabatic-L}),~(\ref{puming-adiabatic-k}) for adiabatic growth of turbulence analytically from two requirements: (i) the pumping term must be small compared to the dispersion and nonlinearity terms of the 1D-NLSE and (ii) the initial state must be very weakly nonlinear and, therefore, practically stationary. 
In Section~\ref{Sec:Results1}, we have checked numerically that when these conditions are satisfied the turbulence indeed grows adiabatically, and if not, then a non-adiabatic regime is realized. 
In particular, according to condition~(\ref{puming-adiabatic-k}), the adiabatic regime is very difficult to implement for large basins, since the upper bound on the pumping coefficient decreases as inverse square of the basin length. 
In such cases, a non-adiabatic regime may be realized, which is characterized by a stronger presence of rogue waves with their frequency increasing with the pumping coefficient or basin length. 
We also have found that, for a sufficiently small initial noise amplitude, the adiabatic regime turns out to be universal, as its growth trajectory does not depend on the noise amplitude. 
For such a regime, apparently, an additional condition~(\ref{no-solitons-in-noise}) is needed for the absence of solitons in the initial noise. 

In the universal adiabatic regime, different realizations of the grown-up wavefield demonstrate a very close values of the 1D-NLSE invariants~(\ref{integrals_rec1})-(\ref{integrals_rec2}). 
This is rather surprising, since, as we have discussed in Section~\ref{Sec:TheProblem:C}, evolution equations for the higher-order invariants at the pumping stage contain functions dependent on specific wavefield; see e.g. Eq.~(\ref{energy-1-t}) for the evolution of total energy, which contains potential energy. 
The fact that, in the adiabatic regime, the invariants of different realizations actually coincide at all stages of the pumping apparently means that, during the time evolution, such wavefield-dependent functions are effectively averaged over time to identical time-dependent functions and lead to the same corrections to the evolving invariants. 
If true, this observation allows one to consider the conditions for adiabatic regime from another point of view, namely, that the time necessary for such an averaging must be much smaller than the characteristic pumping time. 
In terms of the analogy with an ideal gas in a box, to which one molecule is added from time to time, this condition is equivalent to the time interval between additions being sufficient for the system to reach a modified thermodynamic equilibrium. 
In turn, this allows to understand the self-similarity of the adiabatic regime with time $2p_{0}(t-t_{0})$ discussed in Section~\ref{Sec:Results1:D}: indeed, if this time interval is sufficient, then a twice larger interval (equivalent to twice smaller pumping coefficient) will lead to the same equilibriums and growth trajectory.

In our experiments, we have observed that, as the average intensity increases in the universal adiabatic regime, the wavefield tends to a bound-state soliton gas, in which solitons have zero velocities. 
Moreover, different realizations of the grown-up wavefield show (i) almost identical sets of soliton amplitudes and velocities and (ii) very similar statistical distributions of soliton positions and phases, see Fig.~\ref{fig:fig10}. 
Hence, we can suggest that the universal adiabatic regime always grows the same soliton gas, which is defined by the Fourier spectrum of the initial noise and does not depend on specific realization of the Fourier phases $\phi_{k}$ in Eq.~(\ref{IC}). 

The IST analysis of the non-universal adiabatic regime (which depends on the noise amplitude $A_{0}$) and non-adiabatic regime reveals coinciding patterns: for both such regimes, soliton amplitudes have a significantly wider distribution by amplitude, soliton velocities are noticeable, and different realizations of the grown-up wavefield show rather different sets of soliton amplitudes and velocities. 
We think that these distinctions from the universal adiabatic regime are caused by (i) the presence of solitons in the initial noise for the non-universal adiabatic regime and (ii) the process of self-organization of solitons for the non-adiabatic regime. 
A more comprehensive analysis of this problem can be done in the future within the IST perturbation theory~\cite{KivsharRMP1989,gerdjikov1996asymptotic,gerdjikov2001adiabatic}, which allows one to predict the successive small changes in the soliton eigenvalues during the pumping stage.

Since the adiabatic regime differs from the non-adiabatic one in that the former evolves through the statistically stationary states of the integrable 1D-NLSE, and the latter through the non-stationary ones, we have suggested that the phase correlation between different solitons within the wavefield is indeed absent in the adiabatic regime and is present in the non-adiabatic one. 
To verify this hypothesis and exclude random coincidences of phases, we are going to explore in detail the time evolution of soliton parameters in a separate publication.


\begin{center}
\textbf{Acknowledgements}
\end{center}

Simulations have been performed at the Novosibirsk Supercomputer Center (NSU). 
The work of D.A., A.G. and V.Z. on the main part of the paper (Sections~\ref{Sec:TheProblem},~\ref{Sec:NumMethods},~\ref{Sec:Results1},~\ref{Sec:Results2:A},~\ref{Sec:Results2:B}) has been supported by the Russian Science Foundation (Grant No. 19-72-30028). 
A.G. thanks the support of the Russian Foundation for Basic Research (grant No. 19-31-60028) for the work on section \ref{Sec:Results2:C} and Appendix~\ref{Sec:Annex-super-Gaussian}.
The development of numerical IST methods by R.M. (see Appendix~\ref{Sec:Annex-IST-numerics}) has been supported by the state contract with IT SB RAS.


%


\appendix

\section{Theoretical formalism of the IST method}
\label{Sec:Annex-IST-theory}

In this Appendix, we outline the concept of scattering data and the IST method for localized wavefields. 

For the 1D-NLSE~(\ref{NLSE}), the scattering data is found with the direct scattering transform (DST) procedure based on the Zakharov--Shabat (ZS) system of equations~\cite{zakharov1972exact}. 
At a fixed moment of time, the latter represents the following auxiliary linear system for a vector wave function $\mathbf{\Phi}(x, \lambda) = (\phi_1,\phi_2)^\mathrm{T}$,
\begin{eqnarray}
	\mathbf{\Phi}_{x} &=& \frac{1}{\sqrt{2}}\begin{pmatrix}
	-i \lambda & \psi \\ -\psi^* & i \lambda
	\end{pmatrix}
	\mathbf{\Phi},
	\label{ZSsystem1}
\end{eqnarray}
where $\lambda = \xi + i \eta$ is a complex-valued spectral parameter (the eigenvalue) and the superscript $\mathrm{T}$ stands for the matrix transpose. 
Without loss of generality, we consider the spectral parameter in the upper half of the complex plane only, $\eta = \mathrm{Im}\,\lambda\ge 0$, see e.g.~\cite{novikov1984theory} for detail.

Similarly to quantum mechanics~\cite{landau1958quantum}, the scattering problem~(\ref{ZSsystem1}) for the wave function $\mathbf{\Phi}$ is introduced with the following asymptotics at infinity (the so-called ``right'' scattering problem, in contrast to the left scattering problem; see e.g.~\cite{LambBook1980}):
\begin{eqnarray}
	\lim_{x\to -\infty}\biggl\{ \mathbf{\Phi} &-& \begin{pmatrix} e^{-i \lambda x} \\ 0 \end{pmatrix}\biggr\} = 0, \label{ScatteringProblem1}\\
	\lim_{x \to +\infty}\biggl\{\mathbf{\Phi} &-& \begin{pmatrix} a(\lambda)\, e^{-i \lambda x} \\ b(\lambda)\, e^{i \lambda x} \end{pmatrix}\biggr\} = 0. \label{ScatteringProblem2}
\end{eqnarray}
In this problem, the wavefield $\psi(x)$ of the 1D-NLSE is considered as a potential for the scattering wave $\mathbf{\Phi}$, while $a(\lambda)$ and $b(\lambda)$ represent scattering coefficients. 
Its bounded solutions exist for real-valued spectral parameter, $\lambda = \xi\in\mathbb{R}$, and also for complex-valued $\lambda$, $\eta=\mathrm{Im}\,\lambda>0$, if and only if $a = 0$.

In the present paper, we consider only the wavefields with compact support, i.e., non-zero in a finite region of space. 
Then, the eigenvalues $\lambda$ of the ZS system are presented by a finite number of discrete points $\lambda_j$ (discrete spectrum) with $\eta_j=\mathrm{Im}\,\lambda_j > 0$, $j = 1,...,n_{s}$, and the real line $\lambda=\xi\in\mathbb{R}$ (continuous spectrum), see~\cite{novikov1984theory}. 
Using the scattering coefficients $a$ and $b$ defined at $\lambda_j$ and $\lambda=\xi\in\mathbb{R}$, we can construct functions $a(\lambda)$ and $b(\lambda)$ as analytic continuations to the upper half of the $\lambda$-plane~\cite{faddeev2007hamiltonian}. 
Then, the function $a(\lambda)$ has simple zeros at the eigenvalue points, $a(\lambda_n)=0$ (we do not consider the degenerate case when an eigenvalue point represents a multiple zero). 

The full set of the scattering data is a combination of the discrete $\{\lambda_j, \rho_j \}$ and continuous $\{ r \}$ spectra,
\begin{eqnarray}
	&& \bigg\{
	\lambda_j \,\,|\,\, a(\lambda_j) = 0, \,\, \mathrm{Im}\,\lambda_j > 0
	\bigg\}, \nonumber\\
	&& \rho_j = \frac{b(\lambda_j)}{a'(\lambda_j)}, \quad
	r(\xi) = \frac{b(\xi)}{a(\xi)}, \label{ScatteringData}
\end{eqnarray}
where $a'(\lambda)$ is complex derivative of $a(\lambda)$ with respect to $\lambda$, $\rho_j$ are the so-called norming constants associated with the eigenvalues $\lambda_j$, and $r(\xi)$ is the reflection coefficient defined at the real line $\xi\in\mathbb{R}$. 
Most importantly, if the potential $\psi$ is governed by the 1D-NLSE~(\ref{NLSE}), then the time evolution of the scattering data~(\ref{ScatteringData}) is trivial,
\begin{eqnarray}\nonumber
	&& \forall j: \lambda_j=\mathrm{const},
	\\\label{ScattData(t)}
	&& \rho_j (t) = \rho_j (0) e^{2i \lambda_j^2 t},
	\\\nonumber
	&& r(\xi,t) = r(\xi,0)e^{2i \xi^2 t},
\end{eqnarray}
and the wavefield $\psi$ can be recovered from it with the inverse scattering transform (IST) by solving the integral GLM equations~\cite{novikov1984theory}. 
Note that, in the general case, this can only be done numerically, asymptotically at large time, or in the semi-classical approximation~\cite{lewis1985semiclassical,jenkins2014semiclassical}.

The reflection coefficient $r(\xi)$ corresponds to nonlinear dispersive waves, while the discrete eigenvalues $\lambda_j$ together with the norming constants $\rho_j$ represent solitons. 
In particular, for $j = 1,...,n_{s}$, the eigenvalues $\lambda_j = \xi_j + i \eta_j$ contain information about soliton amplitudes $\chi_j = 2\eta_j$ and velocities $V_j = -2\xi_j$, while the soliton norming constants are connected to their positions in space $x_{j}'\in\mathbb{R}$ and phases $\theta_{j}'\in[0,2\pi)$.

In the present paper, we also use soliton norming constants corresponding to the formalism alternative to the IST procedure -- the so-called dressing method (DM)~\cite{novikov1984theory}, also known as the Darboux transformation~\cite{akhmediev1991extremely,matveev1991darboux}. 
The DM norming constants are more favorable for the scattering data analysis, see ~\cite{gelash2018strongly,gelash2019bound,gelash2021solitonic}. 
For the reflectionless case $r(\xi)=0$, the nonlinear dispersive waves are absent and both the IST and DM procedures can be performed analytically, leading to an exact $n_{s}$-soliton solution ($n_{s}$-SS) $\psi_{n_{s}}(x,t)$. 
The DM norming constants $C_j$ are then related to the IST norming constants $\rho_j$ as follows (see~\cite{aref2016control,gelash2020anomalous}),
\begin{eqnarray}
	C_j(t) = \frac{1}{\rho_j(t)} \prod_{m=1}^{n_{s}} (\lambda_j - \lambda_m^*) \times \prod_{l \ne j}^{n_{s}} \frac{1}{\lambda_j - \lambda_l}. \label{rhok_Ck_conn}
\end{eqnarray}
For an $n_{s}$-SS, the dressing method represents a pure algebraic recursive procedure~\cite{novikov1984theory,zakharov1978relativistically,matveev1991darboux}, with the outcome which can be written as ratio of two determinants,
\begin{eqnarray}
	\psi_{n_{s}}(x,t) = -2i\frac{\mathrm{det} \widetilde{\mathbf{M}}}{\mathrm{det} \mathbf{M}}, \quad
	M_{ml}=\frac{(\mathbf{q}_{m}\cdot \mathbf{q}^*_{l})}{\lambda_{m} - \lambda^*_l}, \nonumber\\
	\mathbf{\widetilde{M}}=
	\left(\begin{array}{cc}
	        0 & \begin{array}{ccc}
	              q_{1,2} & \cdots & q_{N,2}
	            \end{array}
	         \\
	        \begin{array}{c}
	          q^*_{1,1} \\
	          \vdots \\
	          q^*_{N,1}
	        \end{array}
	         &  \begin{array}{c}
	              \mathbf{M}^{T}
	            \end{array}
	      \end{array}\right),
	\label{Ndet_SS}
\end{eqnarray}
where $\mathbf{q}_j$ are two-component vectors,
\begin{eqnarray}\nonumber
	\mathbf{q}_{j}(x,t) = \left(\begin{array}{c} q_{j,1} \\ q_{j,2} \end{array}\right) =
	\left(\begin{array}{c} C_j^{-1/2}\, e^{i\lambda_j x} \\ C_j^{1/2}\, e^{-i\lambda_j x} \end{array}\right),
\end{eqnarray}
and the dot in $(\mathbf{q}_{m}\cdot \mathbf{q}^*_{l}) = q_{m,1}q^*_{l,1} + q_{m,2}q^*_{l,2}$ corresponds to the real-symmetric vector scalar product.

Within the DM formalism, the norming constants are related to the soliton positions and phases as
\begin{eqnarray}
	C_j = -\exp\bigg[2i\lambda_{j}x'_{j} + i\theta_{j}'\bigg], \label{Ck_param}
\end{eqnarray}
and evolve with time as
\begin{eqnarray}
	C_{j}(t) = C_{j}(0) e^{-2i\lambda^2_j t}. \label{DM-norming-constants-evolution}
\end{eqnarray}
The parameters $x_{j}'$ and $\theta_{j}'$ equal the observed in the physical space position and phase of a soliton only for the one-soliton solution~(\ref{1-SS}). 
In presence of other solitons or dispersive waves, the observed position and phase may differ considerably from $x_{j}'$ and $\theta_{j}'$.

Finally, we note that, for a pure $n_{s}$-SS, the IST formalism allows one to find the 1D-NLSE invariants~(\ref{integrals_rec1})-(\ref{integrals_rec2}) straightforwardly via the soliton eigenvalues~\cite{novikov1984theory}, 
\begin{eqnarray}\label{integrals}
	\mathcal{I}_{m} = \frac{(2i)^{m}}{mL}\sum_{j=1}^{n_{s}} [(\lambda_j^m)^* - \lambda_j^m].
\end{eqnarray}


\begin{figure}[!t]\centering
	\includegraphics[width=8.9cm]{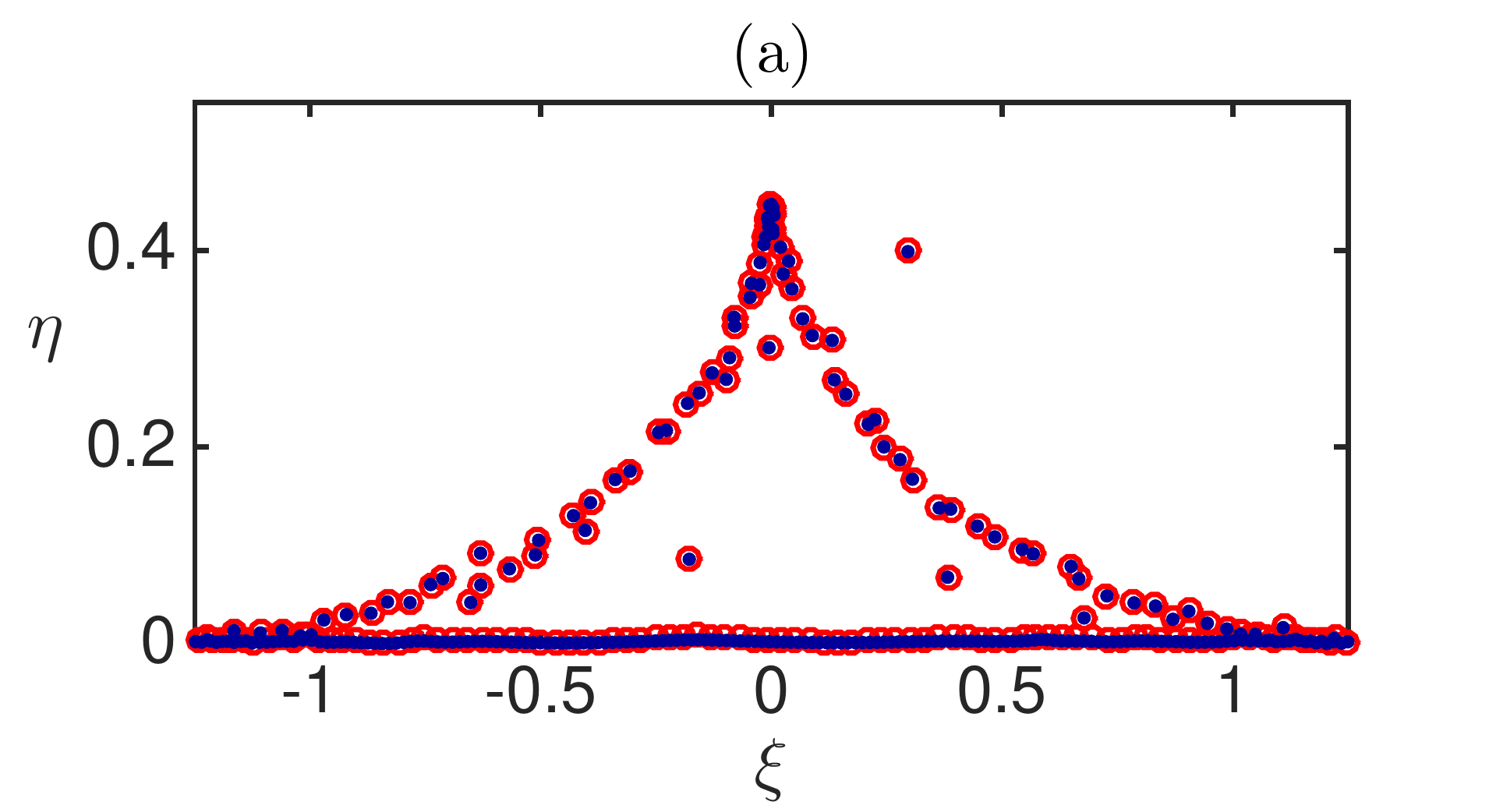}\\
	\includegraphics[width=8.9cm]{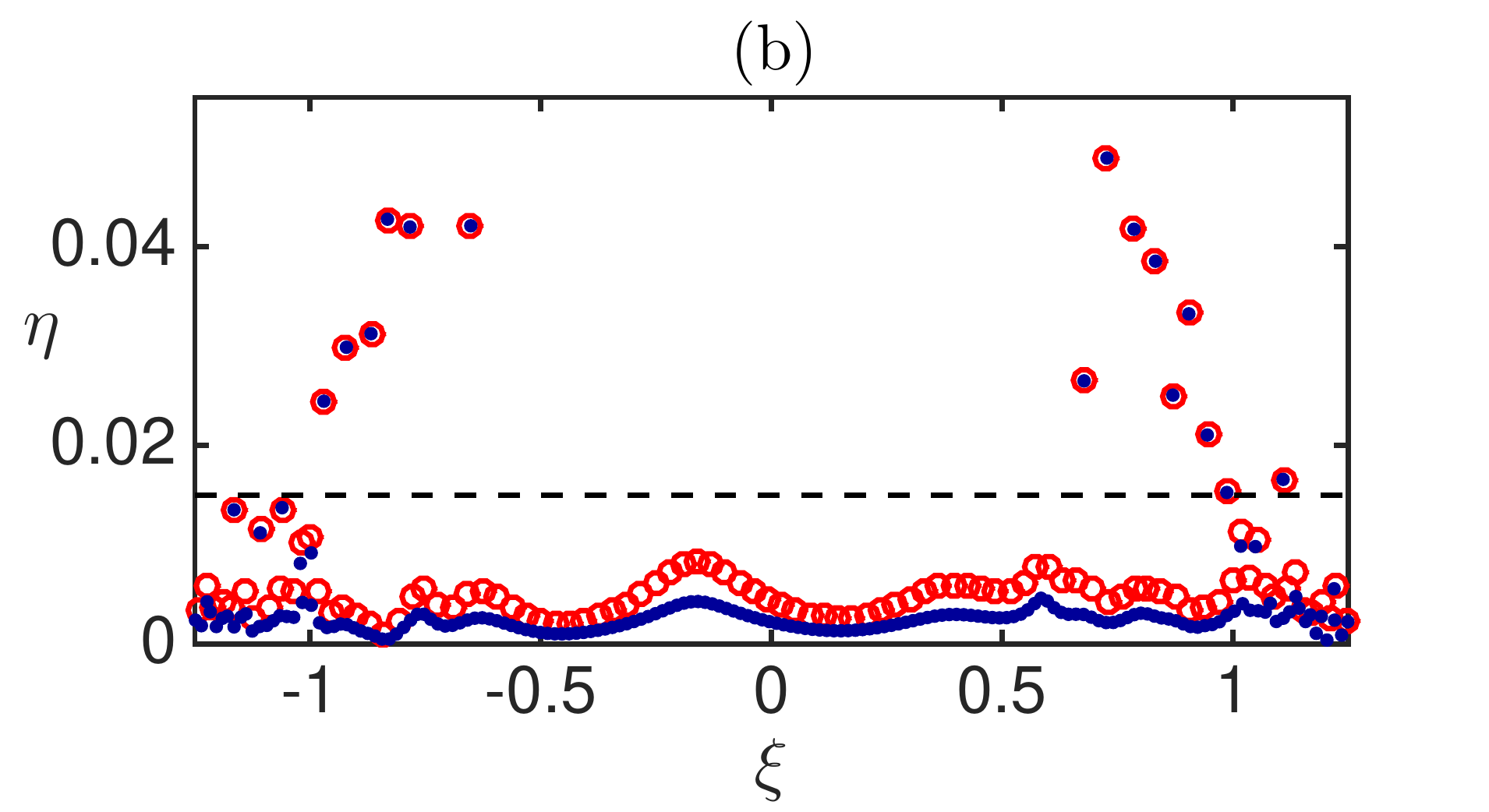}
	
	\caption{\small Eigenvalues $\lambda = \xi + i\eta$ of the ZS system~(\ref{ZSsystem1}) computed with the FC method in the regions $x\in[-3L/4, 3L/4]$ (red circles) and $x\in [-L, L]$ (blue points): (a) full eigenvalue spectrum for a single realization of wavefield grown with parameters $L=128\pi$, $s=2$, $A_{0}=10^{-2}$, $A_{f}=0.5$ and $p_{0}=10^{-5}$ (the same wavefield, as in Fig.\ref{fig:fig08}(e)), and (b) zoom of panel (a) at small values of $\eta$. 
	The eigenvalues that coincide in calculations over the two different regions $[-3L/4, 3L/4]$ and $[-L, L]$ correspond to solitons, while the other eigenvalues represent the continuous spectrum artificially shifted to the upper half-plane as a result of wavefield periodization. 
	Black dashed line in panel (b) indicates threshold above which we assume the eigenvalues as belonging to the discrete spectrum. 
	}
\label{fig:fig13}
\end{figure}

\begin{figure*}[!t]\centering
	\includegraphics[width=17.8cm]{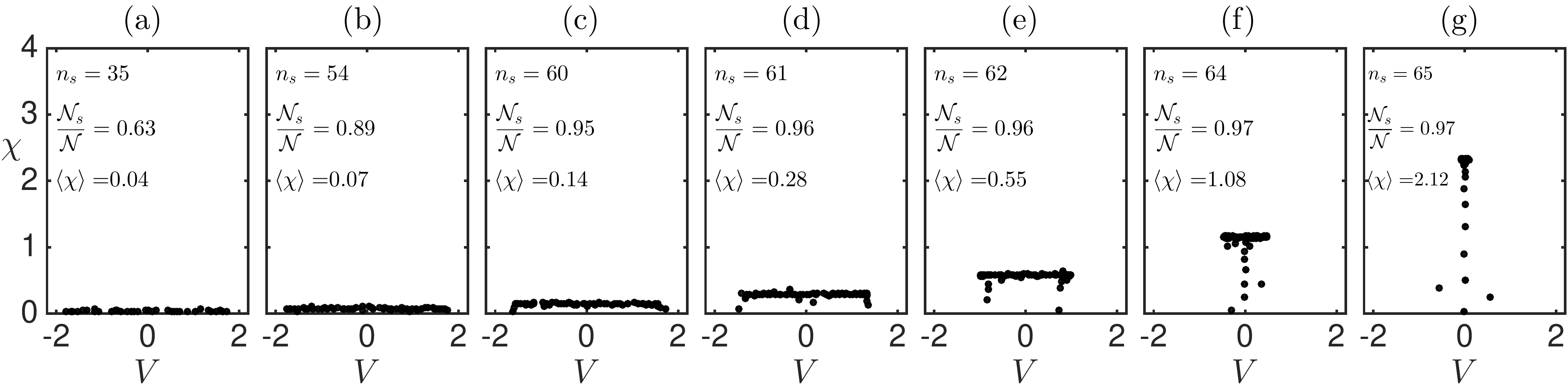}
	
	\caption{\small 	Amplitude-velocity dot diagrams of solitons for simulation from a single realization of initial conditions at different stages of the pumping: (a) $A_{f}=0.125$, (b) $A_{f}=0.177$, (c) $A_{f}=0.25$, (d) $A_{f}=0.35$, (e) $A_{f}=0.5$, (f) $A_{f}=0.71$ and (g) $A_{f}=1$. 
	The simulation parameters are: $L=128\pi$, $s=32$, $A_{0}=10^{-2}$ and $p_{0}=10^{-5}$. 
	Each dot represents a soliton within the studied wavefield, $n_{s}$ shows total number of solitons detected, $\mathcal{N}_{s}/\mathcal{N}$ demonstrates ratio between the wave action of the solitonic part $\mathcal{N}_{s}$, see Eq.~(\ref{Is1}), and the total wave action $\mathcal{N}$, and $\langle\chi\rangle$ indicates the mean soliton amplitude.
	}
\label{fig:fig14}
\end{figure*}

\begin{figure*}[!t]\centering
	\includegraphics[width=17.8cm]{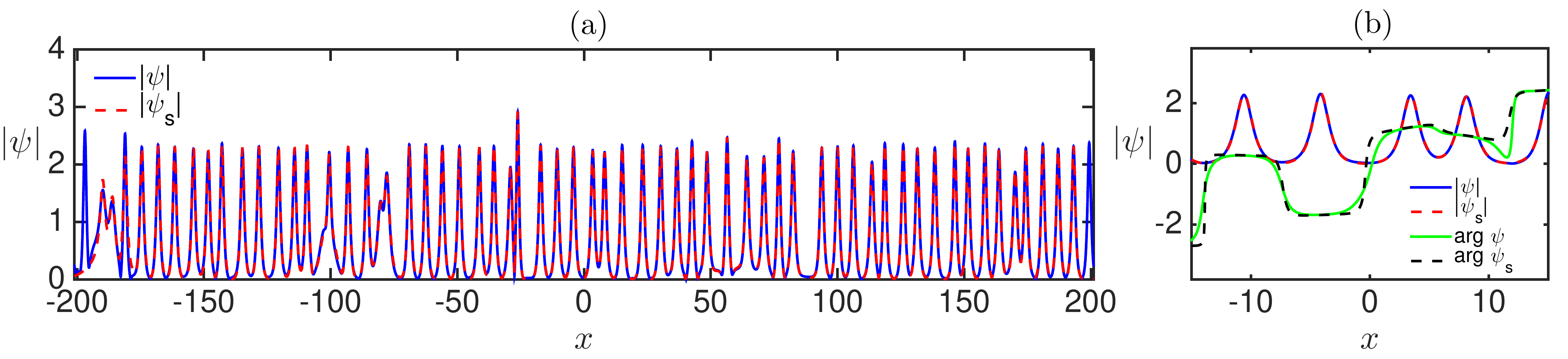}
	
	\caption{\small Typical grown-up wavefield and its approximation with the multi-soliton solution $\psi_{s}$ generated from the discrete part of the scattering data, for the numerical experiment with parameters $L=128\pi$, $s=32$, $A_{0}=10^{-2}$, $A_{f}=1$ and $p_{0}=10^{-5}$. 
	Panel (a) shows absolute value $|\psi|$ of the grown-up wavefield (solid blue) and multi-soliton solution (dashed red), while panel (b) represents zoom of panel (a) at the central region. 
	Also, panel (b) illustrates complex phase $\mathrm{arg}\,\psi$ of the grown-up wavefield (solid green) and multi-soliton solution (dashed black).
	}
\label{fig:fig15}
\end{figure*}


\section{Numerical approaches to the IST method}
\label{Sec:Annex-IST-numerics}

We compute the discrete spectrum~(\ref{ScatteringData}) numerically using the standard DST algorithms~\cite{BofOsb1992, Burtsev1998,YangBook2010} supplemented by our latest studies~\cite{mullyadzhanov2019direct,gelash2020anomalous,mullyadzhanov2021magnus}. 
Although the soliton eigenvalues can be found with many variations of the DST methods~\cite{YangBook2010, turitsyn2017nonlinear,vasylchenkova2019direct}, calculation of the norming constants is a challenging problem leading to several types of numerical instabilities, including the so-called anomalous errors~\cite{gelash2020anomalous}. 
Here we use the following scheme, which guarantees an accurate identification of the full discrete spectrum.

First, we endow a wavefield with a compact support, assuming it to be zero outside the simulation box $[-L/2,L/2]$. 
Due to such a formulation, a few solitons located at the box edges split into two parts and introduce an inaccuracy in the scattering data. 
The dissected solitons affect the continuous spectrum and may lead to the appearance of artificial solitons smaller than them. 
However, the total number of solitons in each of our wavefields is relatively large, $n_{s}\gtrsim 50$, and we neglect this error. 
Also, to mitigate the discontinuity at $x = \pm L/2$, we smooth the wavefield at the box edges. 
The smoothing window is comparable with the characteristic width of solitons, i.e., it is of the same order as region where solitons are dissected by the edge splitting. 
This means that the smoothing affects accuracy in a similar way as endowing wavefield with a compact support, that allows us to neglect the corresponding inaccuracy as well.

Second, we compute soliton eigenvalues with the standard Fourier collocation (FC) method~\cite{YangBook2010}, which is fast and does not require more than $\simeq 2^{13}$ sampling points in our data processing. 
The FC method uses Fourier decomposition of wavefield, which leads to an artificial shift of the continuous spectrum eigenvalues to the upper half-plane as a result of the wavefield periodization. 
Additionally, the FC method does not distinguish between the eigenvalues of discrete and continuous spectra, leading to the problem of identifying low-amplitude solitons. 

Third, to cope with this problem, we consider wavefield in two different regions. 
The first region $[-3L/4, 3L/4]$ is constructed by adding zeros $\psi=0$ in the intervals $[-3L/4, -L/2]$ and $[L/2, 3L/4]$. 
The second region $[-L, L]$ is obtained similarly, only the width of the added intervals is larger. 
Then, we execute the FC method over both regions and select those eigenvalues as belonging to the discrete spectrum, which coincide in these calculations, see Fig.~\ref{fig:fig13}. 
To optimize computational resources, we consider several realizations of the grown-up wavefield for each numerical experiment and determine a threshold $\eta_{th}$, above which all eigenvalues belong to the discrete spectrum; in Fig.~\ref{fig:fig13} it is at $\eta_{th} = 0.015$. 
For subsequent realizations, we no longer check the eigenvalues using two different regions above, but simply ignore all eigenvalues below the threshold and assume all eigenvalues above as belonging to the discrete spectrum. 

The FC method provides a good approximation of the eigenvalues of the ZS system, i.e., zeros of the coefficient $a(\lambda)$. 
However, it has fundamental limitations due to the use of the Fourier approximation for a spatially localized wavefield with a compact support. 
To mitigate the anomalous DST errors~\cite{gelash2020anomalous}, the subsequent steps of our procedure require knowledge of roots $a(\lambda_n)=0$ to hundreds of digits. 
The latter is not possible within the FC method, so that we use the calculated eigenvalues as seeding values for a high-accuracy find-root procedure.

Thus, at the fourth step, we perform the main DST procedure using the standard second-order Boffetta--Osborne (BO) method~\cite{BofOsb1992}, supplemented by high-precision arithmetic operations executed on a fine grid, as suggested in~\cite{mullyadzhanov2019direct, gelash2020anomalous}. 
For high-precision operations, we use the {\it Wolfram Mathematica} software. 
High precision and a fine grid allow us to avoid (i) round-off errors when calculating the wave function $\mathbf{\Phi}$ for the ZS system, (ii) anomalous errors in the calculation of the norming constants, and (iii) numerical instability of the wave scattering through a large potential; see~\cite{mullyadzhanov2019direct,gelash2020anomalous} for detail. 
With the BO method, we can find $a(\lambda)$ and $b(\lambda)$ for any value of $\lambda$ by discrete integration of the ZS system on the interval $[-L/2,L/2]$ and with boundary condition~(\ref{ScatteringProblem1}). 
We run the BO method together with the Newton method, using the eigenvalues obtained by the FC method as seeding values, and find the set of zeros $\lambda_j$ of the function $a(\lambda)$ with the requested precision.

Finally, we compute the soliton norming constants according to their definition given in Eq.~(\ref{ScatteringData}). 
Note that instead of calculating the derivative $a'(\lambda_{j})$ straightforwardly, we find it from the extended scattering matrix of the BO method, see~\cite{BofOsb1992}. 

Our grid for the BO method consists of $5 \times 10^4$ points, obtained using interpolation of the grown-up wavefield with the standard built-in procedures of the {\it Wolfram Mathematica} software. 
Numerical precision of all our arithmetic operations, including the accuracy goal for the root-finding, corresponds to $700$ digits. 
We have checked that reducing the grid to $4\times 10^4$ points, as well as reducing the accuracy to $500$ digits, does not change our results for the soliton norming constants.


\section{Universal adiabatic regime with super-Gaussian initial spectrum}
\label{Sec:Annex-super-Gaussian}

Repeating our IST analysis for the experiment with super-Gaussian initial spectrum $s=32$, see Eq.~(\ref{IC}), and all other parameters the same as for the base experiment $L=128\pi$, $A_{0}=10^{-2}$, $A_{f}=1$ and $p_{0}=10^{-5}$, we come to all the same conclusions as has been discussed in Section~\ref{Sec:Results2:B} for the universal adiabatic regime of the turbulence growth with only one exception. 
Namely, as demonstrated in Fig.~\ref{fig:fig14}, the amplitude-velocity dot diagrams indicate a different distribution of solitons in amplitude and velocity.

First, the total number of solitons $n_{s}$ evolves differently during the growth stage. 
In particular, at $A_{f}\le 0.1$ we do not detect solitons within the wavefield, at $A_{f}=0.125$ we find already $n_{s}=35$ of them, at $A_{f}=0.177$ this number increases to $54$ and then stays at $60$-$65$ for all later stages, see Fig.~\ref{fig:fig14}. 
As shown in Fig.~\ref{fig:fig08}, the process of appearance of solitons for the $s=2$ case is smoother and longer. 
Different number of solitons at the end of the growth stage leads to different mean soliton amplitudes, $\langle\chi\rangle = 2.1$ versus $1.5$, and different soliton densities, $\rho = 0.16$ versus $0.24$, for the experiments $s=32$ and $s=2$ respectively.

Second, in contrast to the $s=2$ case where solitons are fairly densely distributed in amplitude from zero to maximum amplitude, almost all solitons for the $s=32$ experiment have practically identical amplitude. 
For the $s=2$ case, starting from sufficiently large average amplitude $A_{f}\ge 0.5$, large solitons form a bound state with practically zero velocities, see Fig.~\ref{fig:fig08}, while for the $s=32$ experiment we observe that solitons are approximately uniformly distributed in the velocity interval, which shrinks towards zero velocity with increasing $A_{f}$, see Fig.~\ref{fig:fig14}. 

Thus, during the growth stage of the experiment $s=32$, the wavefield tends to a bound-state soliton gas which consists of identical solitons. 
The latter resembles the asymptotic stationary state that develops from the noise-induced modulational instability (MI) of a cnoidal wave, in the limit when this cnoidal wave represents a uniform lattice of large and narrow solitons which practically do not overlap with each other. 
In particular, a typical realization of wavefield at the end of the growth stage for the $s=32$ experiment, shown in Fig.~\ref{fig:fig15} together with its reconstruction with a pure multi-soliton solution, is quite similar to the wavefield demonstrated in Fig.~16 of~\cite{agafontsev2016integrable} for the MI of a cnoidal wave at long time. 
For the $s=32$ experiment, the PDF of relative wave intensity in Fig.~\ref{fig:fig4}(b) (red line) is also quite similar to the PDF reported in Fig.~21(a) of~\cite{agafontsev2016integrable}. 
However, the wave-action spectrum in Fig.~\ref{fig:fig4}(a) (red line) does not demonstrate the characteristic peaks at equally spaced wavenumbers, which mark the asymptotic stationary state developed from the MI of a cnoidal wave, see Fig.~19(a) in~\cite{agafontsev2016integrable}. 

\end{document}